\documentclass[a4paper,11pt]{article}
\pdfoutput=1 

\usepackage{jheppub} 
\usepackage[normalem]{ulem}
\usepackage[T1]{fontenc} 
\usepackage{mathtools}
\newcommand{\fDfDsCchi}{ -0.697(29)}
\newcommand{\fDfDsCCL}{ 0.003(19)}
\newcommand{\fDfDsCh}{ -0.0650(62)}
\newcommand{\fDfDsdof}{16}
\newcommand{\fDfDschidof}{1.17}
\newcommand{\fDfDspval}{0.28}

\newcommand{\xiCchi}{ -0.749(39)}
\newcommand{\xiCCL}{ 0.051(24)}
\newcommand{\xiCh}{ -0.048(12)}
\newcommand{\xidof}{16}
\newcommand{\xichidof}{0.56}
\newcommand{\xipval}{0.91}
\newcommand{\bagCchi}{ -0.087(26)}
\newcommand{\bagCCL}{ 0.062(15)}
\newcommand{\bagCh}{ 0.0246(66)}
\newcommand{\bagdof}{16}
\newcommand{\bagchidof}{0.87}
\newcommand{\bagpval}{0.61}
\newcommand{\fDfDsCs}{ 0.10(11)}

\newcommand{\xiCs}{ 0.09(15)}
\newcommand{\bagCs}{ 0.033(80)}
\newcommand{\fDsfDcentral}{1.1740}

\newcommand{\fDsfDstatint}{51}

\newcommand{\fDsfDsystotpint}{+68}

\newcommand{\fDsfDsystotmint}{-68}

\newcommand{\fBsfBcentral}{1.1949}

\newcommand{\fBsfBstatint}{60}

\newcommand{\fBsfBsysmH}{0.0056}

\newcommand{\fBsfBsystotpint}{+95}

\newcommand{\fBsfBsystotmint}{-175}

\newcommand{\xicentral}{1.1939}

\newcommand{\xistatint}{67}

\newcommand{\xisysmH}{0.0042}

\newcommand{\xisystotpint}{+95}

\newcommand{\xisystotmint}{-177}
\newcommand{\xidircentral}{1.1887}

\newcommand{\xidirstatint}{85}

\newcommand{\bagratcentral}{0.9984}

\newcommand{\bagratstatint}{45}

\newcommand{\bagratsystotpint}{+80}

\newcommand{\bagratsystotmint}{-63}

\newcommand{\CKMfDsfDexccentral}{0.1830}

\newcommand{\CKMfDsfDexcexperrint}{51}
\newcommand{\VcdVcsexccentral}{0.2148}

\newcommand{\VcdVcsexcexperrint}{60}
\newcommand{\VcdVcsexclaterrpint}{+12}
\newcommand{\VcdVcsexclaterrmint}{-12}

\newcommand{\CKMfDsfDinccentral}{0.1843}

\newcommand{\CKMfDsfDincexperrint}{48}
\newcommand{\VcdVcsinccentral}{0.2164}

\newcommand{\VcdVcsincexperrint}{57}
\newcommand{\VcdVcsinclaterrpint}{+12}
\newcommand{\VcdVcsinclaterrmint}{-12}

\newcommand{\CKMxicentral}{0.17026}

\newcommand{\CKMxiexperrint}{34}
\newcommand{\VtdVtscentral}{0.20329}

\newcommand{\VtdVtsexperrint}{41}
\newcommand{\VtdVtslaterrpint}{+162}
\newcommand{\VtdVtslaterrmint}{-301}

\usepackage{lineno}

\newcommand{\mc}[1]{\ensuremath{\mathcal{ #1 }}}

\newcommand{\abs}[1]{\left| #1 \right|}

\newcommand{\avg}[1]{\left< #1 \right>}

\renewcommand{\v}[1]{\ensuremath{\mathbf{ #1 }}}

\newcommand{\matrixel}[3]{\left< #1 \vphantom{#2#3} \right| #2 \left| #3 \vphantom{#1#2} \right>} 
\newcommand{\matel}[2]{\ensuremath{{M_{#1}^{#2}}}}

\title{SU(3)-breaking ratios for $D_{(s)}$ and $B_{(s)}$ mesons}

\author[a,d]{P. A. Boyle,}
\author[a]{L. Del Debbio,}
\author[b]{N. Garron,}
\author[c]{A. J{\"u}ttner,}
\author[d]{A. Soni,}
\author[a,e]{J. T. Tsang \footnote{Corresponding author.}}
\author[a,f]{and O. Witzel}

\collaboration{RBC and UKQCD Collaborations}

\affiliation[a]{Higgs Centre for Theoretical Physics, School of Physics \& Astronomy, University of Edinburgh, EH9 3FD, United Kingdom}
\affiliation[b]{Theoretical Physics Division, Department of Mathematical Sciences, University of Liverpool, Liverpool L69 3BX, United Kingdom}
\affiliation[c]{School of Physics and Astronomy, University of Southampton,  Southampton, SO17 1BJ, United Kingdom}
\affiliation[d]{Physics Department, Brookhaven National Laboratory, Upton, NY 11973, United States}
\affiliation[e]{CP3-Origins and IMADA, University of Southern Denmark, Campusvej 55, 5230 Odense M, Denmark}
\affiliation[f]{Department of Physics, University of Colorado Boulder, Boulder, CO 80309, United States}

\emailAdd{paboyle@ed.ac.uk}
\emailAdd{luigi.del.debbio@ed.ac.uk}
\emailAdd{nicolas.garron@liverpool.ac.uk}
\emailAdd{juettner@soton.ac.uk}
\emailAdd{adlersoni@gmail.com}
\emailAdd{tsang@imada.sdu.dk}
\emailAdd{oliver.witzel@colorado.edu}

\abstract{We present results for the $SU(3)$ breaking ratios of decay constants
  $f_{D_s}/f_D$ and $f_{B_s}/f_B$ and - for the first time with physical pion
  masses - the ratio of bag parameters $B_{B_s}/B_{B_d}$, as well as the ratio
  $\xi$, forming the ratio of the non-perturbative contributions to neutral
  $B_{(s)}$ meson mixing. Our results are based on Lattice QCD simulations with
  chirally symmetric 2+1 dynamical flavours of domain wall fermions. Eight
  ensembles at three different lattice spacings in the range $a = 0.11 -
  0.07\,\mathrm{fm}$ enter the analysis, two of which feature physical light
  quark masses.  Multiple heavy-quark masses are simulated ranging from below
  the charm quark mass to half the bottom-quark mass. The $SU(3)$ breaking
  ratios display a very benign heavy-mass behaviour allowing for extrapolation
  to the physical bottom-quark mass.

  The results in the continuum limit including all sources of systematic errors
  are $f_{D_s}/f_D = \fDsfDcentral(\fDsfDstatint)_\mathrm{stat}
  \left(^{\fDsfDsystotpint}_{\fDsfDsystotmint}\right)_\mathrm{sys}$,
  $f_{B_s}/f_B = \fBsfBcentral(\fBsfBstatint)_\mathrm{stat}
  \left(^{\fBsfBsystotpint}_{\fBsfBsystotmint}\right)_\mathrm{sys}$,
  $B_{B_s}/B_{B_d} = \bagratcentral(\bagratstatint)_\mathrm{stat}
  \left(^{\bagratsystotpint}_{\bagratsystotmint}\right)_\mathrm{sys}$ and $\xi =
  \xicentral(\xistatint)_\mathrm{stat}
  \left(^{\xisystotpint}_{\xisystotmint}\right)_\mathrm{sys}$. Combining these
  with experimentally measured values we extract the ratios of CKM matrix
  elements $\abs{V_{cd}/V_{cs}} = \VcdVcsinccentral \left( \VcdVcsincexperrint
  \right)_\mathrm{exp}
  \left(^{\VcdVcsinclaterrpint}_{\VcdVcsinclaterrmint}\right)_\mathrm{lat}$ and
  $\abs{V_{td}/V_{ts}} = \VtdVtscentral \left( \VtdVtsexperrint
  \right)_\mathrm{exp}
  \left(^{\VtdVtslaterrpint}_{\VtdVtslaterrmint}\right)_\mathrm{lat}$.
}

\begin{document} 
\maketitle
\flushbottom

\section{Introduction} \label{sec:Intro}
In the Standard Model (SM) one can parameterise the QCD contribution
to weak decays of charged pseudoscalar mesons (e.g.~$B^\pm$, $D^\pm$
and $D_s^\pm$) into a lepton and a neutrino via the leptonic decay
constants $f_{B^\pm}$ and $f_{D^\pm_{(s)}}$. Similarly the mass
difference between the two mass eigenstates of neutral mesons, which
mix under the weak interaction, (e.g.~$B^0-\bar{B}^0$ and
$B_s^0-\bar{B}_s^0$ mixing) can be parametrised in terms of Standard
Model free parameters and experimentally known quantities. Both these
parametrisations involve elements of the CKM
matrix~\cite{CKM_Cabibbo,CKM_Kobayashi}, which are not known a
priori. However, the structure of the SM constrains this matrix to be
unitary, so by independent precise determinations of the elements of
this matrix, its unitarity can be tested and hence tests of the SM
performed.

For charged pseudoscalar mesons $P$ with quark content $\bar{q}_2 q_1$
experiments measure the decay rates $\Gamma(P\to l\nu_l)$ which can be expressed
as
\begin{equation}
  \Gamma(P\to l\nu_l) = \abs{V_{q_2q_1}}^2 f^2_P
  \,\mathcal{K}_1 + \mc{O}(\alpha_{EM}).
  \label{eq:decayconstants}
\end{equation}
Here $\mathcal{K}_1$ are perturbatively known expressions, $V_{q_2 q_1}$ is the
relevant CKM matrix element and $f_P$ is the decay constant. When
electromagnetic effects are neglected (c.f.~equation \eqref{eq:decayconstants}),
the decay rate factorises and hence precise knowledge of the non-perturbative
quantity $f_P$ allows for an extraction of the CKM matrix element under
consideration.

These decay rates have been measured for $P^\pm=D_{(s)}^\pm$ and $B^\pm$ by
CLEO-c~\cite{Artuso:2005ym, Eisenstein:2008aa, Artuso:2007zg, Alexander:2009ux,
  Naik:2009tk, Ecklund:2007aa, Onyisi:2009th}, BaBar~\cite{delAmoSanchez:2010jg,
  Lees:2012ju}, Belle~\cite{Zupanc:2013byn, Kronenbitter:2015kls} and
BESIII~\cite{Ablikim:2013uvu, Ablikim:2018jun}. After accounting for the
perturbative contributions $\mc{K}_1$, we can identify the product of the
relevant CKM matrix element and the charged decay constants ($f_P$) as
summarised by the Particle Data Group (PDG)~\cite{Tanabashi:2018oca} leading to
the following global averages:
\begin{equation}
  \begin{aligned}
    \abs{V_{cd}} f_{D^+} &= 45.91(1.05)\,\mathrm{MeV} & \text{\cite{Artuso:2005ym, Eisenstein:2008aa, Ablikim:2013uvu}}\\
    \abs{V_{cs}} f_{D_s^+} &= 250.9(4.0)\,\mathrm{MeV} & \text{\cite{Artuso:2007zg, Alexander:2009ux, Zupanc:2013byn, Naik:2009tk, Ecklund:2007aa, Onyisi:2009th, delAmoSanchez:2010jg}}\\
    \abs{V_{ub}} f_{B^+} &= 0.77(7)\,\mathrm{MeV} & \text{\cite{Kronenbitter:2015kls, Lees:2012ju}}.
    \label{eq:expdecayconstants}
  \end{aligned}
\end{equation}
We note that the very recent result by BESIII~\cite{Ablikim:2018jun} quoting
$\abs{V_{cs}} f_{D_s^+} = 246.2(3.6)_\mathrm{stat}(3.5)_\mathrm{sys}$ is not
included in this average yet. Adding this into the average, by treating
\cite{Ablikim:2018jun} and the average presented in \cite{Tanabashi:2018oca} as
uncorrelated, we obtain
\begin{equation}
  \abs{V_{cs}} f_{D_s^+} = 249.1(3.2)\,\mathrm{MeV} \qquad
  \text{\cite{Artuso:2007zg, Alexander:2009ux, Zupanc:2013byn, Naik:2009tk,
      Ecklund:2007aa, Onyisi:2009th, delAmoSanchez:2010jg, Ablikim:2018jun}},
  \label{eq:expdecayconstantsBES}
\end{equation}
in full agreement with the PDG value, but with a slightly reduced error.

\begin{figure}
  \begin{center}
    \includegraphics[width=.48\textwidth]{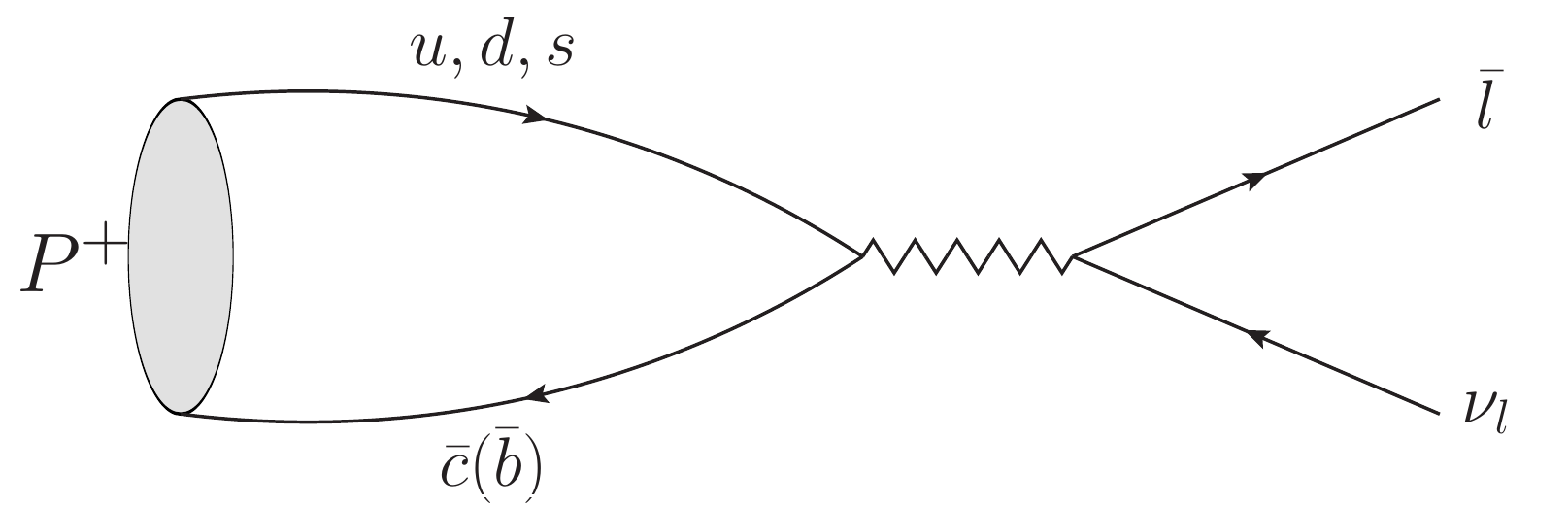}
    \includegraphics[width=.48\textwidth]{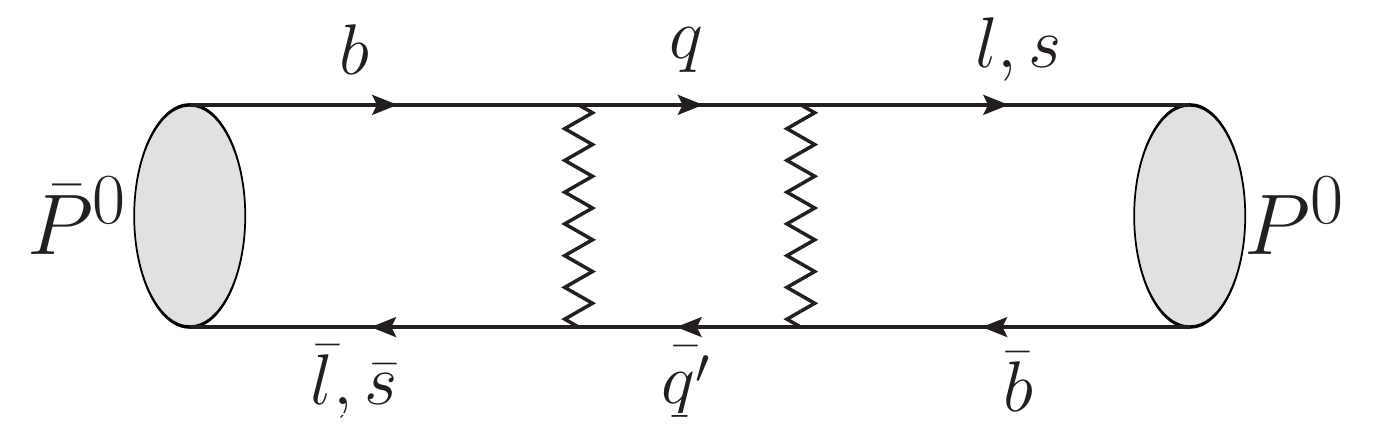}
  \end{center}
  \caption{\emph{Left}: Quark flow diagrams for the decay of a charged
    pseudoscalar meson. \emph{Right}: Quark flow diagram for neutral meson
    mixing. For the shown diagram (i.e.~for $P = B_{(s)}$), the quarks $q$ and
    $q'$ in the loop have charge +2/3 (i.e.~u,c,t).  For the case
    $D^0-\bar{D}^0$ mixing the bottom quark would be replaced by a charm quark
    and $q, q'$ would have charge -1/3.}
  \label{fig:corr_schematics}
\end{figure}

Similarly, the mass differences between the mass eigenstates of the
$B^0-\bar{B}^0$ and $B_s^0-\bar{B}_s^0$ systems can be measured to great
precision as oscillation frequencies. When considering the mixing of $B^0_{(s)}$
mesons in the SM, the right diagram in Figure \ref{fig:corr_schematics} is
dominated by top loops (i.e.~$q=q'=t$) and therefore by short distance
contributions. The SM prediction of the mass differences $\Delta m_d$ and
$\Delta m_s$ (for $P=B^0, B_s^0$, respectively) can again be expressed as a
function of known perturbative factors ($\mathcal{K}_2$), CKM matrix elements
and non-perturbative quantities such as decay constants $f_{P}$ and
renormalisation group invariant bag parameters $\hat{B}_{P}$, i.e.
\begin{equation}
  \Delta m_q =  \abs{V_{tq}^* V_{tb}}^2  \mathcal{K}_2 f_P^2 m_P \hat{B}_P.
  \label{eq:deltamdef}
\end{equation}
The mass difference $\Delta m_d$ has been measured by
ALEPH~\cite{Buskulic:1996qt}, BaBar~\cite{Aubert:2001te, Aubert:2002rg,
  Aubert:2001tf, Aubert:2002sh, Aubert:2005kf}, Belle~\cite{Hastings:2002ff,
  Zheng:2002jv, Abe:2004mz}, CDF~\cite{Abe:1997qf, Abe:1998sq, Abe:1999pv,
  Abe:1999ds, Affolder:1999cn}, D0~\cite{Abazov:2006qp},
DELPHI~\cite{Abreu:1997xq, Abdallah:2002mr}, OPAL~\cite{Ackerstaff:1997vd},
L3~\cite{Acciarri:1998pq}, LHCb~\cite{Aaij:2012nt, Aaij:2016fdk, Aaij:2011qx,
  Aaij:2013gja}, whilst $\Delta m_s$ has only been measured by
CDF~\cite{Abulencia:2006ze} and
LHCb~\cite{Aaij:2014zsa,Aaij:2011qx,Aaij:2013gja,Aaij:2013mpa}. The values for
both observables have been summarised and averaged in
Ref.~\cite{Tanabashi:2018oca} leading to the global averages
\begin{equation}
  \begin{aligned}
    \Delta m_d &= 0.5065(16)(11)\,\mathrm{ps}^{-1} &
    \text{\cite{Abdallah:2002mr, Abe:1997qf, Abe:1998sq, Abe:1999pv, Abe:1999ds,
        Affolder:1999cn, Abazov:2006qp, Buskulic:1996qt, Abreu:1997xq,
        Acciarri:1998pq, Ackerstaff:1997vd, Aaij:2011qx, Aaij:2012nt,
        Aaij:2013gja, Aaij:2016fdk, Hastings:2002ff, Zheng:2002jv, Abe:2004mz,
        Aubert:2001te, Aubert:2002rg, Aubert:2001tf, Aubert:2002sh,
        Aubert:2005kf}},\\ \Delta m_s &= 17.757(20)(07)\,\mathrm{ps}^{-1} &
    \text{\cite{Abulencia:2006ze, Aaij:2014zsa, Aaij:2011qx, Aaij:2013gja,
        Aaij:2013mpa}},
  \label{eq:deltamexp}
  \end{aligned}
\end{equation}
where the first error is statistical and the second systematical. Note that the
perturbative factor $\mc{K}_2$ in \eqref{eq:deltamdef} cancels in the ratio $\Delta
m_s/\Delta m_d$ leading to
\begin{equation}
  \frac{\Delta m_s}{\Delta m_d} = \abs{\frac{V_{ts}}{V_{td}}}^2
  \frac{m_{B^0_s}}{m_{B^0}} \frac{f^2_{B^0_s} \hat{B}_{B^0_s}}{f^2_{B^0} \hat{B}_{B^0}}.
  \label{eq:deltamratios}
\end{equation}
Similar to the case of leptonic decays, precise predictions of the
non-perturbative quantities $f_P$ and $\hat{B}_P$ (for $P = B^0_{(s)}$) enables
the extraction of $\abs{V_{ts}/V_{td}}$.

The current central values and one $\sigma$ error band for the CKM matrix
elements as determined by the CKMfitter group~\cite{Charles:2004jd, CKMfitter}
(left) and the UTfit~\cite{UTfit} group (right) are
\begin{equation}
  \begin{aligned}
    &0.224608\left(^{+254}_{-\hphantom{0}60}\right)   & = \abs{V_{cd}} &= 0.22500(54)\\
    &0.973526\left(^{+50}_{-61}\right)&= \abs{V_{cs}}  &= 0.97344(12)\\
    &0.008710\left(^{+\hphantom{0}86}_{-246}\right) &= \abs{V_{td}}  &= 0.00869(14)\\
    &0.04169\left(^{+\hphantom{0}28}_{-108}\right)  &= \abs{V_{ts}}  &= 0.04124(56).
  \end{aligned}
  \label{eq:CKMprec}
\end{equation}
The CKMfitter~\cite{Charles:2004jd, CKMfitter} (left) and UTfit~\cite{UTfit} (right) groups quotes their
current best estimate for the ratios $\abs{V_{cd}/V_{cs}}$ and
$\abs{V_{td}/V_{ts}}$\footnote{We thank S{\'e}bastien Descotes-Genon,
  J{\'e}r{\^o}me Charles and Marcella Bona for private communication of these
  results.} to be
\begin{equation}
  \begin{aligned}
    &0.230730\left(^{+280}_{-\hphantom{0}74}\right) &=  \abs{V_{cd}/V_{cs}} &\\
    &0.2088\left(^{+16}_{-30}\right) &= \abs{V_{td}/V_{ts}} &= 0.211\left(3\right).
  \end{aligned}
  \label{eq:CKMfitterratios}
\end{equation}
Further detail on how the numbers in equations \eqref{eq:CKMprec} and
\eqref{eq:CKMfitterratios} are obtained are given in
Ref.~\cite{Charles:2004jd,CKMfitter,UTfit}.

The non-perturbative quantities $f_P$ and $\hat{B}_P$ can be calculated in
lattice QCD. The bare decay constants and bare bag parameters are defined as
\begin{equation}
  \matrixel{0}{A_{q_1q_2}^\mu}{P(p)} = i f_P \,p^\mu_P
\end{equation}
and
\begin{equation}
  B_P = \frac{\matrixel{\bar{P}^0}{O_{VV+AA}}{P^0}}{8/3 f_P^2 m_P^2},
\end{equation}
where $P$ is the pseudoscalar meson under consideration with four-momentum
$p^\mu$ and mass $m_P$. In particular we will consider $P=D_{(s)}, B_{(s)}$,
i.e.~$q_2=c,b$ and $q_1=u/d,s$. $A^\mu_{q_1q_2}$ is the axial vector current
defined by $A^\mu_{q_1q_2} = \bar{q}_2 \gamma^\mu \gamma_5 q_1$ and the
four-quark operator $O_{VV+AA}$ is given by $\left(\bar{q}_2 \gamma^\mu
(1-\gamma_5) q_1\right)\left(\bar{q}_2 \gamma^\mu (1-\gamma_5)
q_1\right)$. Quark flow diagrams that describe these processes are shown in
Figure \ref{fig:corr_schematics}.

In this paper we consider the leptonic weak decays of charged mesons ($D^\pm$,
$D_s^\pm$ and $B^\pm$) as well as the mixing of the neutral $B^0_{(s)}$-meson
with its antiparticle $\bar{B}^0_{(s)}$.\footnote{We use the notation $B^0_{(s)}$
  to simultaneously refer to $B^0\equiv B^0_d$ and $B^0_s$.} More
specifically, we will consider ratios which are typically more precise since
common factors and parts of the systematic errors and of the statistical noise
cancel. In particular we will consider the $SU(3)$ breaking ratios
$f_{D_s}/f_D$, $f_{B_s}/f_B$, $\hat{B}_{B_s}/\hat{B}_{B_d}\equiv B_{B_s}/B_{B_d}$
and
\begin{equation}
  \xi  \equiv \frac{f_{B_s}\sqrt{B_{B_s}}}{f_B \sqrt{B_{B_d}}}.
  \label{eq:xi}
\end{equation}
As was first pointed out in Ref.~\cite{Bernard:1998dg}, precise knowledge of
SU(3) breaking ratios, such as $B_{B_s}/B_{B_d}$, $f_{B_s}/f_B$ and $\xi$, can be
combined with the measured mass differences to extract the ratio
$\abs{V_{td}/V_{ts}}$ from
\begin{equation}
  \abs{\frac{V_{td}}{V_{ts}}} = \sqrt{\left(\frac{\Delta m_d}{\Delta m_s}\,
  \frac{m_{B^0_s}}{m_{B_d^0}}\right)}_\mathrm{exp} \left(\xi^{\vphantom{A}}_{\vphantom{A}}\right)_\mathrm{lat}.
\end{equation}
As a result, we present constraints for the ratios $\abs{V_{cd}/V_{cs}}$ and
$\abs{V_{td}/V_{ts}}$.

A summary of relevant lattice results for $f_{D_s}/f_D$,
$f_{B_s}/f_B$, $\xi$ and $B_{B_s}/B_{B_d}$ was presented by the
Flavour Lattice Averaging Group (FLAG)~\cite{Aoki:2019cca}.\footnote{In
  version 1 of this work we referred to the previous version of FLAG,
  i.e.~FLAG16~\cite{Aoki:2016frl}.} Whilst lattice computations of
heavy-light decay constants have become more mature over the last few
years, there are still only few results for direct simulations at the
physical pion mass~\cite{Boyle:2017jwu, Bazavov:2017lyh,
  Bazavov:2014wgs}. For the case of neutral meson mixing
($B_{B_s}/B_{B_d}$ and $\xi$) this is the first result that is
obtained from simulations including physical pion
masses.\footnote{Since the first version of this work a new lattice
  computation which also uses physical pion masses has
  appeared~\cite{Dowdall:2019bea}.}

For the ratio $f_{D_s}/f_D$ FLAG averaged the results presented in
Refs.~\cite{Boyle:2017jwu,Bazavov:2017lyh,Na:2012iu, Bazavov:2014wgs,
  Carrasco:2014poa, Follana:2007uv, Aubin:2005ar, Dimopoulos:2011gx,
  Blossier:2009bx, Carrasco:2013zta, Bazavov:2011aa}. Similarly, the ratio of
decay constants $f_{B_s}/f_B$ have also been computed by various lattice
groups~\cite{Bernardoni:2014fva, Carrasco:2013zta, Christ:2014uea,
  Bazavov:2011aa, Dowdall:2013tga, Bazavov:2017lyh, Bussone:2016iua,
  Hughes:2017spc, Aoki:2014nga}. For $\xi$ and $B_{B_s}/B_{B_d}$ only a few
collaborations have published results~\cite{Carrasco:2013zta, Aoki:2014nga,
  Bazavov:2016nty, Gamiz:2009ku,
  Bazavov:2012zs,Dowdall:2019bea,King:2019lal}.\footnote{We note that the lattice
  result ~\cite{Dowdall:2019bea} and the QCD sum rule result
  ~\cite{King:2019lal} have appeared after the first version of this paper has
  been posted.} For results in the $b$-sector, the lattice formulations of the
heavy quark vary widely, leading to differing systematic errors. The results
presented in this paper are obtained from a chirally symmetric action which
renormalises multiplicatively and therefore is free of renormalisation
uncertainties. A more detailed discussion of these results is presented in
Section \ref{sec:comparison}.

The remainder of this paper is organised as follows. In Section \ref{sec:Setup}
we describe our ensembles, our choice of heavy quark discretisation and our
strategy to obtain correlation functions. In Section \ref{sec:CorrlatorFitting}
we describe our correlation function analysis to deduce the required energies
and matrix elements, before addressing the global fit and the full error budget
in Section \ref{sec:GFs}. Section \ref{sec:comparison} provides a comparison of
our results with the known literature. Section \ref{sec:CKM} assesses the
phenomenological implications, such as the determination of ratios of CKM matrix
elements before we conclude in Section \ref{sec:summary}. The status of this
calculation was previously reported in~\cite{Boyle:2015kyy,Boyle:2017kli}.

\section{Numerical simulations} \label{sec:Setup}
We are performing this calculation in isospin symmetric lattice QCD
with $N_f=2+1$ flavours, thereby capturing the dynamical effects of
light (degenerate up and down) and strange quarks in the sea. We
utilise RBC/UKQCD's $N_f=2+1$ ensembles with physical light quarks at
$a^{-1} \sim 1.7, 2.4\,\mathrm{GeV}$~\cite{Blum:2014tka} and one
ensemble with a finer lattice spacing of $a^{-1}\sim
2.7\,\mathrm{GeV}$ and $m_\pi \approx
230\,\mathrm{MeV}$~\cite{Boyle:2017jwu}. We complement our dataset
with RBC/UKQCD's heavier pion mass ensembles
~\cite{Allton:2008pn,Aoki:2010pe,Aoki:2010dy}, to guide the small
correction of the fine ensemble towards the physical pion mass.

For the heavy quarks we adopt a similar strategy to our previous
work~\cite{Boyle:2017jwu} by simulating a range of heavy quark masses $m_h$ from
slightly below the charm quark mass to approximately half the $b$ quark
mass. For the neutral meson mixing computation, we only consider the charge
assignment suitable for $B_{(s)}$ meson mixing (cf. Figure
\ref{fig:corr_schematics}), so that in the limit $m_h \to m_b$, we recover the
correct quantities for $B_{(s)}$-meson mixing.

As we will lay out in Sections \ref{subsec:ensembles} and
\ref{subsec:charm}, our discretisation differs between the
light/strange and the heavy quark sector, resulting in a mixed
action. In this work we solely focus on results for observables where
the renormalisation constants cancel. Work is in process to calculate
the required mixed-action renormalisation factors (as laid out in
\cite{Boyle:2017jwu}) in order to also obtain results for the
individual decay constants and bag parameters, rather than their
ratios.

\subsection{Ensemble properties} \label{subsec:ensembles}
\begin{table}
  \begin{center}
    \begin{tabular}{l c c c c c c }
      \hline\hline\\[-4mm]
      Name & $L/a$ & $T/a$ &  $a^{-1}[\mathrm{GeV}]$ & $m_\pi[\mathrm{MeV}$] & $m_\pi L$ & hits $\times N_\mathrm{conf}$ \\\hline
      C0   & 48    & 96     & 1.7295(38)   & 139.17(0.35) & 3.86 & $48 \times  90$\\
      C1   & 24    & 64     & 1.7848(50)   & 339.76(1.22) & 4.57 & $32 \times  100$\\
      C2   & 24    & 64     & 1.7848(50)   & 430.63(1.38) & 5.79 & $32 \times  101$\\\hline
      M0   & 64    & 128    & 2.3586(70)   & 139.34(0.46) & 3.78 & $64 \times  82$ \\
      M1   & 32    & 64     & 2.3833(86)   & 303.56(1.38) & 4.08 & $32 \times  83$ \\
      M2   & 32    & 64     & 2.3833(86)   & 360.71(1.58) & 4.84 & $32 \times  76$ \\
      M3   & 32    & 64     & 2.3833(86)   & 410.76(1.74) & 5.51 & $32 \times  81$ \\\hline
      F1M   & 48&96 & 2.708(10)\hphantom{0} & 232.01(1.01) & 4.11 & $48 \times  72$ \\
      \hline \hline
    \end{tabular}
  \end{center}
  \caption{This table summarises the main parameters of the ensembles used for
    the presented calculation. All ensembles have $N_f=2+1$ flavours in the
    sea. C stands for coarse, M for medium and F for fine. The columns
    \emph{hits} and $N_\mathrm{conf}$ give the number of measurements on a given
    configuration and the total number of configurations used, respectively.}
  \label{tab:ensembles}
\end{table}

All ensembles use the Iwasaki gauge action~\cite{Iwasaki:2011np} and
the domain wall fermion action~\cite{Kaplan:1992bt, Blum:1996jf,
  Blum:1997mz, Shamir:1993zy}. The ensembles with heavier pion masses
(C1-2, M1-3) use the Shamir action approximation to the sign
function~\cite{Shamir:1993zy, Furman:1994ky}, the remaining ensembles
(C0, M0, F1M) the M\"obius action approximation with the same $H_T$
kernel~\cite{Brower:2012vk}.\footnote{For more detail on the F1M
  ensemble, please refer to Appendix \ref{sec:F1M}.}  In the
convention of Ref.~\cite{Brower:2012vk}, the Shamir action is
simulated with $b_5=1$, $c_5=0$, whereas for the M{\"o}bius action
$b_5=1.5$, $c_5=0.5$ is used. The parameters of both of these actions
are chosen such that they lie on the same scaling trajectory, allowing
for a combined continuum limit~\cite{Blum:2014tka}. Details of the
main parameters of these ensembles are summarised in Tables
\ref{tab:ensembles} and \ref{tab:DWFls}.
\begin{table}
  \begin{center}
    \begin{tabular}{l@{\hspace{2mm}} c@{\hspace{2mm}} c@{\hspace{2mm}} c@{\hspace{2mm}} l@{\hspace{2mm}} l@{\hspace{2mm}} c@{\hspace{2mm}} c@{\hspace{2mm}} c}
      \hline\hline\\[-4mm]
      Name & DWF & $M_5$  & $L_s$ & $am_l^\mathrm{sea, val}$ & $am_s^{sea}$ & $am_s^{\mathrm{phys}}$ & $\sigma$ & $N_\mathrm{\sigma}$\\\hline
      C0  & M    & 1.8    &  24   & 0.00078  & 0.0362   & 0.03580(16) & 4.5 & 400\\
      C1  & S    & 1.8    &  16   & 0.005    & 0.04     & 0.03224(18) & 4.5 & 400\\
      C2  & S    & 1.8    &  16   & 0.01     & 0.04     & 0.03224(18) & 4.5 & 100\\\hline
      M0  & M    & 1.8    &  12   & 0.000678 & 0.02661  & 0.02539(17) & 6.5 & 400\\
      M1  & S    & 1.8    &  16   & 0.004    & 0.03     & 0.02477(18) & 6.5 & 400\\
      M2  & S    & 1.8    &  16   & 0.006    & 0.03     & 0.02477(18) & 6.5 & 100\\
      M3  & S    & 1.8    &  16   & 0.008    & 0.03     & 0.02477(18) & 6.5 & 100\\\hline
      F1M  & M    & 1.8    &  12   & 0.002144 & 0.02144  & 0.02217(16) & - & -\\
      \hline \hline
    \end{tabular}
  \end{center}
  \caption{Domain wall parameters for the light and strange quarks. All quoted
    values for $am_{l,s}$ are bare quark masses in lattice units. The column DWF
    corresponds to the chosen domain wall fermion formulation, i.e.~M(\"obius)
    or S(hamir) domain wall fermions.}
  \label{tab:DWFls}
\end{table}

Table \ref{tab:DWFls} also describes the light and strange valence
parameters. All light quarks are simulated at their unitary value
$am_l^\mathrm{sea}=am_l^\mathrm{val}$ whilst the valence strange quark
masses were tuned to their physical values as determined in Refs.
\cite{Blum:2014tka,Boyle:2017jwu} with the exception of the F1M
ensemble where we simulated at the unitary strange quark mass. All
propagators were generated using $Z_2$-wall
sources~\cite{Foster:1998vw,Boyle:2008rh,McNeile:2006bz}. For the
light and strange quark propagators on the coarse and medium
ensembles, we used Gaussian
smearing~\cite{Gusken:1989ad,Alexandrou:1990dq,Allton:1993wc} to
achieve a better overlap with the ground state. The smearing
parameters $\sigma$ and $N_\sigma$ are listed in Table
\ref{tab:DWFls}.

\subsection{Heavy quark discretisation} \label{subsec:charm}
In our previous work~\cite{Boyle:2016imm,Boyle:2017jwu} the
limitations of our formalism prohibited the direct simulation of the
physical charm quark mass on the coarse ensembles. We therefore
required a slight extrapolation in the heavy quark mass to reach the
physical charm quark mass on our coarsest ensembles. We found that it
is possible to increase the heavy-quark mass reach by stout
smearing~\cite{Morningstar:2003gk} the gauge fields prior to
performing the charm quark
inversions~\cite{Cho:2015ffa,Boyle:2015kyy}. A comparison of the
effect on the residual chiral symmetry breaking parameter
$m_\mathrm{res}$ was presented in Ref.~\cite{Boyle:2015kyy}. We found
that three hits of stout smearing with the standard parameter
$\rho=0.1$ extends the reach in the heavy quark mass compared to our
previous work~\cite{Cho:2015ffa,Boyle:2015kyy}. Table \ref{tab:heavy}
lists the domain wall parameters as well as the quark masses that were
used on the various ensembles. Since the charm quark is quenched in
our calculations this has no additional unitarity implications which
are not already present.

\begin{table}
  \begin{center}
    \begin{tabular}{lcccc}
      \hline
      Name & DWF & $L_s$ & $M_5$ & $am_h$ \\\hline\hline
      C0  & M & 12 & 1.0 & 0.51, 0.57, 0.63, 0.69 \\
      C1  & M & 12 & 1.0 & 0.50, 0.58, 0.64, 0.69 \\
      C2  & M & 12 & 1.0 & 0.51, 0.59, 0.64, 0.68 \\\hline
      M0  & M & 12 & 1.0 & 0.41, 0.50, 0.59, 0.68 \\
      M1  & M & 12 & 1.0 & 0.41, 0.50, 0.59, 0.68 \\
      M2  & M & 12 & 1.0 & 0.41, 0.50, 0.59, 0.68 \\
      M3  & M & 12 & 1.0 & 0.41, 0.50, 0.59, 0.68 \\\hline
      F1M  & M & 12 & 1.0 & 0.32, 0.41, 0.50, 0.59, 0.68\\
      \hline 
    \end{tabular}
  \end{center}
  \caption{Bare heavy quark masses in lattice units.}
  \label{tab:heavy}
\end{table}

\subsection{Measurement strategy}
\begin{figure}
  \begin{center}
    \includegraphics[width=.48\textwidth]{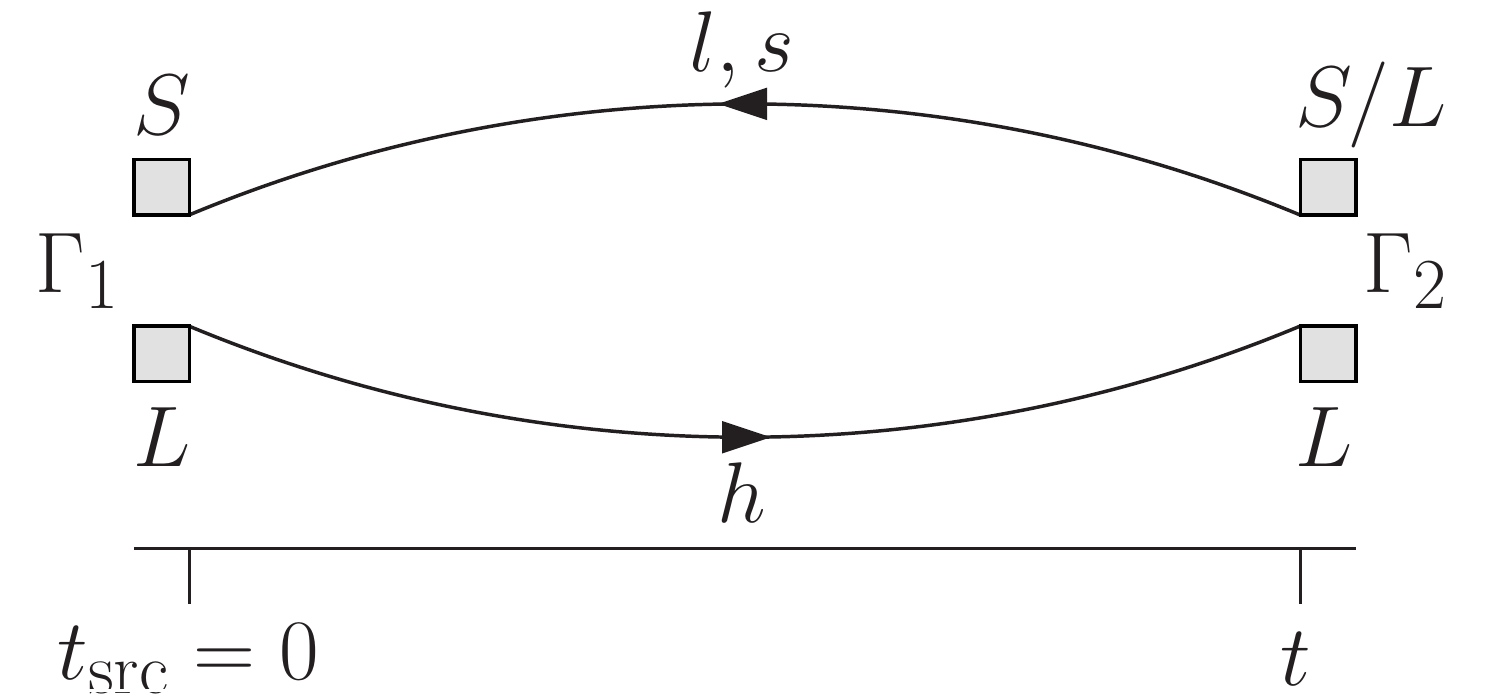}
    \includegraphics[width=.48\textwidth]{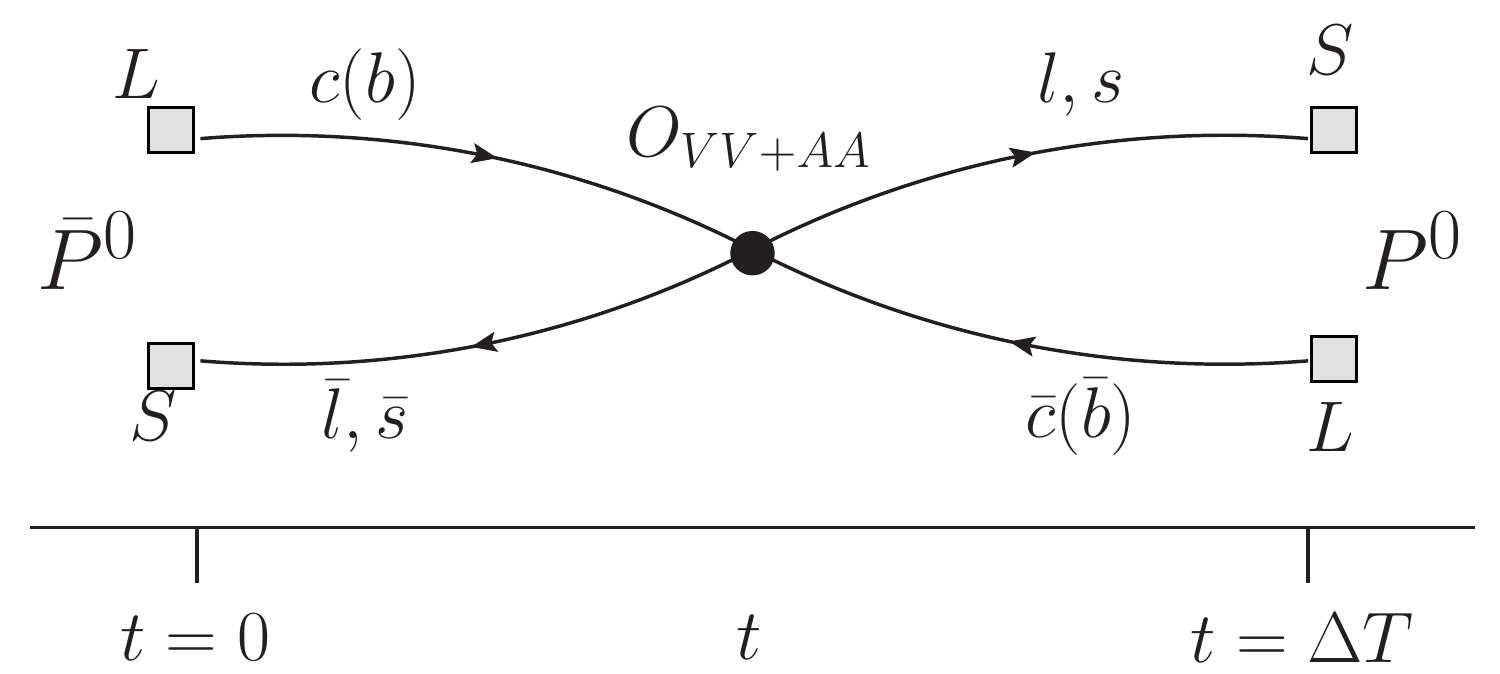}\\
  \end{center}
  \caption{Schematic description of the set-up of our two-point (left) and
    three-point (right) correlation functions.}
  \label{fig:corr_setup}
\end{figure}

The left panel of Figure \ref{fig:corr_setup} shows our set-up for
the computation of two-point functions. These take the form
\begin{equation}
  \begin{aligned}
  C_{\Gamma_1,\Gamma_2}^{s_1,s_2}(t) &\equiv \sum_{\v{x}} \avg{\left(O_{\Gamma_2}^{s_2}(\v{x},t)\right) \left({O_{\Gamma_1}^{s_1}}(\v{0},0)\right)^\dagger}\\
  & = \sum^{N=\infty}_{n=0} \frac{\left(\matel{\Gamma_2}{s_2}\right)_n \left(\matel{\Gamma_1}{s_1}\right)_n^*}{2E_n} \left(e^{-E_nt} \pm e^{-E_n(T-t)}\right),
  \label{eq:C2fitform}
  \end{aligned}
\end{equation}
where the interpolation operators $O_{\Gamma_i}^{s}$ define the quantum numbers
of the meson under consideration and are given by
\begin{equation}
  O_{\Gamma_i}^{s}(t,\v{x}) =
  \left(\bar{q}_2(t,\v{x}) \sum_{\v{y}} \omega_s(\v{x},\v{y}) \Gamma_i q_1(t,\v{y}) \right).
\end{equation}
Here $q_1$ and $q_2$ give the quark content of the meson and we consider the
cases $\Gamma_i = \gamma_5\equiv P$ (pseudoscalar) and
$\Gamma_i=\gamma_0\gamma_5\equiv A$ (axial vector). $\omega_s$ denotes that each
propagator can be smeared (S) or local (L) at both the source and the sink. For
the local case, $\omega$ reduces to a Kronecker-delta (i.e.~$\omega_L =
\delta_{\v{x},\v{y}}$). In principle we consider the cases $s\in\{LL,SL,LS,SS\}$
for each of the two operators (where the first entry corresponds to the smearing
of the source and the second entry to that of the sink). For the smeared case,
$\omega_s$ is obtained by Gaussian smearing via Jacobi
iteration~\cite{Gusken:1989ad,Alford:1995dm,Allton:1993wc}, the parameters of
which are given in Table \ref{tab:DWFls}. In practice, we never smear the heavy
quark propagator. On the coarse and medium ensembles we always smear the source
of the light and strange quark propagators and alow both options for the
sink. On the fine ensemble (F1M), both source and sink of all quark propagators
are kept local. For the heavy-light systems under consideration we thus consider
$SL$ and $SS$ only for the coarse and medium ensembles and $LL$ on the fine
ensemble, where in both cases we have dropped the indices corresponding to the
heavy propagators. The overlap coefficients $\matel{\Gamma_i}{s_i}$ for state
$n$ are given by
\begin{equation}
  \left(\matel{\Gamma_i}{s_i}\right)_n = \matrixel{X_n}{O_{\Gamma_i}^{s_i}}{0},
\end{equation}
where $X_n$ is the $n$th excited meson state $X$ with the correct quantum
numbers. In the remainder of this paper we will omit the label for the state if
only one state is considered.

The right panel of Figure \ref{fig:corr_setup} shows how we obtain the
three point functions from which the bag parameters are determined.
We create a state with the quantum numbers of $\bar{P}$ at $t=0$, let
it propagate to the operator insertion $t$, where it is transformed to
the state $P$ and then annihilate this state at $\Delta T$. Noting
that for the coarse and medium (fine) ensembles the external states
are always build from a smeared (local) light or strange propagator
there is no need to label the smearing combination for the three point
function $C_3(t,\Delta T)$. Considering the zero momentum projected
three point function, we can rewrite the correlation functions as
\begin{equation}
  \begin{aligned}
    C_3(t,\Delta T) &\equiv \avg{P(\Delta T) O_{VV+AA}(t) \bar{P}^\dagger(0)}\\
    &= \sum_{n,n'} \frac{1}{4m_nm_{n'}} \left(\matel{P}{i}\right)_n \matrixel{n}{O_{VV+AA}(t)}{n'} \left(\matel{P}{i}\right)^*_{n'}\\
    & \qquad \times \left(e^{-(\Delta T - t)m_n} + e^{-(T-\Delta T + t)m_n} \right) \left(e^{-tm_{n'}} + e^{-(T-t) m_{n'}}\right)\\
    &\approx \frac{1}{4m^2} \left(\matel{P}{i}\right)_0  e^{-(\Delta T - t)m} \matrixel{P}{O_{VV+AA}(t)}{P} \left(\matel{P}{i}\right)_0^* e^{-tm},
  \end{aligned}
  \label{eq:C3fitform}
\end{equation}
where $i=S$ ($i=L$) for the coarse and medium (fine) ensembles. In the
final line, we assumed that only the ground state contributes and that
``around-the-world'' contributions are negligible.

The signal-to-noise ratio quickly deteriorates for large
times so obtaining a signal in the low $t$ region is favourable. Hence a trade
off between choosing $\Delta T$ as small as possible without pollution from
excited states is required, which will be discussed in Section
\ref{subsec:bagfits}.

We place a $Z_2$-wall source on every second time slice across the lattice,
hence produce all required correlation functions $(T/a)/2$ times per
configuration (cf.~column hits in Table \ref{tab:ensembles}). These correlation
functions are translated in time and binned into one effective measurement per
configuration prior to any statistical analysis. In addition to improving the
statistical signal, this allows us to compute the bag parameters for many
source-sink separations $\Delta T$ (compare Figure \ref{fig:corr_setup})
without needing to invert additional propagators. For a given $\Delta T$ these
three point functions are obtained by contracting the propagators originating
from different wall source positions with the four-quark operator. Finally, this
multi-source approach allows us to make efficient use of the HDCG
algorithm~\cite{Boyle:2014rwa}, rendering this computation affordable.

\section{Correlator analysis} \label{sec:CorrlatorFitting}
We bin all measurements on a given configuration into one effective
measurement. Prior to any analysis step, we make use of the last lines of
equations \eqref{eq:C2fitform} and \eqref{eq:C3fitform} and symmetrise all two
and three point correlation functions with respect to $T/2$ and $\Delta T/2$,
respectively and before restricting the data to the temporal extent from $ t \in
[0,T/2]$ and $t \in [0,\Delta T/2]$, respectively.

We conservatively choose to illustrate all correlator fits for the heaviest mass
point on the M0 ensemble, since this is a worst case scenario given the large
difference between the physical light quark mass and the heavier-than-charm
quark mass. The error propagation is carried out by using bootstrap resampling
using 2000 bootstrap samples. We use different seeds for the random number
generator for different ensembles, to avoid the introduction of any spurious
correlations.

\subsection{Two point function fits} \label{subsec:corfits}
\begin{figure}
  \begin{center}
    \includegraphics[width=.49\textwidth]{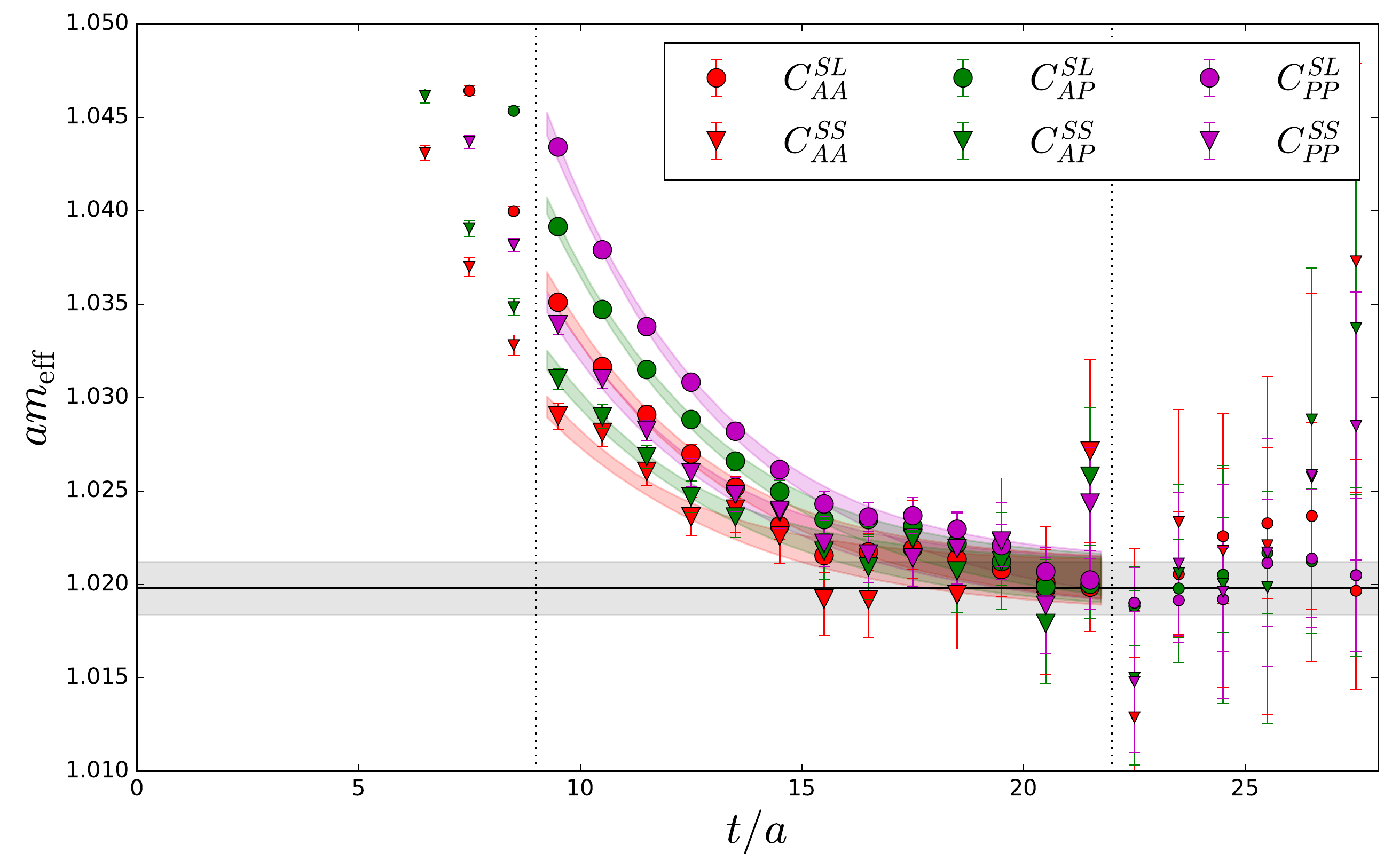}
    \includegraphics[width=.49\textwidth]{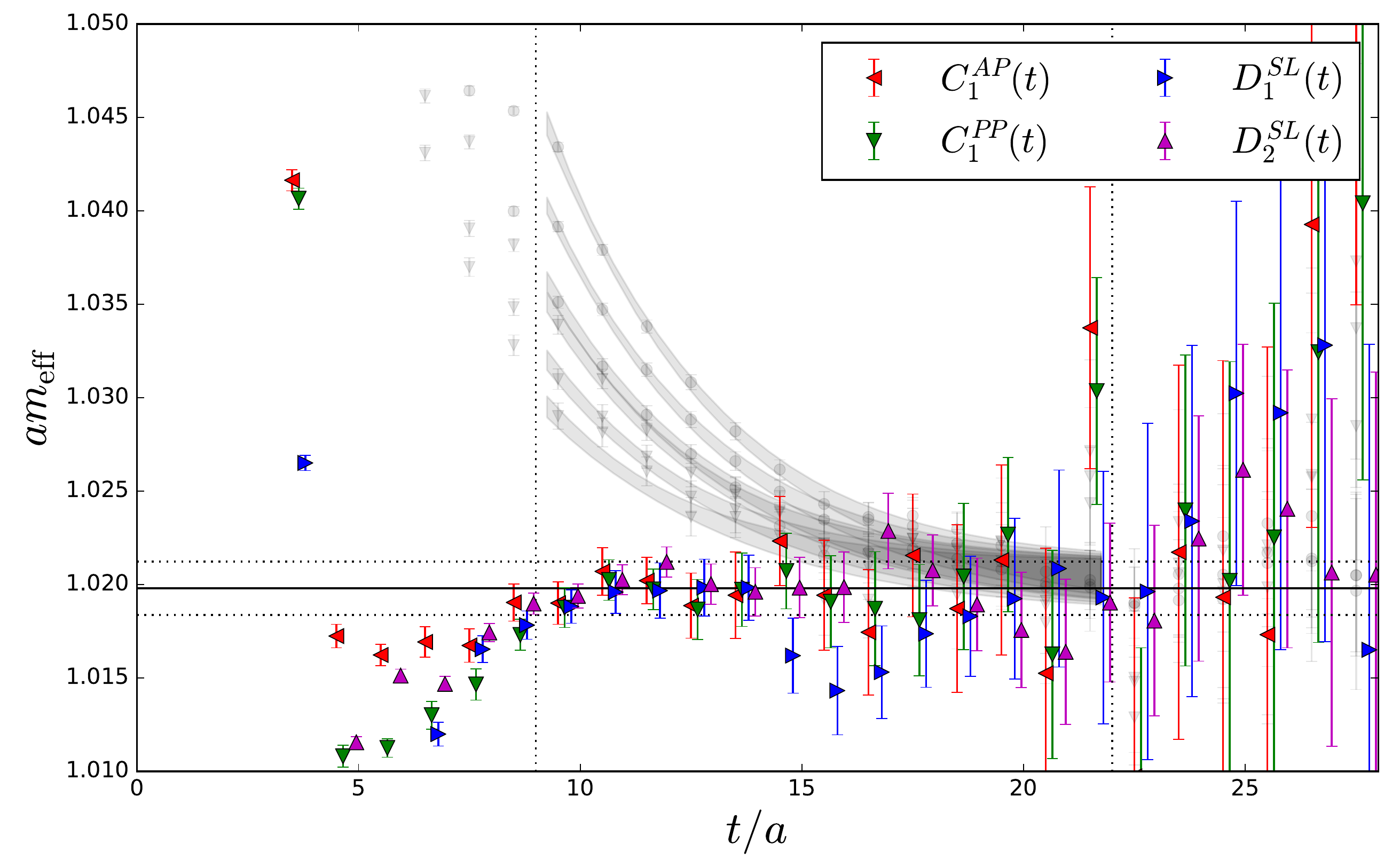}
  \end{center}
  \caption{Example correlation function fit for heaviest $D$-like meson on the
    M0 ensemble. The data points in the left panel show the effective masses of
    the correlation functions that enter the fit. On top of this, the effective
    mass of the fit results is superimposed. The grey horizontal band shows the
    ground state fit result obtained in this way. The right panel shows the
    effective mass of the linear combinations of correlation functions that are
    mentioned in the text. The dashed vertical lines correspond to the values of
    $t_\mathrm{min}$ and $t_\mathrm{max}$.}
  \label{fig:2pointexamplefit}
\end{figure}

For the coarse and medium ensembles we extract values for the masses
and matrix elements by performing a simultaneous double-exponential
fit (i.e.~$n=0,1$ in \eqref{eq:C2fitform}) to six correlation
functions in the interval $t\in
\vphantom{(}[t_\mathrm{min},t_\mathrm{max})\vphantom{]}$. In
particular we simultaneously fit the correlation functions
$C^{SL}_{AA}$, $C^{SS}_{AA}$, $C^{SL}_{AP}$, $C^{SS}_{AP}$,
$C^{SL}_{PP}$ and $C^{SS}_{PP}$. From this we obtain the mass $m_i$ as
well as the bare matrix elements $\matel{P,n}{L}$, $\matel{P,n}{S}$,
$\matel{A,n}{L}$ and $\matel{A,n}{S}$, where $n=0,1$ refers to the
ground state and the first excited state, respectively. The result of
such a fit is shown in the left hand panel of Figure
\ref{fig:2pointexamplefit}. The coloured data points (circles and
squares) show the effective mass of the correlation functions that
enter the fit, the grey horizontal band depicts the ground state mass
that is obtained from a fit to the data in the range
$[t_\mathrm{min},t_\mathrm{max})$ (indicated by the vertical dotted
  lines). The coloured shaded bands show the effective mass obtained
  by reconstructing the respective correlation functions from the
  fit-results. We can see that the data is well described by these
  fits. In the case of the F1M ensemble this situation simplifies due
  to the absence of source and sink smearing and the above reduces to
  a joint fit of $C_{AA}^{LL}$, $C_{AP}^{LL}$ and $C_{PP}^{LL}$ to
  extract the matrix elements $\matel{P,n}{L}$ and $\matel{A,n}{L}$.
  The results to all correlation functions fits are tabulated in the
  appendix in Table \ref{tab:corrfits}.

Whilst for a pure ground state fit, the effective mass provides a visual
cross-check of a \emph{plateau range} in which one can approximate the
correlation function as a single state, this is more difficult for fits
including excited states. We circumvent this in the following: Assuming we are
in a range where only the ground state and the first excited states contribute
(and for simplicity restricting ourselves to $t\ll T/2$), we can write
\begin{equation}
  \begin{aligned}
    C_{ab}^{pq}(t) &\equiv \sum_{n=0}^\infty \frac{e^{-E_nt}}{2E_n} \left(\matel{a}{p}\right)_n \left(\matel{b}{q}\right)_n \approx \frac{e^{-E_0t}}{2E_0} \left(\matel{a}{p}\right)_0 \left(\matel{b}{q}\right)_0 +\frac{e^{-E_1t}}{2E_1} \left(\matel{a}{p}\right)_1 \left(\matel{b}{q}\right)_1\,. \\
  \end{aligned}
\end{equation}
We now consider linear combinations of the form
\begin{equation}
  \mc{E}_{X,Y}(t) = \mc{C}(t) X - \mc{D}(t) Y\,,
\end{equation}
where $\mc{C}$ and $\mc{D}$ are two of the original correlation functions and
$X,Y$ are some constants. Assuming we have carried out a fit to determine the
matrix elements $\left(\matel{A}{L}\right)_n$, $\left(\matel{A}{S}\right)_n$,
$\left(\matel{P}{L}\right)_n$ and $\left(\matel{P}{S}\right)_n$ for $n=0,1$, we
can now choose $\mc{C}(t)$ and $\mc{D}(t)$, such that they have one of the two
matrix element factors in common. Furthermore we identify the factors $X$ and
$Y$ with the excited state matrix element of the respective other correlation
function which they do not have in common. More precisely, we construct the linear
combinations
\begin{equation}
  \begin{aligned}
    C^{AP}_1(t) &\equiv C^{SS}_{AP}(t) {\left(\matel{A}{L}\right)_1}|_\mathrm{fit} - C^{LS}_{AP}(t) {\left(\matel{A}{S}\right)_1}|_\mathrm{fit}\\
    C^{PP}_1(t) &\equiv C^{SS}_{PP}(t) {\left(\matel{P}{L}\right)_1}|_\mathrm{fit} - C^{LS}_{PP}(t) {\left(\matel{P}{S}\right)_1}|_\mathrm{fit}\\
    D^{SL}_1(t) &\equiv C^{SL}_{AP}(t) {\left(\matel{A}{L}\right)_1}|_\mathrm{fit} - C^{SL}_{AA}(t) {\left(\matel{P}{L}\right)_1}|_\mathrm{fit}\\
    D^{SL}_2(t) &\equiv C^{SL}_{PP}(t) {\left(\matel{A}{L}\right)_1}|_\mathrm{fit} - C^{LS}_{AP}(t) {\left(\matel{P}{L}\right)_1}|_\mathrm{fit},
  \end{aligned}
\end{equation}
where ${\left(\matel{P}{S}\right)_1}|_\mathrm{fit}$ and
${\left(\matel{P}{L}\right)_1}|_\mathrm{fit}$ refer to the
\emph{central values} of the fit. We stress that this is applicable to
any pair of two-point correlation functions that have the same
spectrum and one matrix element in common.\footnote{We note that if
  the backwards travelling contribution comes with the opposite sign
  between the two correlation functions in this difference, this only
  holds for values of $t$ where temporal ``around-the-world'' effects
  are negligible. However for heavy-light quantities this contribution
  is suppressed by a factor smaller than $e^{-ET/2}$ where the
  smallest simulated values are $ET/2 \sim 23$. This is therefore
  negligible.}

\emph{If} the fit describes the data well, the excited state contribution
cancels in this difference and such an effective mass plot should show a plateau in
the region of the fit. Furthermore this plateau needs to coincide with the fit
result for the ground state energy. This procedure therefore serves as a strong
\emph{a posteriori} check. 

\begin{figure}
  \begin{center}
    \includegraphics[width=.495\textwidth]{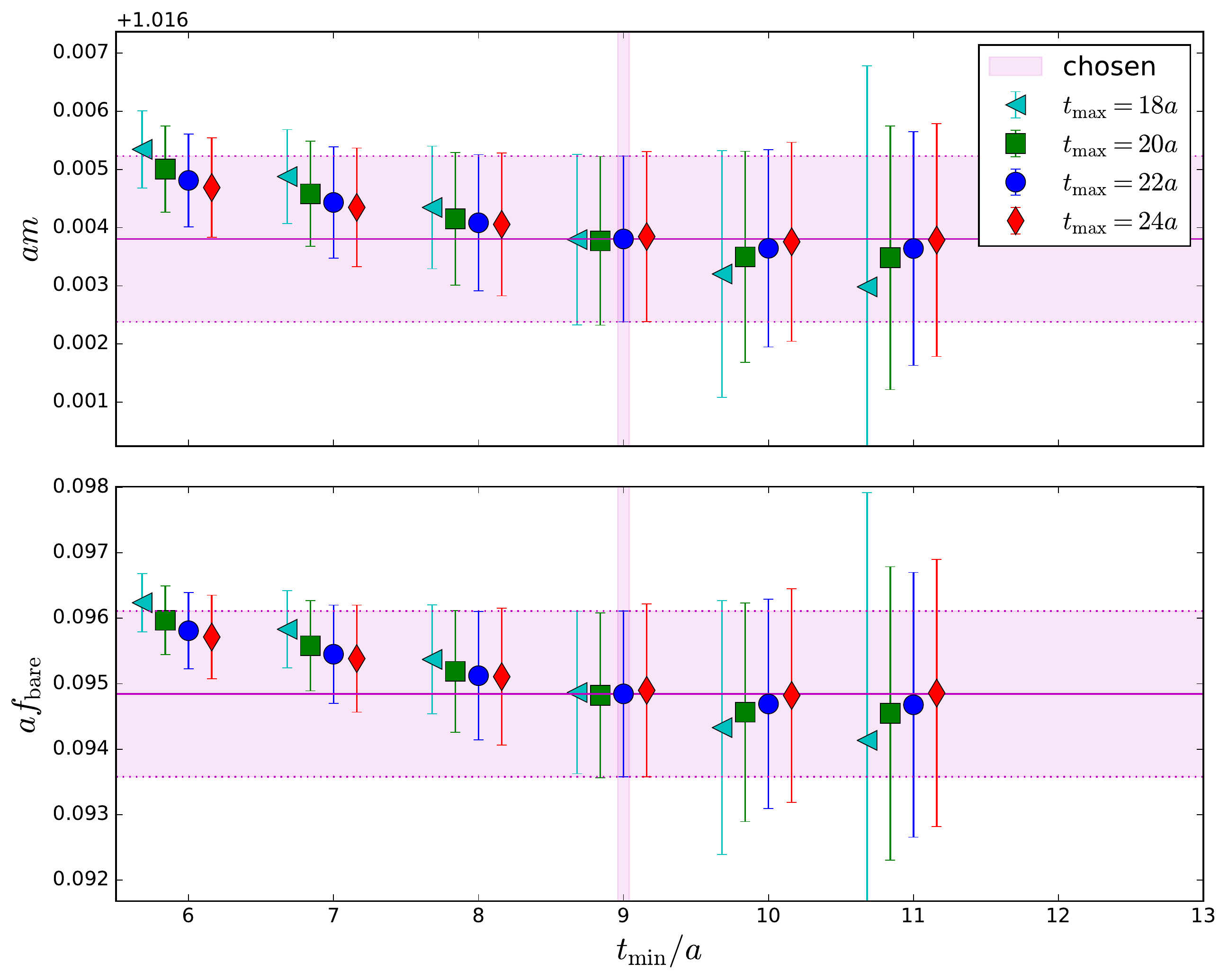}
    \includegraphics[width=.495\textwidth]{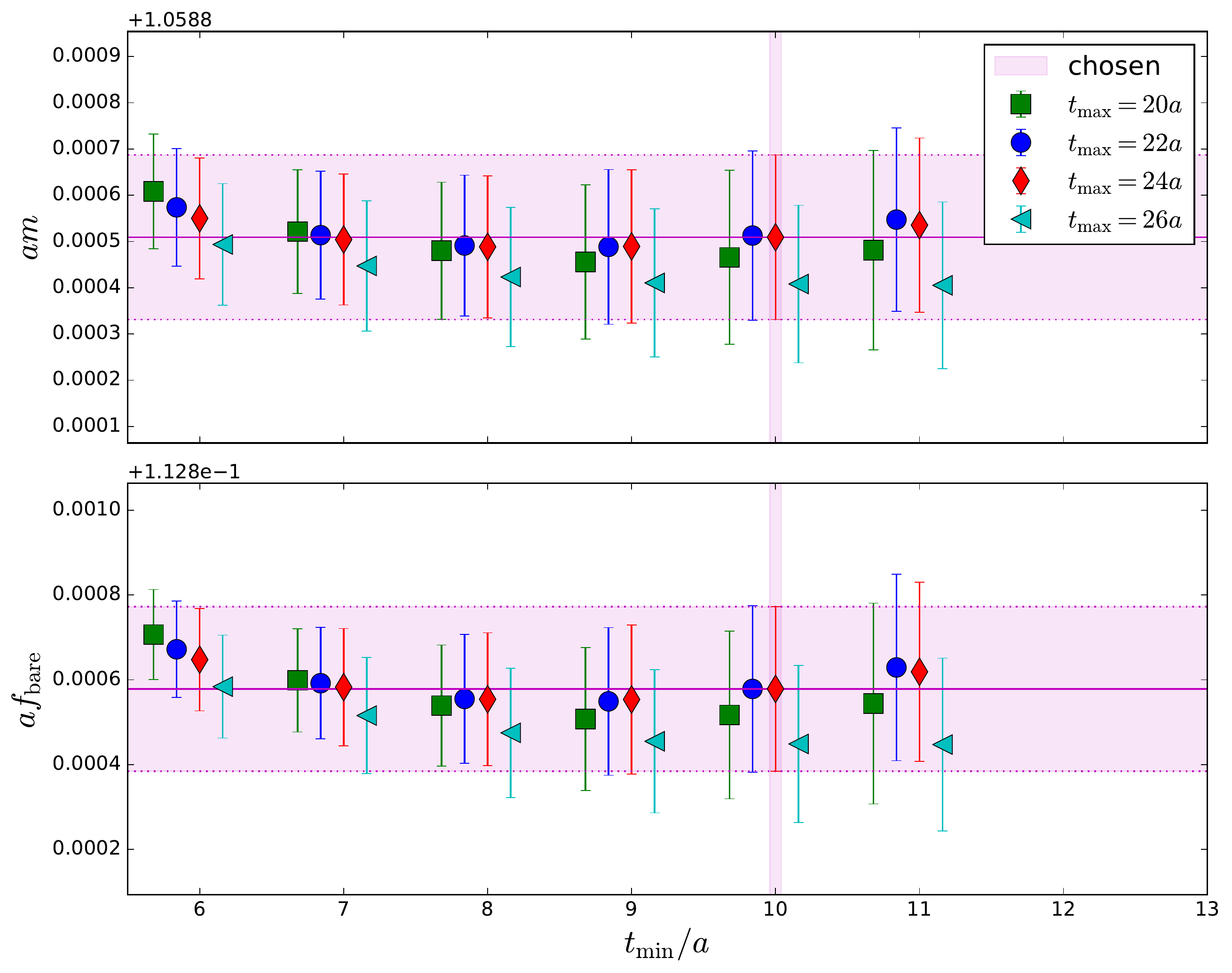}
  \end{center}
  \caption{Impact of the choice of fit range on the observables of interest,
    i.e.~the mass $m$ and the (bare) decay constant $f$. Results are shown for
    the heaviest heavy-light (left) and heavy-strange (right) mesons on the M0
    ensemble.}
  \label{fig:cor_stability}
\end{figure}

The right panel of Figure \ref{fig:2pointexamplefit} shows the effective masses
of some of the linear combinations which can be obtained from the correlation
functions. The grey horizontal band shows the ground state mass which is
obtained from the fit. We note that in between the two vertical lines, the
effective mass of the reconstructed data points lie within the grey band. In
addition to this strong visual check, we also varied $t_\mathrm{min}$ and
$t_\mathrm{max}$ to investigate stability under these changes. This is presented
for the case of the heaviest heavy quark mass on the M0 ensemble in Figure
\ref{fig:cor_stability}. All variations of the fit range are well within the
quoted statistical uncertainty, particularly for the heavy-light case which
dominates the error on all the presented ratios.

\subsection{Bag parameter fits} \label{subsec:bagfits}
\begin{figure}
  \begin{center}
    \includegraphics[width=\textwidth]{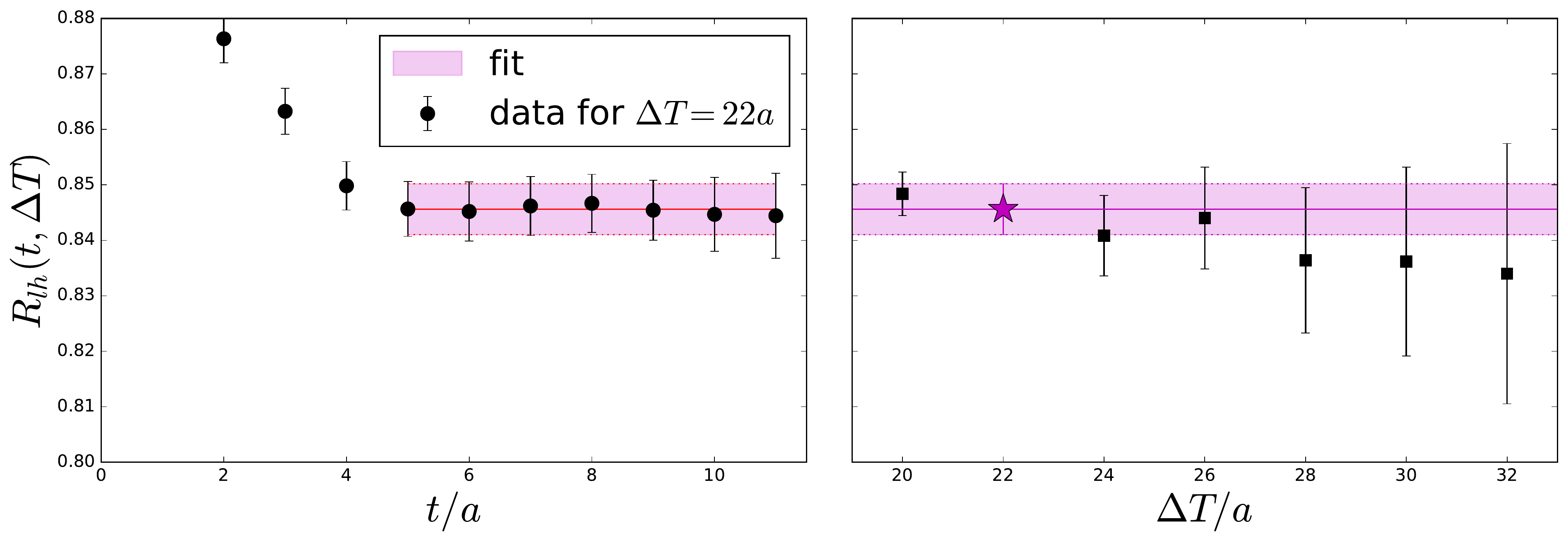}
  \end{center}
  \caption{Example fits for the heavy-light bag parameters on the M0 ensemble
    for the heaviest heavy quark mass for the respective choices of $\Delta
    T$. The left panel shows the fit to a constant for the chosen value of
    $\Delta T/a=22$. The right panels show the stability as a function of the
    source-sink separation $\Delta T$. The magenta star illustrates our chosen
    value for $\Delta T$ and the obtained result.}
  \label{fig:bagexamplefit_l}
\end{figure}

Since we are interested in $B_P$, we construct ratios in which the matrix
elements $\matel{P}{S}$ cancel (c.f.~equation \eqref{eq:C3fitform}). More
precisely we construct ratios $R(t,\Delta T)$ which, in the limit of large $t$
and $\Delta T$, plateau to the value of the bag parameter $B_P$
\begin{equation}
  R(t,\Delta T) = \frac{C_3(t,\Delta T)}{8/3 C^{SL}_{PA}(\Delta T - t)
    C^{LS}_{AP}(t)} \to \frac{\matrixel{P}{O_{VV+AA}(t)}{P}}{8/3 m_P^2 f_P^2}\equiv B_P \quad \mathrm{for} \,\, t, \Delta T \gg 0\,.
\end{equation}

\begin{figure}
  \begin{center}
    \includegraphics[width=\textwidth]{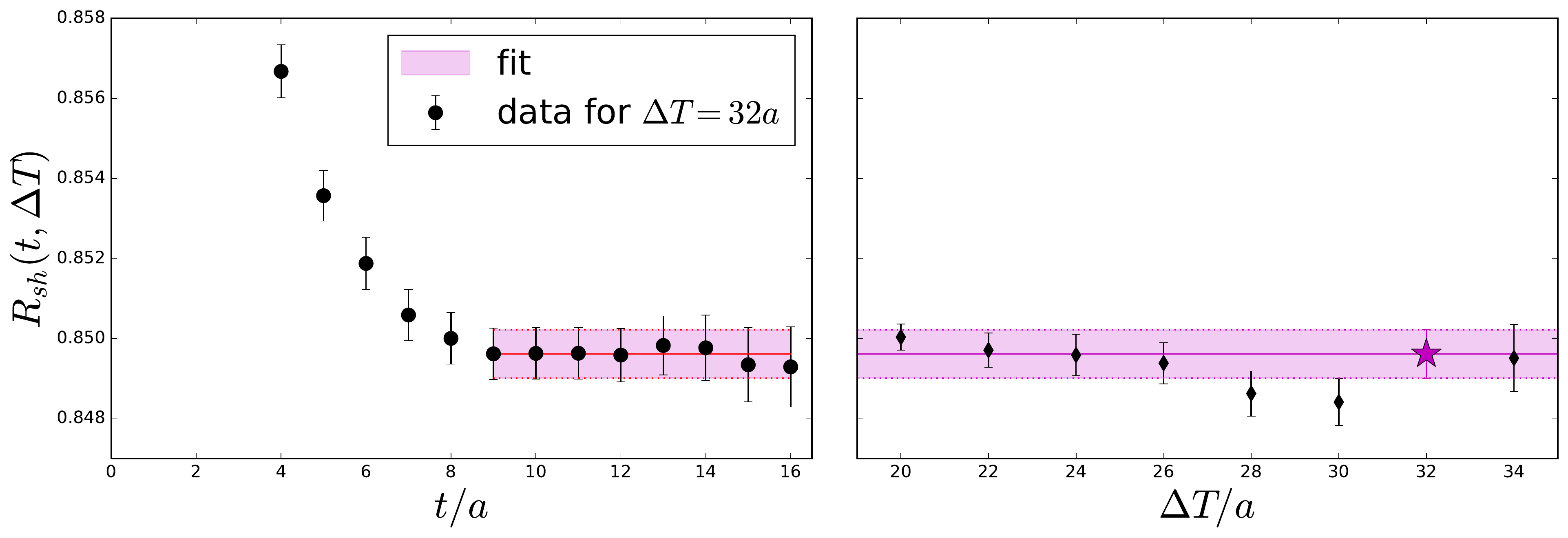}
  \end{center}
  \caption{Same plot as Figure \ref{fig:bagexamplefit_l} but for the
    heavy-strange bag parameter. Note that the heavy-strange quantity is nearly
    an order of magnitude more precise than the heavy-light one.}
  \label{fig:bagexamplefit_s}
\end{figure}

Figures \ref{fig:bagexamplefit_l} and \ref{fig:bagexamplefit_s} show example
fits of such plateaux and the fits to them for the case of the heaviest
heavy-light and heavy-strange mass points on M0, respectively.

\section{Data analysis}\label{sec:GFs}
\begin{figure}
  \begin{center}
    \includegraphics[width=.495\textwidth]{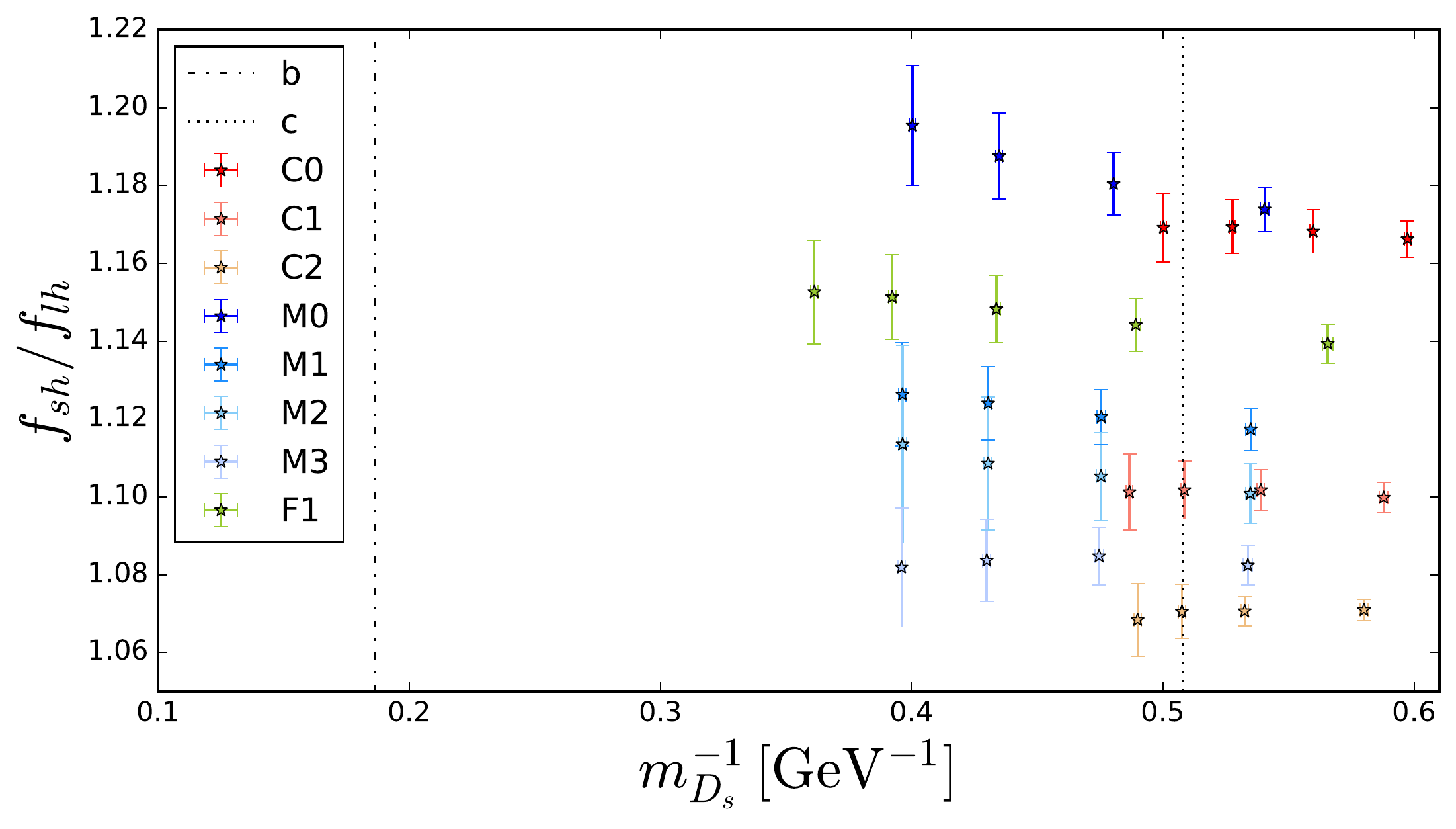}
    \includegraphics[width=.495\textwidth]{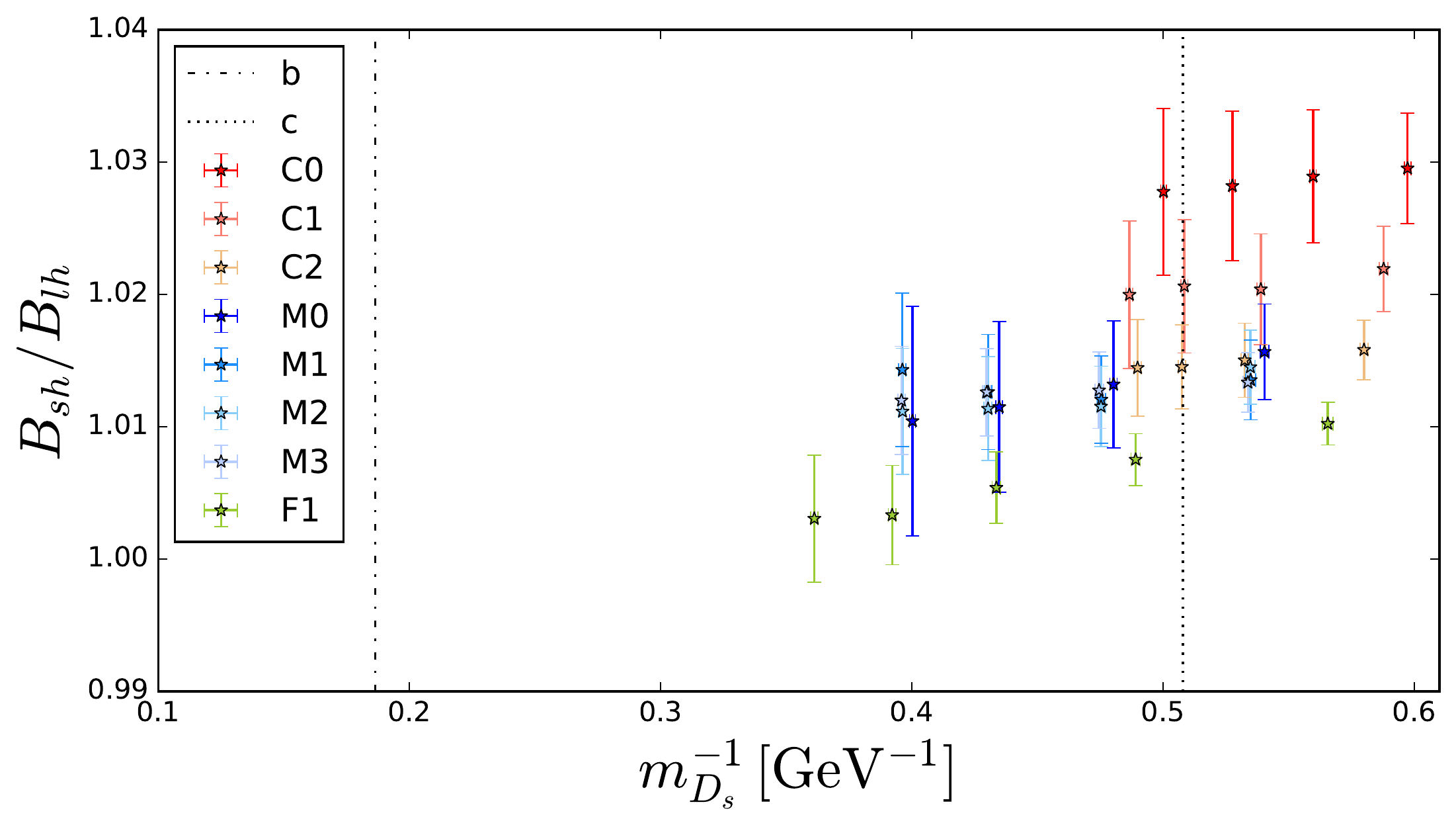}
  \end{center}
  \caption{Summary of the ratio of decay constants (left) and the ratio of bag
    parameters $B_{B_s}/B_B$ (right) as a function of the inverse ${D_s}$ mass.}
  \label{fig:alldata}
\end{figure}

From the fits to the correlation function data we have obtained decay constants
and bag parameters for a range of charm quark and pion masses and lattice
spacings. Due to the use of chiral fermions all of these observables renormalise
multiplicatively. So by constructing the ratios $f_{D_s}/f_D$ and $\xi$ (see
\eqref{eq:xi}) all renormalisation constants cancel, such that we can replace
$\hat{B}_{B_s}/\hat{B}_B$ by $B_{B_s}/B_B$ in the same equation. Some of the
statistical noise and discretisation effects also cancel, making these
observables cleaner. Figure \ref{fig:alldata} shows the ratio of decay constants
(left) and the ratio of bag parameters (right) as a function of the inverse
heavy-strange pseudoscalar meson mass. The behaviour of these ratios as a
function of the heavy quark mass (set via a meson mass containing a heavy quark)
is very benign, lending confidence to the use of inter/extrapolations in the
heavy meson mass. By comparing the C0 and M0 ensembles (which are at the same
pion mass, but differ in lattice spacing), we note that the discretisation
effects appear to also be mild. We notice a stronger dependence on the pion
mass, as is expected for $SU(3)$-breaking ratios, since in the limit of $m_\pi
\to m_K$ they are identically unity. This is the first computation of the ratio
of bag parameters and the $SU(3)$ breaking ratio $\xi$ using ensembles at the
physical pion mass, so that the main reason for this extrapolation is to guide
the small extrapolation of the F1M ensemble towards the physical pion mass.

Recalling how $\xi$ is constructed from the ratio of decay constants
and the square root of the ratio of bag parameters (compare
\eqref{eq:xi}) and noting that $\sqrt{B_{hs}/B_{hl}}$ is very close to
unity, we expect $f_{B_s}/f_B$ and $\xi$ to be very similar in
magnitude, in agreement with a comment made in
Ref.~\cite{Albertus:2010nm}.  This in turn implies that with a high
degree of accuracy the calculation of the SU(3) breaking ratio $\xi$
can be approximated by just studying the ratio of two-point functions
required for determining pseudoscalar decay constants.

\subsection{Fit ansatz} \label{subsec:GFansatz}
From our simulation data we determine observables $\mc{O}$ as a
function of the lattice spacing $a$, the finite volume $V$ and the
quark masses. To combine this with the experimental data, we need to
extra/interpolate our data to the physical values of the quark masses
as well as to the continuum ($a=0$) and infinite volume. Since quark
masses are experimentally not directly accessible quantities, we set
the heavy quark mass by inter/extrapolating the results to the
physical value of appropriate meson masses. We set the light quark
mass by extrapolating to the neutral pion mass of
$135\,\mathrm{MeV}$~\cite{Tanabashi:2018oca}. The charm (bottom)
quark mass is fixed by the heavy-light ($m_{D}$-like), heavy-strange
($m_{D_s}$-like) or heavy-heavy ($\eta_c$-like) pseudoscalar meson
mass. From our previous experience~\cite{Boyle:2017kli}, we find that
the chiral slope in our data and the continuum limit artifacts are
well described by terms linear in $\Delta m_\pi^2 \equiv
m_\pi^2-\left(m_\pi^\mathrm{phys}\right)^2$ and $a^2$, respectively.
In the past, we further found that the heavy quark behaviour is
captured well, by expanding in $\Delta m_H^{-1} \equiv
1/{m_H}-1/{m^\mathrm{expand}_H}$ where $m_H$ is the meson chosen to
set the heavy quark mass, and $m_H^\mathrm{expand}$ is the point
around which the expansion is performed. A small caveat arises from
the fact that the physical strange quark mass on the F1M ensemble was
not known prior to the data production, so the simulation was carried
out at the unitary value $am_s^\mathrm{uni}$ (c.f.~Table
\ref{tab:DWFls} and appendix \ref{sec:F1M}). We compensate for this
slight mistuning by including a term proportional to the mistuning
$\Delta m_s \equiv \left(am_s^\mathrm{phys} -
am_s^\mathrm{uni}\right)/am_s^\mathrm{phys}$. This term is only
non-zero for the F1M ensemble.

We therefore describe the data $\mc{O}(a,m_\pi,m_H)$ at given lattice
spacing $a$, pion mass ($m_\pi$) and heavy meson mass ($m_H$), by the fit ansatz
\begin{equation}
  f(a,m_\pi,m_H) = \mc{O}(0,m^\mathrm{phys}_\pi,m^\mathrm{phys}_H) +
  C_\chi \Delta m_\pi^2 + C_{CL} a^2 + C_H \Delta m_H^{-1} + C_s \Delta m_s\,.
  \label{eq:GFansatz}
\end{equation}
To check the validity and to estimate any systematic errors induced by this
ansatz, we systematically vary this ansatz and the data that enters the fit
(cuts). For example, we consider the impact of various pion mass cuts, the
exclusion of the heaviest data points etc. Finally, we will also estimate higher
order effects that are not captured by this fit form.

\subsection{Global fit strategy} \label{subsec:GFstrat}
In addition to the number of data points $N_\mathrm{obs}$ of the
observable under consideration ($f_{sh}/f_{lh}$, $B_{B_s}/B_{d}$ or
$\xi$), also the parameters that the expression in \eqref{eq:GFansatz}
depends on, enter the fit. These are the $N_\mathrm{obs}$ values of
the heavy meson mass $m_H$ (there is a corresponding meson mass for
each value of the observable), the $N_\mathrm{ens}$ values of the pion
masses (one per ensemble) and the $N_a$ values of distinct lattice
spacings (i.e. C1/2 and M1/2/3 share the same lattice spacings
respectively). We will collectively refer to these $N_x \equiv
N_\mathrm{obs}+ N_\mathrm{ens} +N_a$ values as $x_i$ and note that
their uncertainties have to be taken into account correctly. For the
meson masses $m_\pi$ and $m_H$ these arise from correlator fits and
are therefore fully correlated between the observables and each
other. However, this is not the case for the lattice spacing $a$,
since this was determined from a different analysis including a larger
set of gauge ensembles as described in
Refs.~\cite{Blum:2014tka,Boyle:2017jwu}. To propagate this
uncertainty, we generate a Gaussian bootstrap distribution with the
correct central value and match its width to the error.

The fit is then carried out via $\chi^2$ minimisation, where $\chi^2$ is defined
as

\begin{equation}
  \chi^2 = \sum_{i=1}^{N_\mathrm{tot}} \sum_{j=1}^{N_\mathrm{tot}} \left[y_i -
    F(\overline{x_i})\right] C^{-1}_{ij} \left[y_j - F(\overline{x_j})\right],
  \label{eq:chisqglobal}
\end{equation}
with $N_\mathrm{tot} = N_\mathrm{obs}+N_x$, (i.e.~$N_\mathrm{obs}$ values for the
observable, $N_\mathrm{obs}$ values for the corresponding heavy meson mass,
$N_\mathrm{ens}$ values of the pion mass and $N_\mathrm{a}$ values of the
lattice spacing). The $y_i$ in \eqref{eq:chisqglobal} are given by
\begin{equation}
  y_i =
  \begin{cases*}
    \mc{O}_i(a_i,{m_\pi}_i,{m_H}_i) & \text{for} $i \leq N_\mathrm{obs}$ \\
    x_i         & \text{otherwise}.
  \end{cases*}
\end{equation}
The appropriate values of $f(\overline{x_i})$ are given by
\begin{equation}
  F_i(\overline{x_i}) =
  \begin{cases*}
    f(\overline{a}_i,{\overline{m_\pi}}_i,{\overline{m_H}}_i) & \text{for} $i \leq N_\mathrm{obs}$ \\
    \overline{x_i}         & \text{otherwise.}
  \end{cases*}
\end{equation}
Since $C_{ij}$ is the \emph{full} covariance matrix (i.e.~of size
$N_\mathrm{tot} \times N_\mathrm{tot}$), this procedure takes all correlations
between the various data points (pion masses, heavy meson masses and
observables) into account.

In summary, the fit determines not only the parameters
$\mc{O}(0,m^\mathrm{phys}_\pi,m^\mathrm{phys}_H)$, $C_\chi$, $C_{CL}$ and $C_H$
but also re-determines the $x$ values. We note that this does not add any
degrees of freedom, since the same number of additional parameters that are
added to the fit are also re-determined by it. For the observables considered in
this work, we find that the relative error on the arguments of
\eqref{eq:GFansatz} are sufficiently small that the inclusion of the $x$-errors
only has a negligible effect (i.e.~the effect is far smaller than the
statistical error). We check that the output values ($\overline{x}_i$) are
within errors of the input values ($x_i$).

\subsection{Global fit results} \label{subsec:GFres}
We now present the results of the global fits described in the previous
sections. We choose as our central value the results obtained from a fit to the
data according to \eqref{eq:GFansatz} with a pion mass cut of
$350\,\mathrm{MeV}$ and the heavy mass being set by the heavy-strange
pseudoscalar mass. The central values and statistical errors of these fits are
\begin{equation}
  \begin{aligned}
    f_{D_s}/f_D = \fDsfDcentral\left(\fDsfDstatint \right)_\mathrm{stat} \quad &  f_{B_s}/f_B = \fBsfBcentral\left(\fBsfBstatint \right)_\mathrm{stat} &\quad  B_{B_s}/B_{B_d} =  \bagratcentral\left(\bagratstatint\right)_\mathrm{stat}\,
  \end{aligned}.
  \label{eq:GFres}
\end{equation}
The coefficients obtained from these fits together with the
goodness-of-fit measure $\chi^2/\mathrm{d.o.f.}$ and the associated
$p$-values are listed in Table \ref{tab:GFcoefs}. The coefficient
$C_s$ is small and compatible with zero in all cases, indicating that
the strange quark mass mistuning has no significant effect and we
confirmed that performing the same fit without such a term leads to
fully compatible results. We note that the $\chi^2/\mathrm{d.o.f.}$
values of all three fits are excellent, producing good
$p$-values. This is remarkable, given the small number of fit
parameters (5) and the large number of degrees of freedom (16).
\begin{table}
  \begin{center}
    \begin{tabular}{c|cccc|ccc}
      \hline\hline
      observable & $C_{CL}\,[\mathrm{GeV}^2]$ & $C_\chi\,[\mathrm{GeV}^{-2}]$ &  $C_H\,[\mathrm{GeV}]$ & $C_s$ & $\mathrm{d.o.f.}$ & $\chi^2/\mathrm{d.o.f.}$ & $p$-value \\\hline
      $f_{hs}/f_{hl}$   & \fDfDsCCL & \fDfDsCchi & \fDfDsCh & \fDfDsCs & \fDfDsdof & \fDfDschidof & \fDfDspval\\
      $B_{B_s}/B_{B_d}$  & \bagCCL & \bagCchi & \bagCh & \bagCs & \bagdof & \bagchidof & \bagpval\\
      $\xi$            &\xiCCL & \xiCchi & \xiCh & \xiCs & \xidof & \xichidof & \xipval\\
      \hline \hline
    \end{tabular}
  \end{center}
  \caption{Results of the base fit (i.e.~$m_\pi^\mathrm{max} =
    350\,\mathrm{MeV}$ and $M_H = m_{sh}$). We list the determined
    coefficients for the continuum limit slope ($C_{CL}$), pion mass
    dependence ($C_\chi$), heavy mass dependence ($C_H$) and strange
    mistuning ($C_s$), the number of degrees of freedom in the fit as
    well as the goodness of fit measures $\chi^2/\mathrm{d.o.f.}$ and
    the corresponding $p$-values.}
  \label{tab:GFcoefs}
\end{table}

\begin{figure}
  \begin{center}
    \includegraphics[width=.49\textwidth]{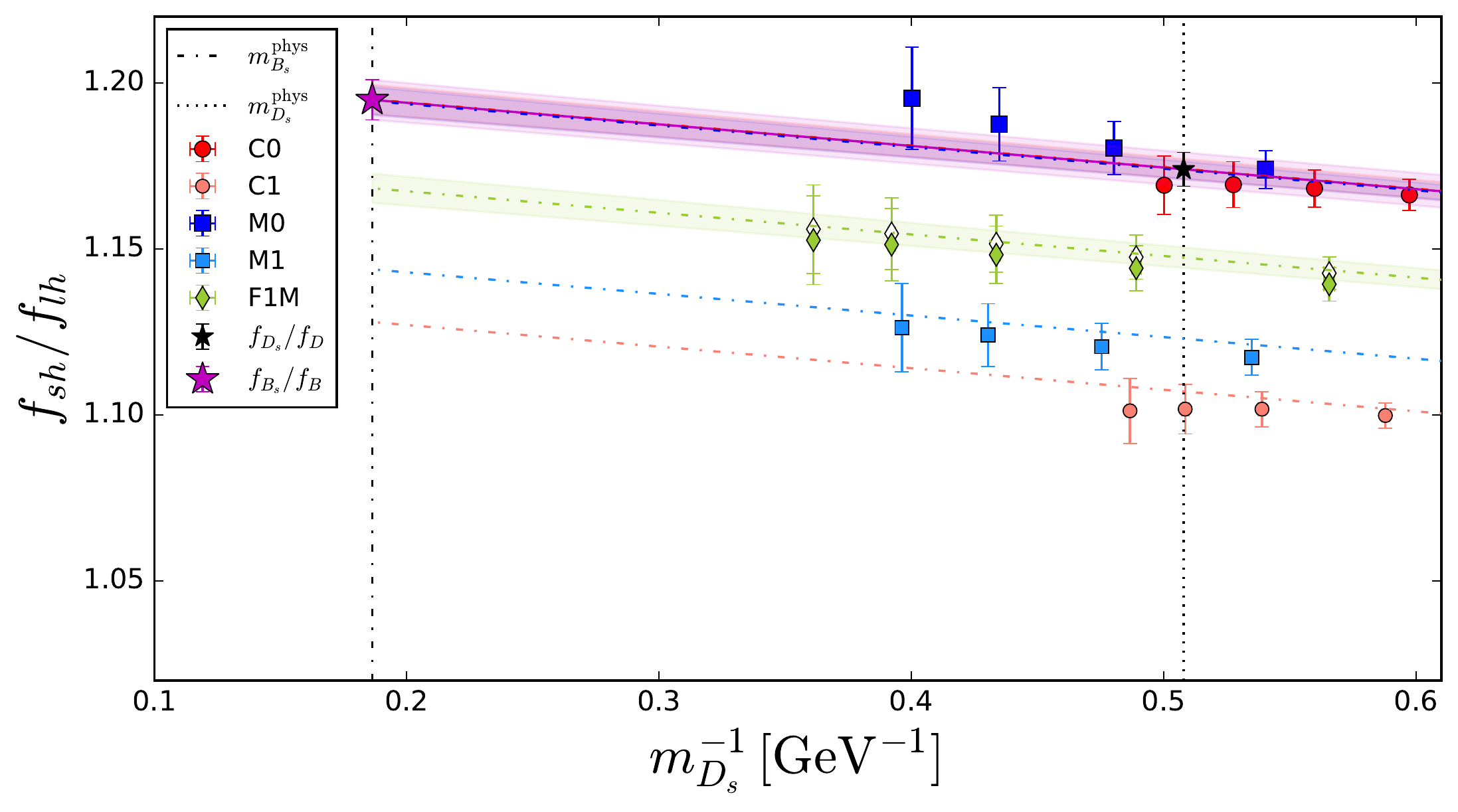}
    \includegraphics[width=.49\textwidth]{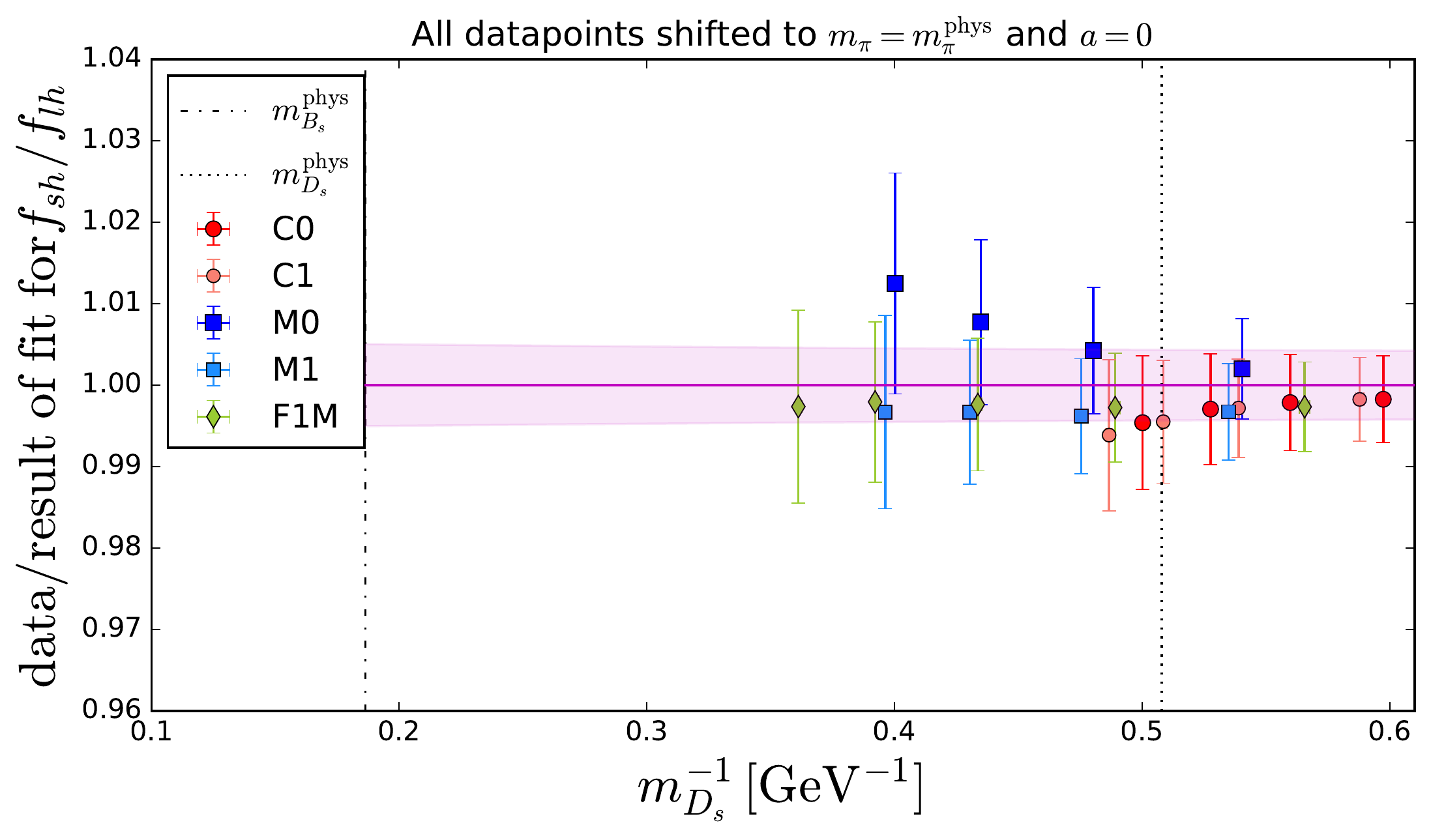}
  \end{center}
  \caption{The global fit result for the ratio of decay constants as a
    function of the inverse heavy meson mass (left) and between the
    data and the fit results (right) as described in equation
    \eqref{eq:GFfraction}. The red circles (blue squares, green
    diamonds) show the data for the coarse (medium, fine) ensembles
    that enter the fit. The open green diamonds show the correction
    due to the strange quark mass mistuning. The magenta line shows
    the fit function evaluated at physical pion masses in the
    continuum. The magenta band illustrates the statistical error. The
    black and magenta stars show the result (statistical error only)
    for $f_{D_s}/f_D$ and $f_{B_s}/f_B$, respectively.}
  \label{fig:GFdecrat_notproj}
  \label{fig:GFdecrat_massrat}
\end{figure}

The left panel of Figure \ref{fig:GFdecrat_notproj} shows the data for
the ratio of decay constants entering our preferred fit together with
the fit result (magenta band) for the case of the ratio of decay
constants. The coloured bands and dashed lines show the fit function
(c.f.~equation \eqref{eq:GFansatz}) evaluated at the physical strange
quark mass and the respective pion masses and lattice spacings for
each ensemble. The closed and open green diamonds show the F1M data
before and after the adjustement for the strange quark mass mistuning,
respectively.

We stress that due to the high degree of correlation of the data points on a
given ensemble, care needs to be taken when trying to consider the contribution
to the value of $\chi^2$ from a given data point. In the right panel of Figure
\ref{fig:GFdecrat_massrat} we present the data corrected to the physical pion
mass and vanishing lattice spacing, normalised by the heavy mass behaviour. More
precisely we show
\begin{equation}
  \frac{\mathcal{O}(a,m_\pi,m_H) - C_{CL} a^2 - C_\chi \Delta m_\pi^2 - C_s \Delta m_s}{f(0,m_\pi^\mathrm{phys},m_H)}.
  \label{eq:GFfraction}
\end{equation}
This illustrates that all data points are compatible with the fit at the $\sim 1
\sigma$ level. We observe the above mentioned correlations by noting that data
points on a given ensemble remain at a roughly constant distance from the fit.
This lends further confidence in our description of the behaviour as the heavy
mass is varied. Whilst the goodness-of-fit for the presented fit is excellent,
we note that the largest contribution to the $\chi^2/\mathrm{d.o.f.}$ arises from
the ensemble M0. This is conservatively addressed in our systematic error
analysis by investigating different choices of pion mass cuts, leading to one of
our dominant systematic errors.

\begin{figure}
  \begin{center}
    \includegraphics[width=.8\textwidth]{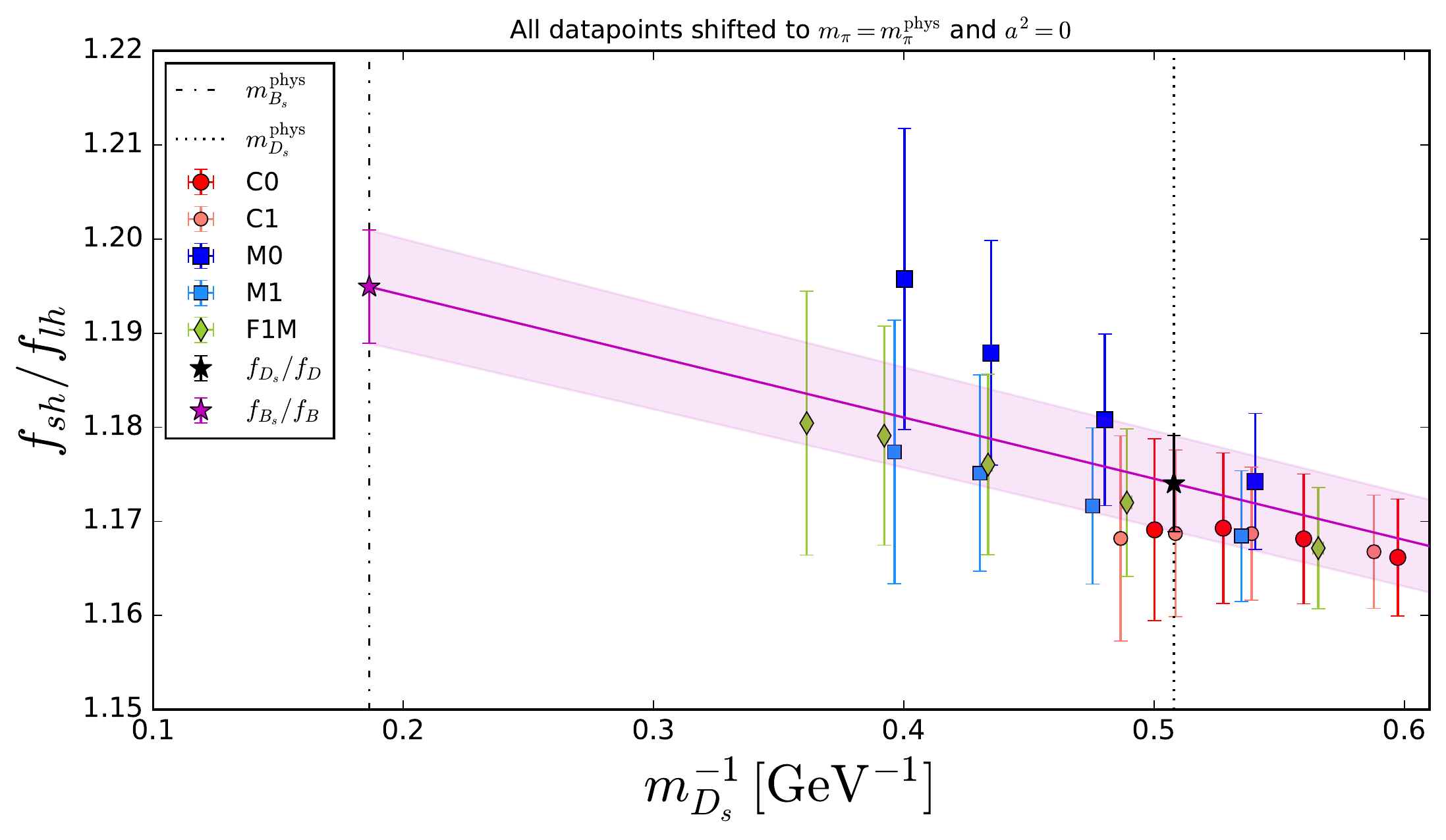}
  \end{center}
  \caption{Chosen global fit for the ratio of decay constants. All data points
    are shifted to the physical pion mass and zero lattice spacing, so the plot
    shows the behaviour as a function of the inverse heavy mass.}
  \label{fig:GFdecrat_mh}
\end{figure}

\begin{figure}
  \begin{center}
    \includegraphics[width=.45\textwidth]{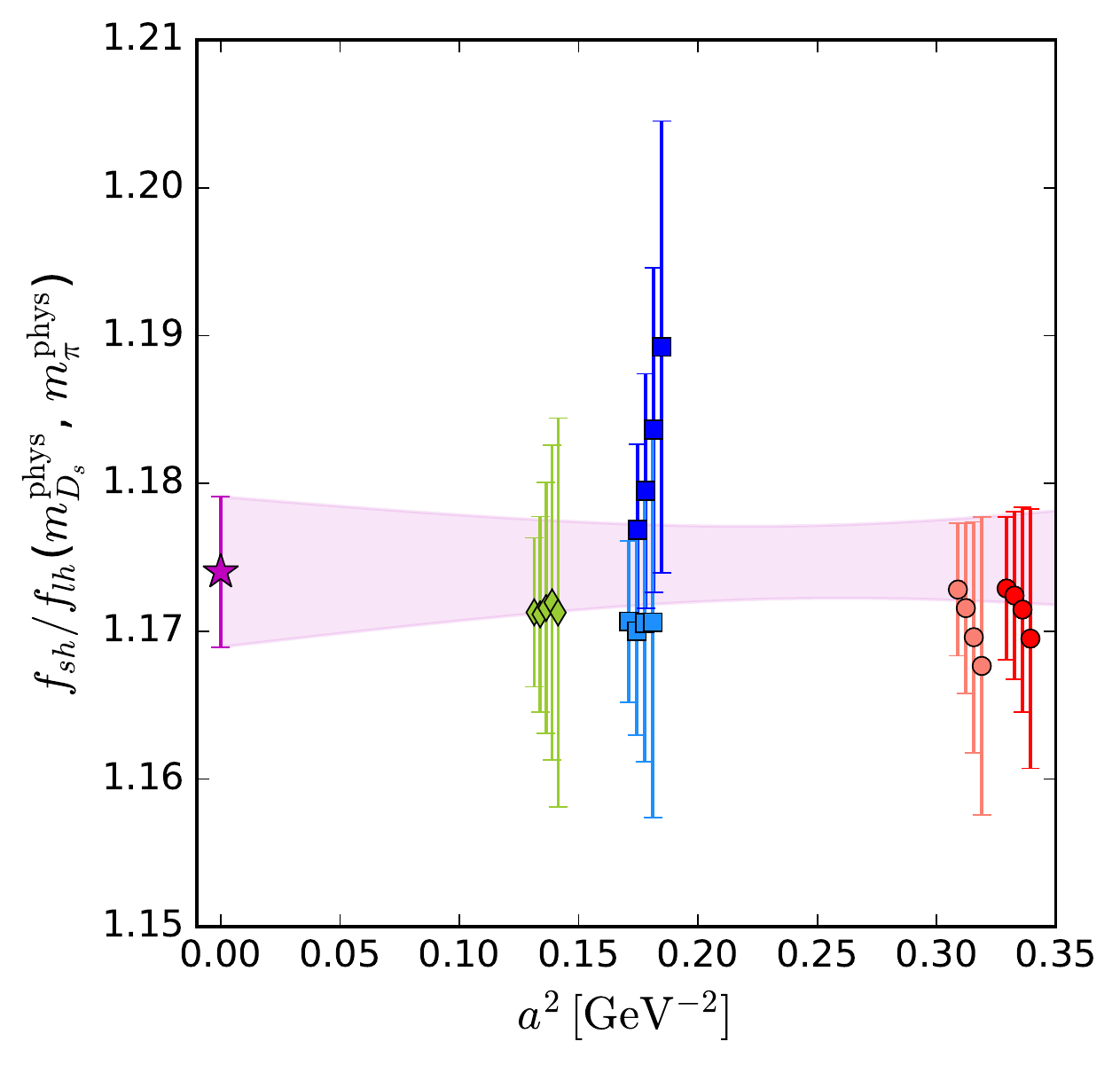}
    \includegraphics[width=.45\textwidth]{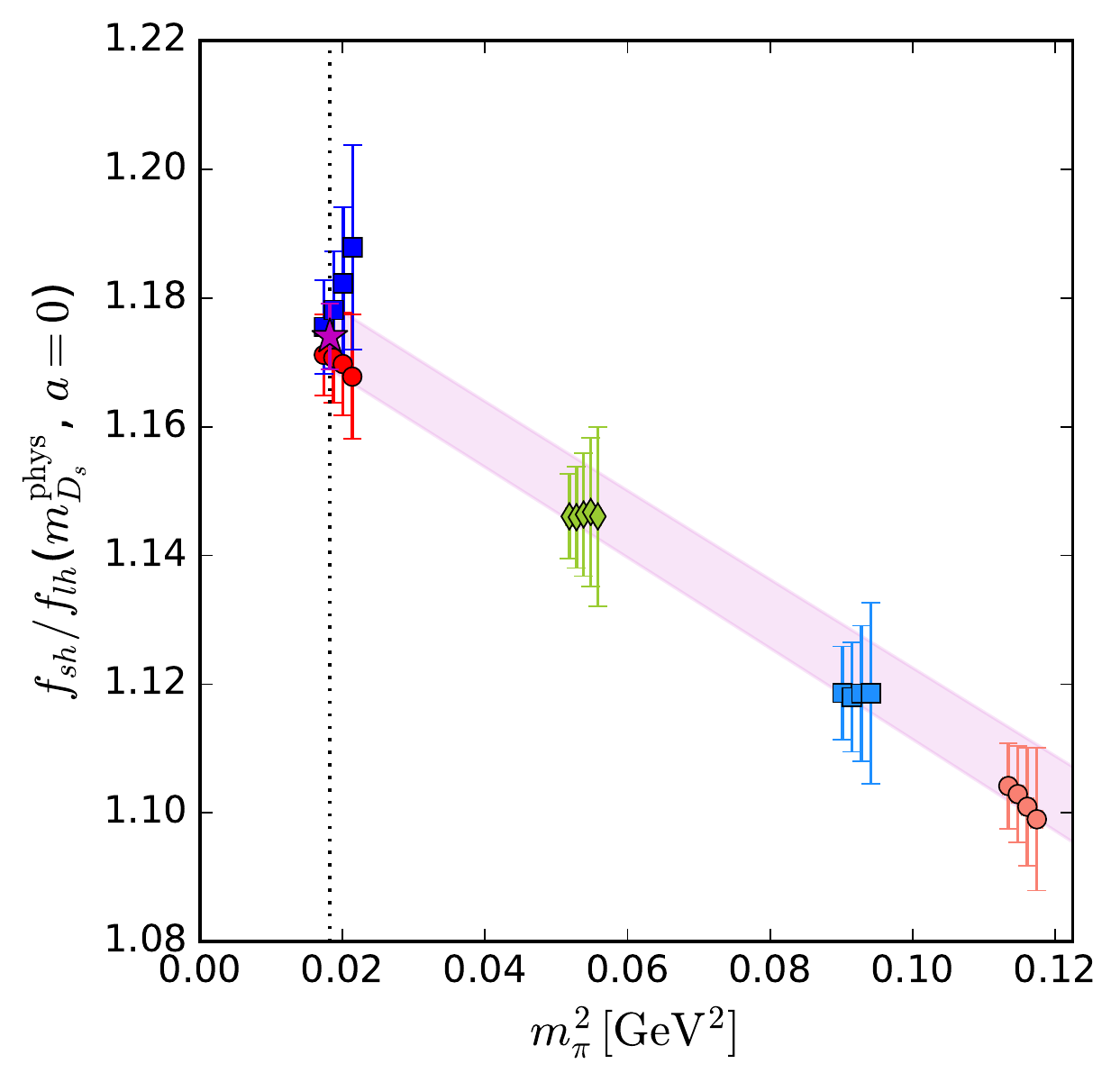}\\
  \end{center}
  \caption{Global fit result for the ratio of decay constants. All data points
    are projected to the physical charm quark mass, set via $m_{D_s}$. The data
    points on the left (right) panels are also shifted to the physical pion mass
    (zero lattice spacing) and hence illustrates the scaling (chiral) behaviour
    of our data. We slightly shift data points along the horizontal axis for
    better visualisation of the different data points.}
  \label{fig:GFdecratprojection}
\end{figure}

To expose the functional behaviour with respect to each of the three parameters
($a^2$, $m_\pi^2$ and $m_H^{-1}$) expression \eqref{eq:GFansatz} depends on, we
shift the data points to their physical values along two of these three
directions, to validate the behaviour in the third. Figure \ref{fig:GFdecrat_mh}
shows the data points shifted to the physical pion mass and after discretisation
effects have been removed. Note the change in the $y$-axis between Figures
\ref{fig:GFdecrat_notproj}, \ref{fig:GFdecrat_mh} and
\ref{fig:GFdecratprojection}. The data points display a very linear behaviour
all the way from the lightest data point (below $D_s$ which is shown by the
vertical dotted line) up to heaviest data point (at approximately half the $B_s$
mass, which is indicated by the vertical dash-dotted line). This linear
behaviour allows us to extrapolate our results to obtain results at the
$b$-quark mass. The difference between the data at the charm mass and at the
bottom mass is only of the order of $\sim 3\%$, making this extrapolation very
benign. We note that this is largely due to the fact that the heavy quark
behaviour cancels in the ratio of decay constants. Figure
\ref{fig:GFdecratprojection} shows the projections of the data points shown in
the left panel of Figure \ref{fig:GFdecrat_notproj} to the physical charm quark
mass, set by the $D_s$ mass. In the left (right) panel the data points are also
shifted to the physical pion mass (zero lattice spacing), so that we can compare
the continuum limit (pion mass) behaviour with the data. We see that the
continuum limit is rather flat (cf. coefficients $C_{CL}$ in Table
\ref{tab:GFcoefs}), with discretisation effects of around one percent for the
coarsest ensemble. The behaviour with $m_\pi^2$ is stronger, as expected for an
$SU(3)$ breaking ratio, with the ensemble at $m_\pi \sim 340\,\mathrm{MeV}$
differing by $\sim -6\%$ compared to the physical pion mass. This is very well
described by the linear ansatz in $\Delta m_\pi^2$. We emphasise that since our
simulation includes two ensembles at the physical pion mass, the main impact of
this slope is to guide the small extrapolation on the fine ensemble to the
physical pion mass. The same behaviour is observed for the projection to the
physical $b$-quark mass, and we refer to these very similar looking plots
(cf.~Figure \ref{fig:GFdecratprojection_atb}) in appendix \ref{sec:globalfits}.

\begin{figure}
  \begin{center}
    \includegraphics[width=.8\textwidth]{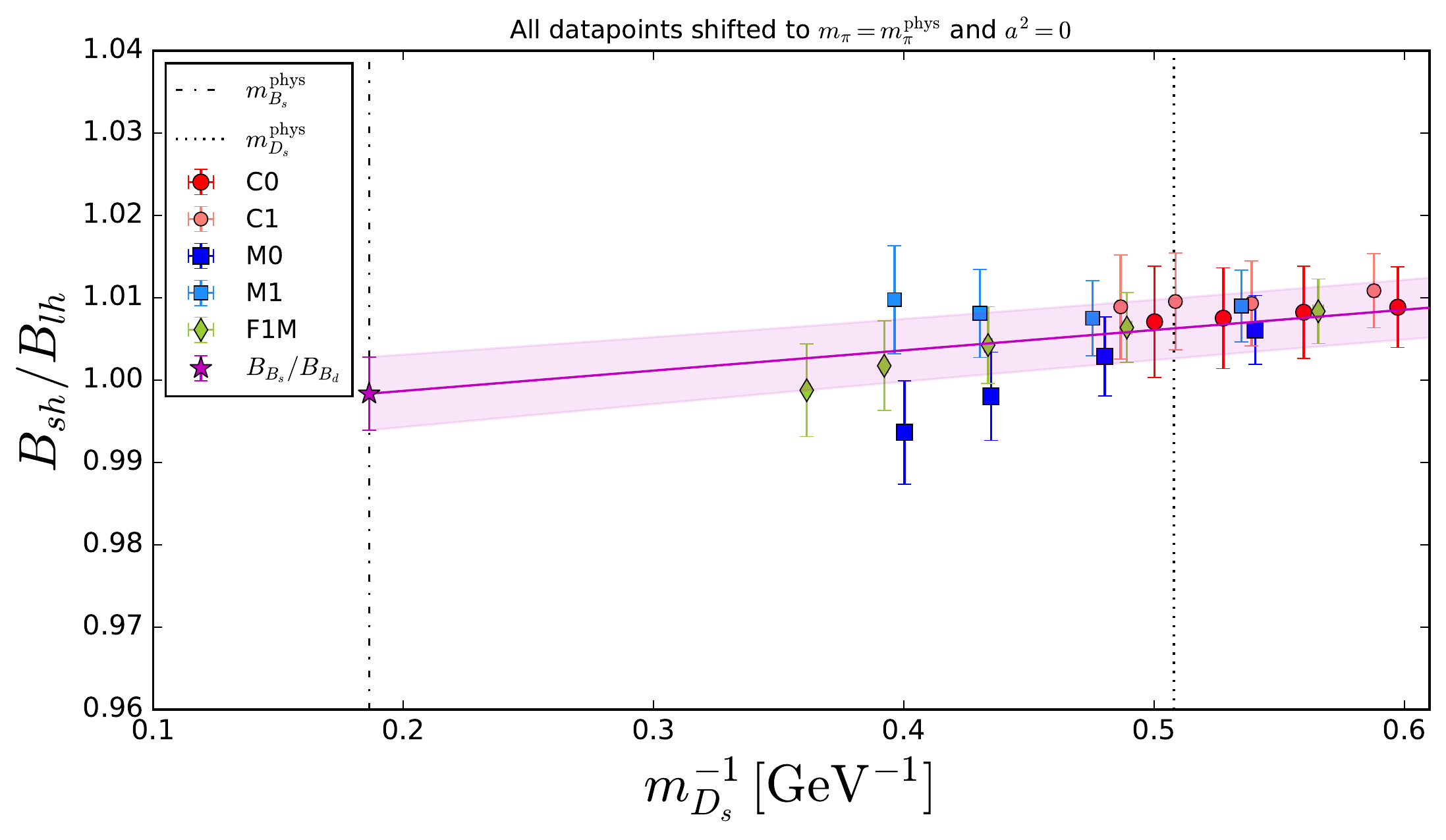}
  \end{center}
  \caption{Global fit result for $B_{B_s}/B_B$. All data points are shifted to the physical pion mass and zero lattice spacing.}
  \label{fig:GFbagrat_mh}
\end{figure}

\begin{figure}
  \begin{center}
    \includegraphics[width=.45\textwidth]{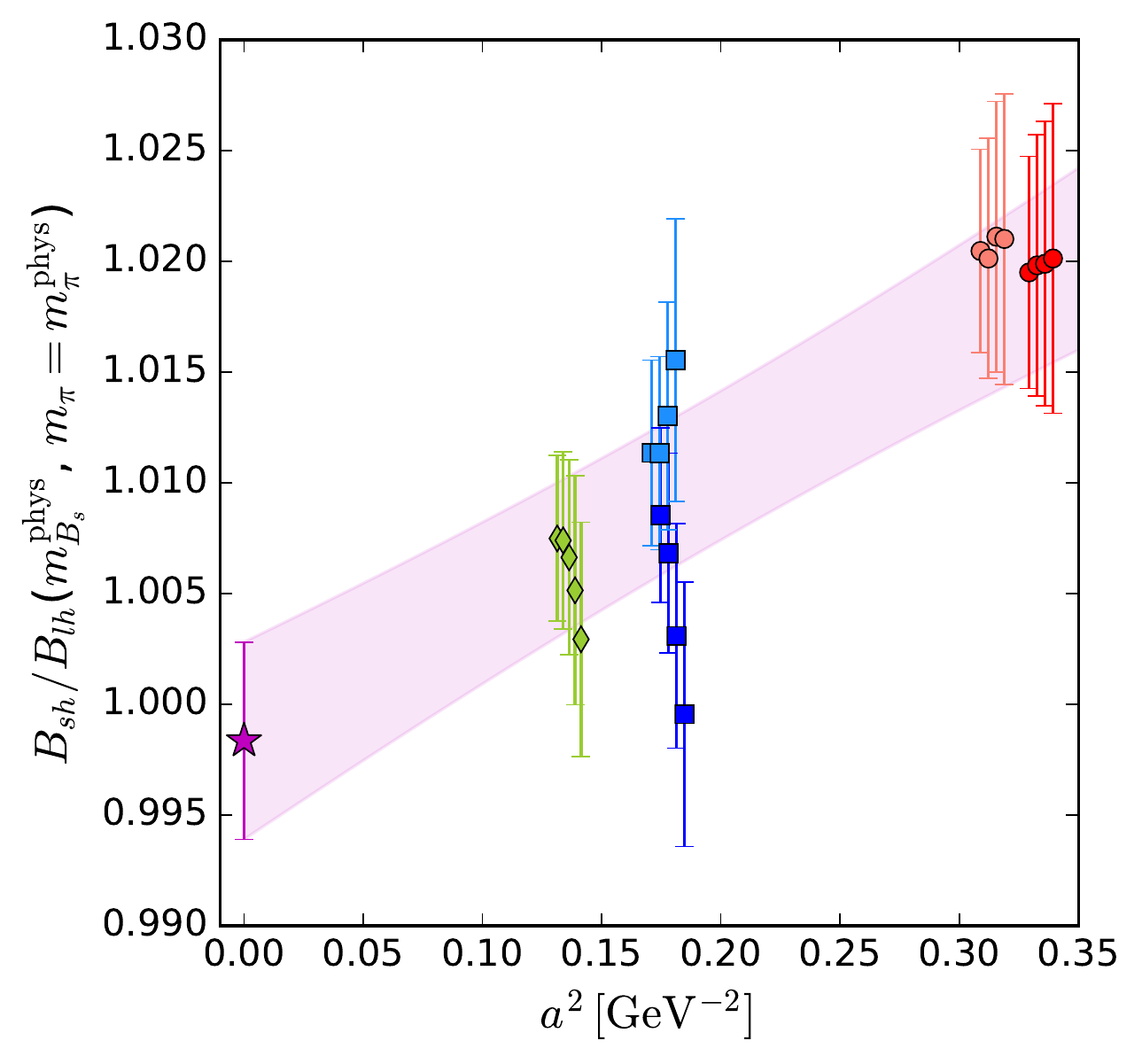}
    \includegraphics[width=.45\textwidth]{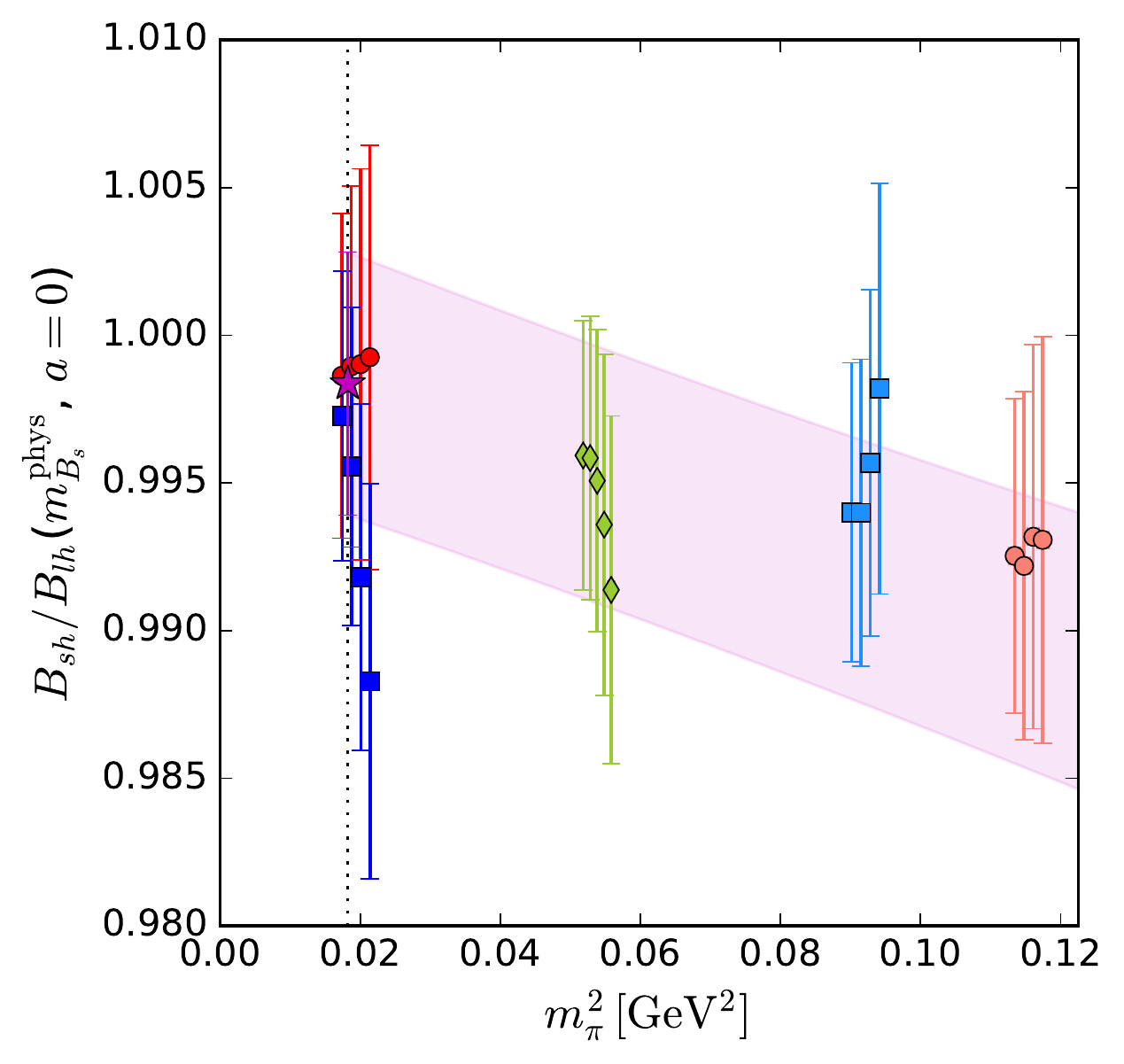}\\
  \end{center}
  \caption{Continuum limit and pion mass dependence obtained from the chosen global fit for $B_{B_s}/B_B$.}
  \label{fig:GFbagratprojection}
\end{figure}

For the ratio of bag parameters $B_{B_s}/B_{B_d}$, the discretisation
effects are very similar to the above. The chiral behaviour is
suppressed with a coefficient that is roughly an order of magnitude
smaller. So a pion mass of $\mathrm{340}\,\mathrm{MeV}$ only leads to
a difference of $\sim 1\%$ compared to the physical value. The
behaviour with the heavy mass, is very benign and similar in magnitude
to the ratio of decay constants, but opposite in sign. These results
are summarised in Figures \ref{fig:GFbagrat_mh} and
\ref{fig:GFbagratprojection}.

We can obtain the observable $\xi$ in two ways: We can construct
$\xi(a,m_\pi,m_H)$ ensemble by ensemble and perform the global fit (see equation
\eqref{eq:GFansatz}) on this quantity. Alternatively we can take the output of
the global fit for $f_{B_s}/f_B$ and $B_{B_s}/B_{B_d}$ and then construct $\xi$
from these outputs (via equation \eqref{eq:xi}) in the continuum limit and after
the extrapolation to physical masses \emph{but including all correlations}. We
will refer to the former as \emph{direct} and latter as \emph{indirect}
determinations. The results with statistical error of these two are
\begin{equation}
  \xi = \xidircentral\left(\xidirstatint\right)_\mathrm{direct} \qquad \qquad \qquad
  \xi = \xicentral\left(\xistatint\right)_\mathrm{indirect}.
  \label{eq:GFxires}
\end{equation}
The central value remains within the statistical uncertainty, but the
statistical error of the indirect determination is reduced by roughly
30\%. This occurs due to stronger cancellations of statistical errors
in the individual ratios $f_{hs}/f_{hl}$ and $B_{sh}/B_{hl}$ as
opposed to the direct construction of $\xi$ where some of these
correlations appear to be washed out. We will therefore take the
indirect determination as our preferred value.

\begin{figure}
  \begin{center}
    \includegraphics[width=.8\textwidth]{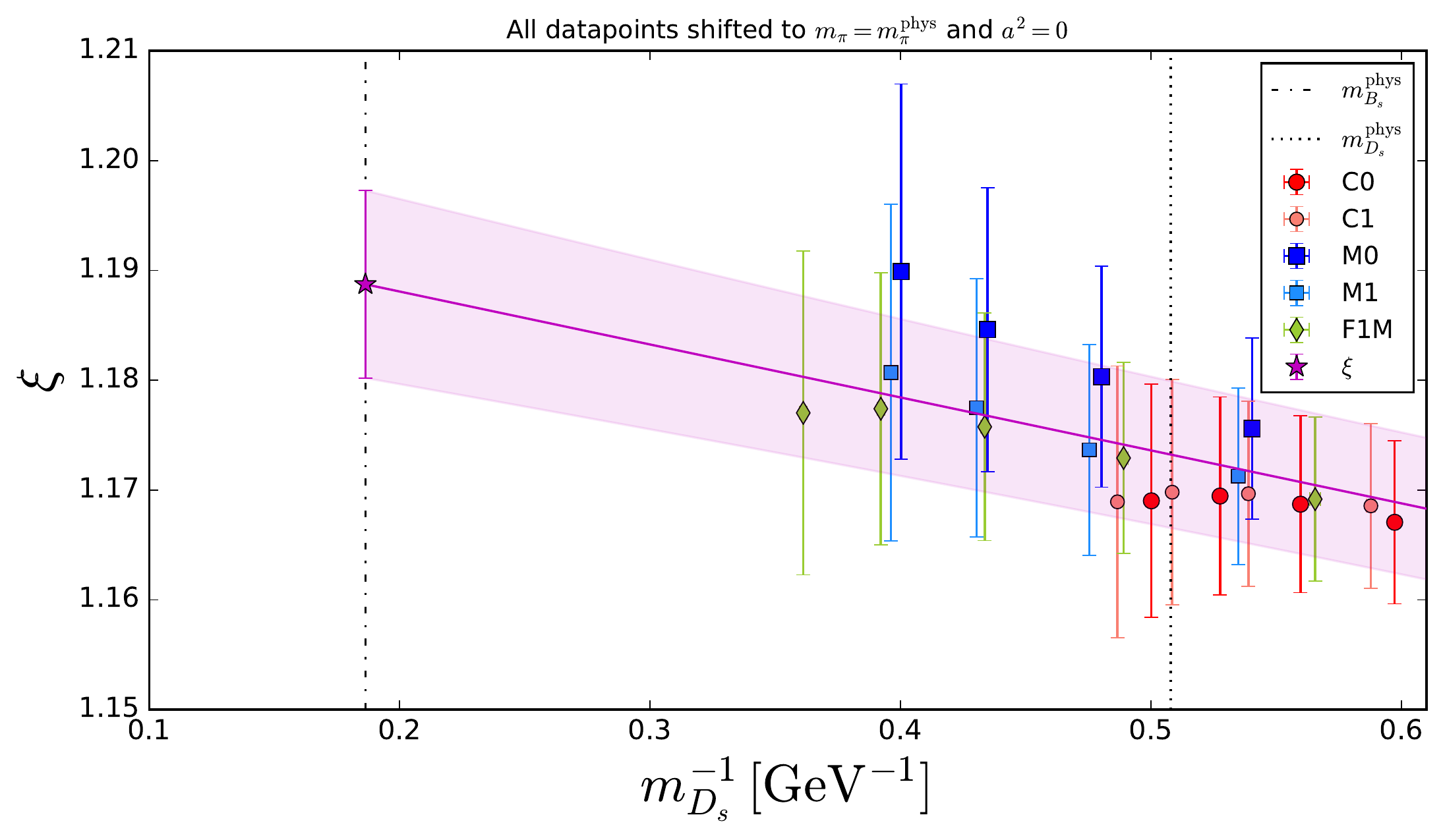}
  \end{center}
  \caption{Global fit result for $\xi$. Similar to Figure \ref{fig:GFdecrat_mh},
    all data points are shifted to the physical pion mass and zero lattice
    spacing.}
  \label{fig:GFxi_mh}
\end{figure}

\begin{figure}
  \begin{center}
    \includegraphics[width=.45\textwidth]{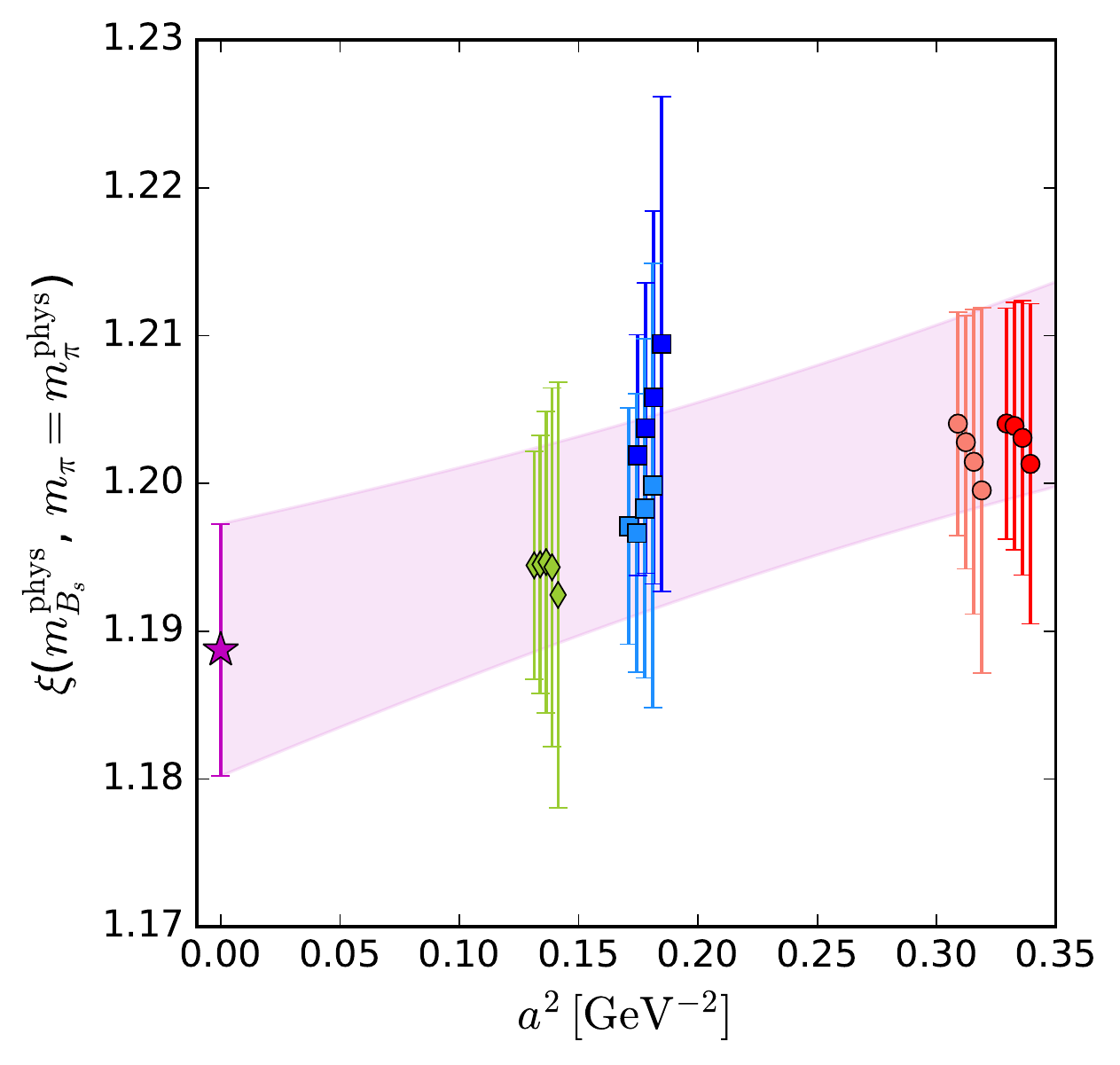}
    \includegraphics[width=.45\textwidth]{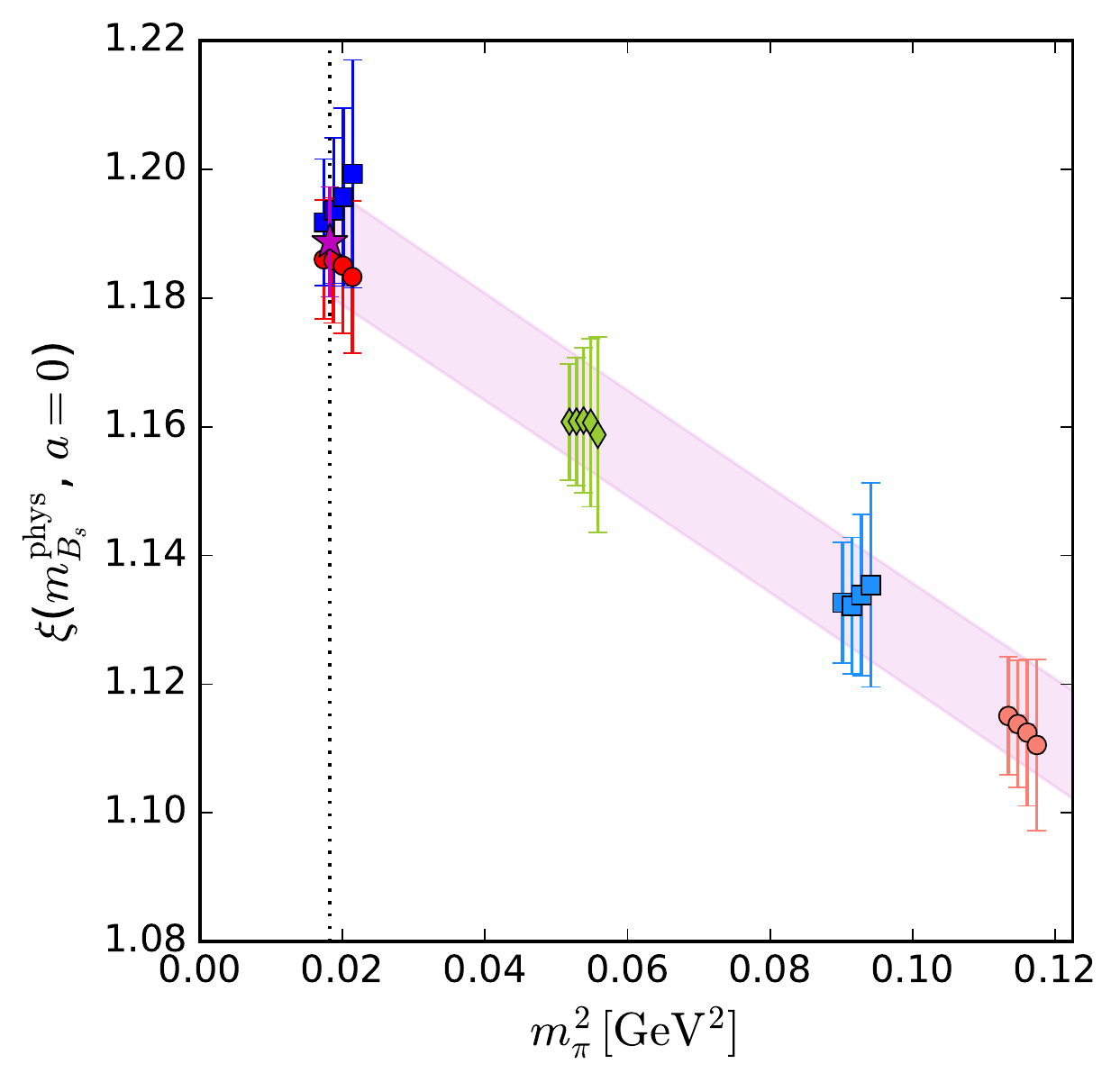}\\
  \end{center}
  \caption{Continuum limit and pion mass dependence obtained from the chosen
    global fit for $\xi$. All data points are shifted to the physical $B_s$
    meson mass and the physical pion mass (left) or vanishing lattice spacing
    (right).}
  \label{fig:GFxiprojection}
\end{figure}

The results of the direct determination are presented in Table \ref{tab:GFcoefs}
and Figures \ref{fig:GFxi_mh} and \ref{fig:GFxiprojection}. Figure
\ref{fig:GFxi_mh} again shows the data points shifted to the physical pion mass
and zero lattice spacing. The heavy mass behaviour displayed is very similar to
the case of the ratio of decay constants, with the data being well described by
a linear term in the inverse heavy meson mass that is chosen (here
$m_{D_s}$-like mesons). The two panels in Figure \ref{fig:GFxiprojection} show
the projections of the data points shown in Figure \ref{fig:GFxi_mh} to the
physical $B_s$ mass. The same observations as in the case of the ratio of decay
constants hold true for this case. However, the approach to the continuum is
slightly steeper with discretisation effects on the coarsest ensemble being
$\sim2\%$ (compare $C_{CL}$ in Table \ref{tab:GFcoefs}).

\subsection{Error budget}
\begin{figure}
  \begin{center}
    \includegraphics[width=.9\textwidth]{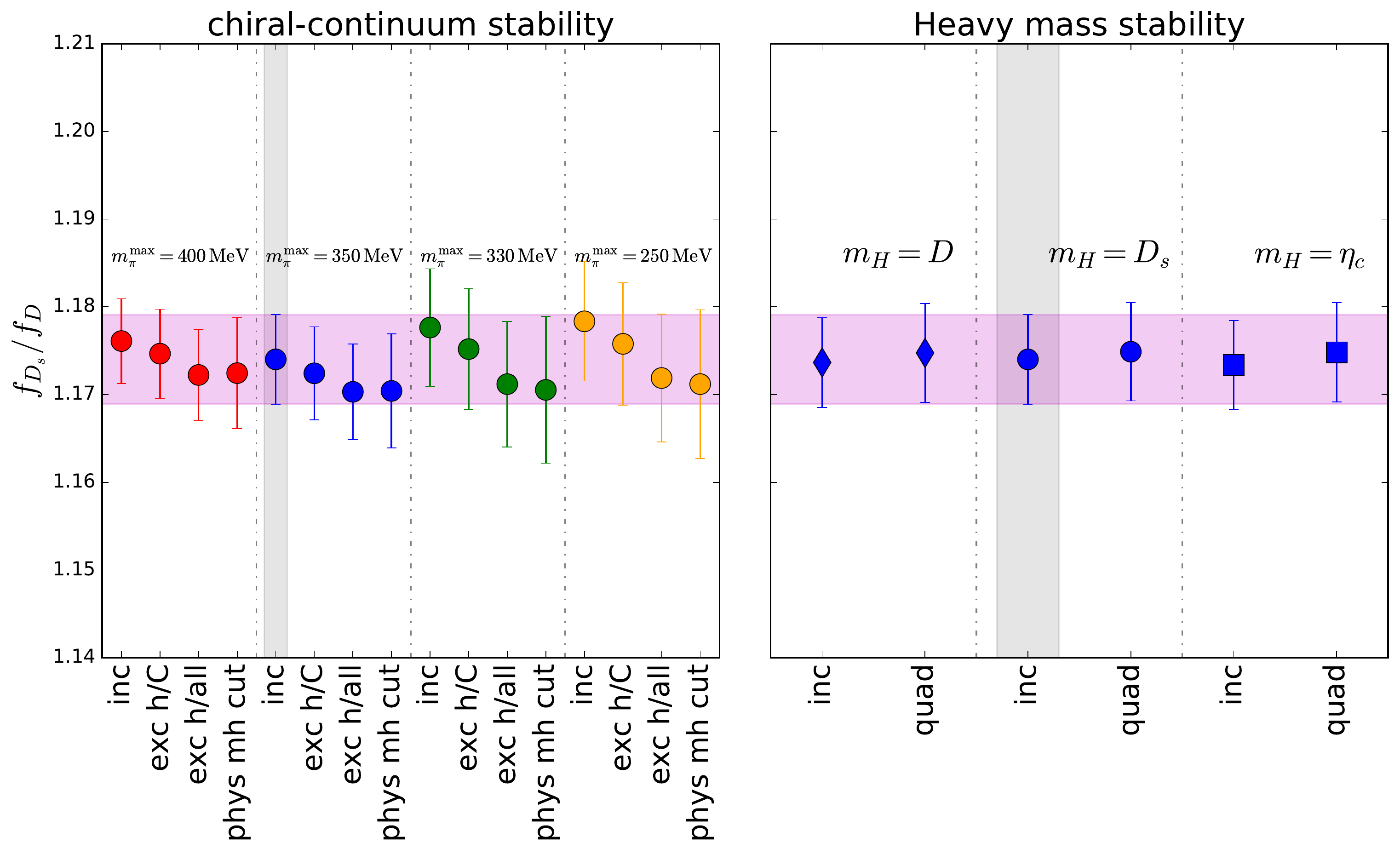}
  \end{center}
  \caption{Stability of fit results for $f_{D_s}/f_D$. The left plot shows
    variations of the pion mass cuts (separated by dotted vertical lines, from
    left to right) between $m_\pi^\mathrm{max}=400, 350, 330$ and
    $250\,\mathrm{MeV}$. The right hand panel compares different approaches for
    the heavy quark interpolation. More detail can be found in the text.}
  \label{fig:GFstability_fDsfD}
\end{figure}

\begin{figure}
  \begin{center}
    \includegraphics[width=.9\textwidth]{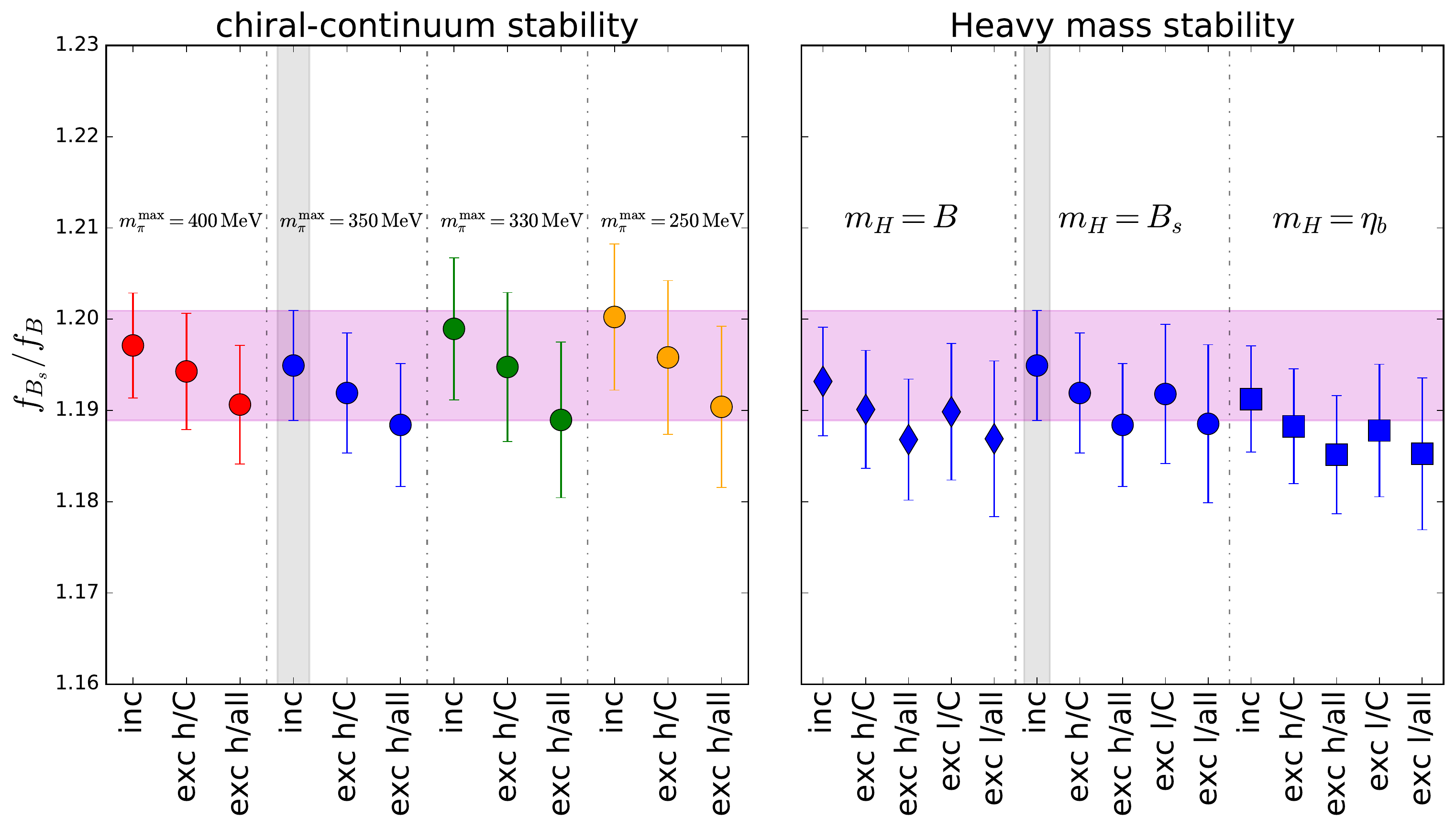}
  \end{center}
  \caption{Stability of fit results for $f_{B_s}/f_B$. The left plot
    shows variations of the pion mass cuts (separated by dotted
    vertical lines, from left to right) between
    $m_\pi^\mathrm{max}=400, 350, 330$ and $250\,\mathrm{MeV}$. The
    right plot shows variations of the fit for different choices of
    the heavy pseudoscalar meson mass used to set the bottom quark
    mass (triangles for $H=B$, circles for $H=B_s$, squares for
    $H=\eta_b$) as well as different different cuts to the
    data which are described in more detail in the text.}
  \label{fig:GFstability_fBsfB}
\end{figure}

\begin{figure}
  \begin{center}
    \includegraphics[width=.9\textwidth]{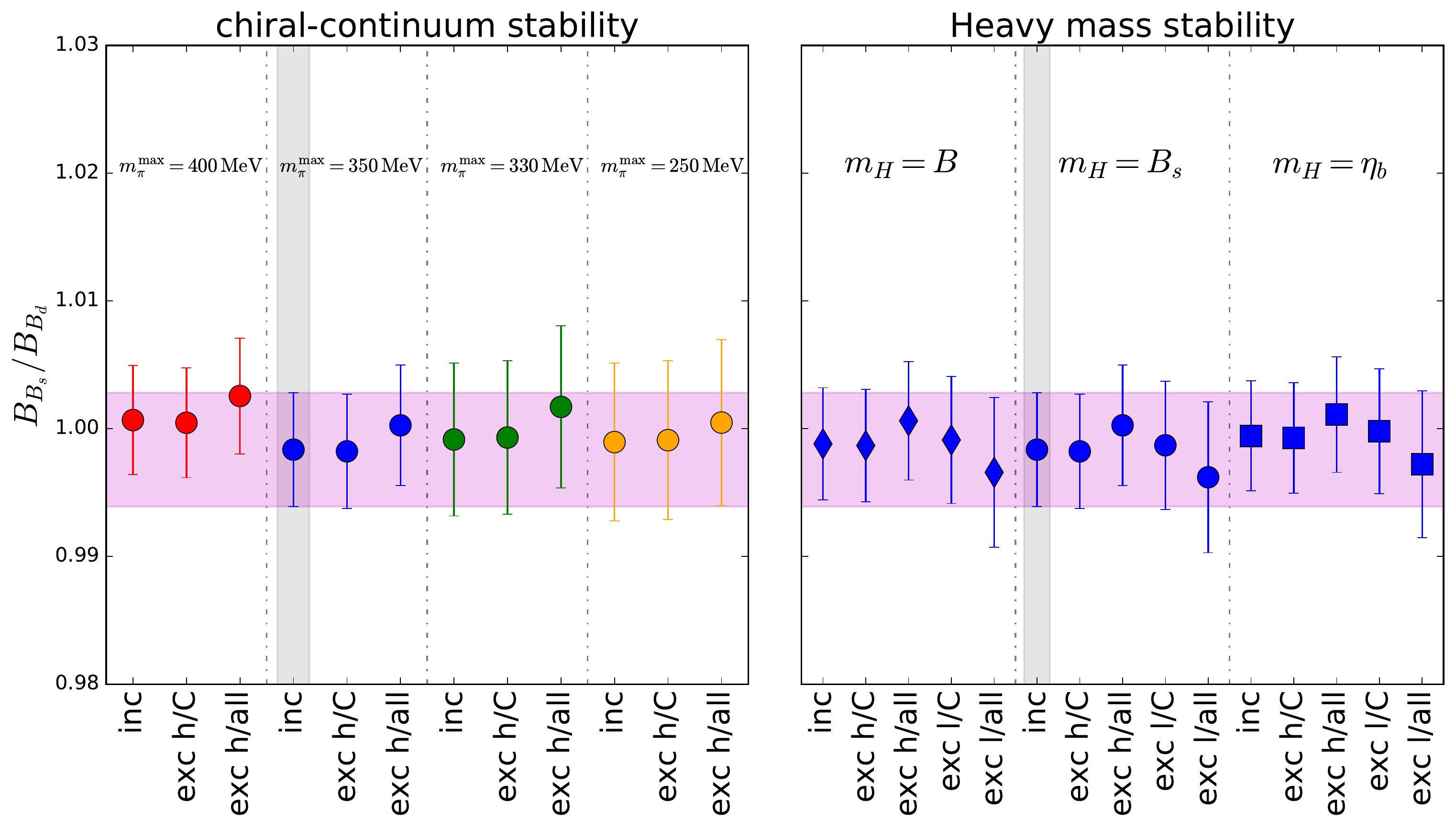}
  \end{center}
  \caption{Variations of the fit result for the ratio of bag parameters
    $B_{B_s}/B_{B_d}$, analogous to Figure \ref{fig:GFstability_fBsfB}}
  \label{fig:GFstability_bagrat}
\end{figure}

\begin{figure}
  \begin{center}
    \includegraphics[width=.9\textwidth]{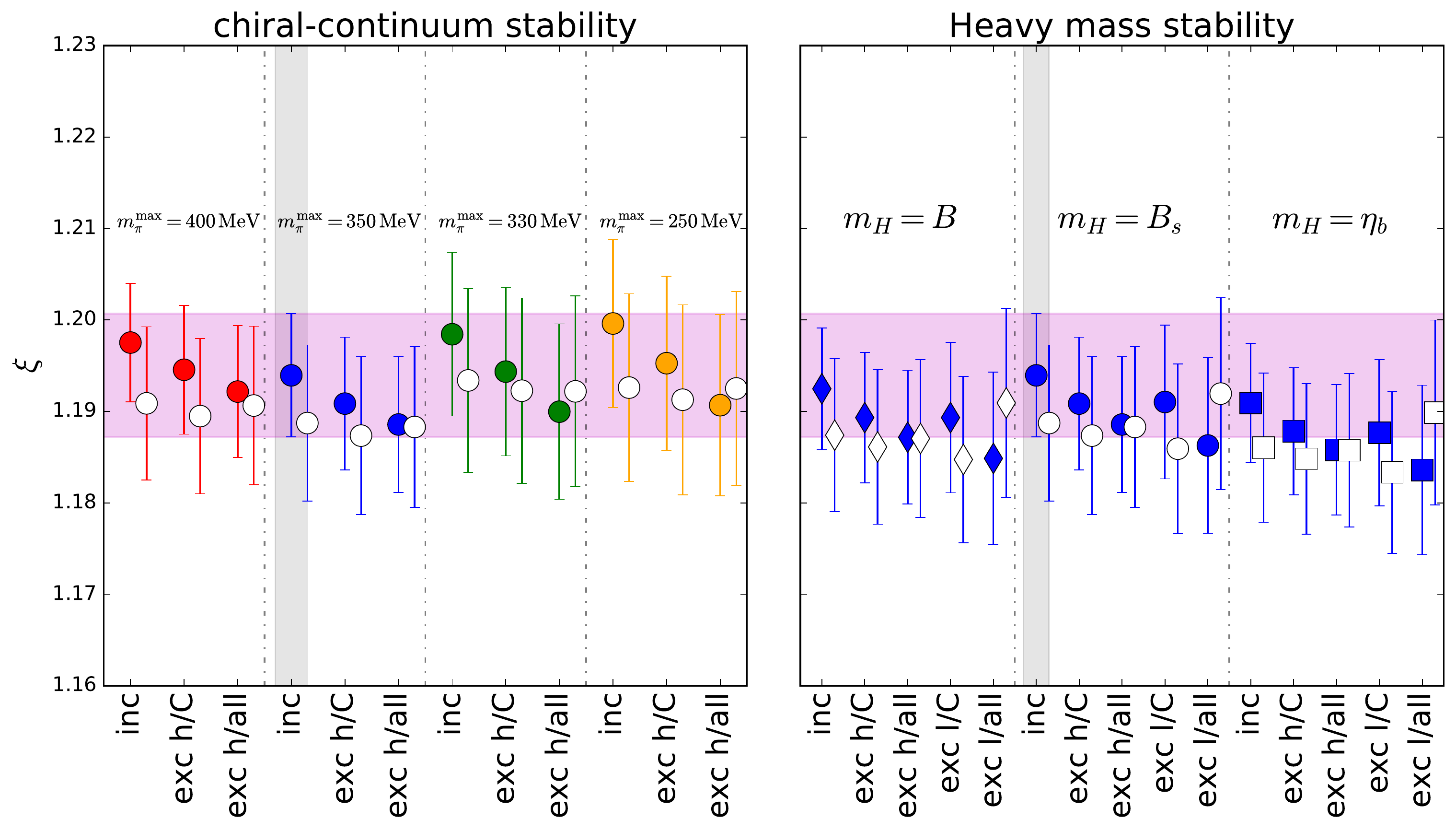}
  \end{center}
  \caption{Stability of fit results for $\xi$ analogous to Figures
    \ref{fig:GFstability_fBsfB} and \ref{fig:GFstability_bagrat}. The closed
    data points display the \emph{indirect} determinations of $\xi$, whilst the
    open symbols show the direct determinations (cf.~equation
    \eqref{eq:GFxires}).}
  \label{fig:GFstability_xi}
\end{figure}

We will now estimate the various systematic errors. These are
tabulated in Table \ref{tab:errorbudget}.  We start by considering the
systematic errors due to the global fitting procedure. To this end, we
consider variations in the fit ansatz. First we compare the results
obtained from the global fit for different pion mass cuts. The results
for all of these fits are listed in Tables \ref{tab:GFfDsfD},
\ref{tab:GFfBsfB}, \ref{tab:GFBBsBB} and \ref{tab:GFxi}. We note some
of the fits including the heavier pion mass ensembles display a poorer
fit quality (some with unacceptable values of
$\chi^2/\mathrm{d.o.f.}$). This occurs in particular for
$m_\pi^\mathrm{max} = 450\,\mathrm{MeV}$ and $m_\pi^\mathrm{max} =
430\,\mathrm{MeV}$. For this reason we only consider the pion mass
cuts of $m_\pi^\mathrm{max}=400, 350, 330$ and $250\,\mathrm{MeV}$
(the red, blue, green and yellow data points in Figures
\ref{fig:GFstability_fDsfD} - \ref{fig:GFstability_xi}, respectively)
for our systematic error estimation. These cuts successively eliminate
ensembles from the fit. For the pion mass cut of $250\,\mathrm{MeV}$
the fit in the form of \eqref{eq:GFansatz} becomes insufficiently
constrained. Noting that the $C_s$ parameter is compatible with zero
(cf.~Table \ref{tab:GFcoefs}) and $C_s$ as well as $\Delta m_s$ are
small numbers, we drop this term from the fits with the
$250\,\mathrm{MeV}$ pion mass cut.

Furthermore we consider including (``inc'') all data points or
excluding the heaviest data point on all coarse (``exc h/C'') or all
(``exc h/all'') ensembles. This choice is justified, since we expect
the heaviest mass points and the coarse ensembles to be most strongly
afflicted by discretisation effects. For the case of $f_{D_s}/f_D$,
for which we have data bracketing the physical value on all ensembles,
we also consider a fit where a physical mass cut of
$1.76\,\mathrm{GeV} \leq m_{D_s}^\mathrm{PDG} \leq 2.15\,\mathrm{GeV}$
is applied to the heavy-strange meson mass (labelled as ``phys mh
cut''). The left hand panels of Figures \ref{fig:GFstability_fDsfD},
\ref{fig:GFstability_fBsfB}, \ref{fig:GFstability_bagrat} and
\ref{fig:GFstability_xi} list the outcomes of these variations for
$f_{D_s}/f_D$, $f_{B_s}/f_{B}$, $B_{B_s}/B_{B_d}$ and $\xi$,
respectively. We conservatively assign a systematic error due to the
chiral-continuum limit part of the fit as the maximum spread of the
central values from the chosen fit. This is labelled ``fit chiral-CL''
in Table \ref{tab:errorbudget}. All of these variations remain within
the quoted statistical error. We use the pion mass cut of
$350\,\mathrm{MeV}$ as our central value, since this better
constrains the coefficients (therefore fully exploiting the third
lattice spacing) whilst giving an excellent goodness-of-fit. This
choice might change if additional ensembles at light quark masses
became available. One desirable choice for such an ensemble would be a
physical pion mass ensemble at $a^{-1}\sim 2.8\,\mathrm{GeV}$ (F0).

Recalling that we have two ways to determine $\xi$, which have different
statistical properties, we choose the indirect determination of $\xi$ as our
central value, as discussed in equation \eqref{eq:GFxires}. For this
determination, we take both, the ratio of decay constants and the ratio of bag
parameters from fits with the specified cuts. For comparison we also show the
results of the direct determinations as open symbols in Figure
\ref{fig:GFstability_xi}.

To assess the systematic errors due to the heavy mass dependence we compare
setting the heavy quark mass via a heavy-light ($D$ and $B$), heavy-strange
($D_s$ and $B_s$) or heavy-heavy ($\eta_c$ and $\eta_b$) pseudoscalar meson
mass. These are respectively shown as diamonds, circles and squares in Figures
\ref{fig:GFstability_fDsfD}-\ref{fig:GFstability_xi}. The physical masses we
use are given by the PDG averages given in Ref.~\cite{Tanabashi:2018oca}
\begin{equation}
  \begin{aligned}
    m_D^0 =1.86483(05)\,\mathrm{GeV} \quad\quad &m_{D_s}^\pm =1.9690(14)\,\mathrm{GeV} &\quad m_{\eta_c} &= 2.9834(05)\,\mathrm{GeV} \\
    m_B^0 =5.27955(26)\,\mathrm{GeV} \quad\quad &m_{B_s}^0 =5.36684(30)\,\mathrm{GeV} & \quad m_{\eta_b}&= 9.3990(23)\,\mathrm{GeV}.
    \label{eq:PDGheavymasses}
  \end{aligned}
\end{equation}
We note that the $\eta_c$ contains a small quark-disconnected contribution which
we neglect in our simulation. In addition to the smallness of this contribution,
its effect is further suppressed due to the very benign behaviour with the heavy
quark mass, displayed in the SU(3) breaking ratios under consideration. Since in
the base fit we choose to fix the heavy quark mass with the heavy-strange meson
mass, this small quark-disconnected contribution does not affect the final
result. For $f_{D_s}/f_D$, we expand around $m_H^\mathrm{expand} =
m_H^\mathrm{PDG}$. We also compare fits where we additionally include a term
$C_h^2 \left(1/m_H- 1/m_H^\mathrm{PDG}\right)^2$ in \eqref{eq:GFansatz}. We note
that we cannot resolve this additional coefficient from zero, since the data
does not display significant curvature. The variations of the results are shown
in the right panels of Figure \ref{fig:GFstability_fDsfD}. We note that for
$f_{D_s}/f_D$, there is no significant variation due to these choices, due to
the presence of precise data in and around the charm region.

For quantities involving a $b$ quark, we require to extrapolate from the region
where we have data to the $B_{(s)}$ or $\eta_b$ mass. Motivated by heavy quark
effective theory (HQET)~\cite{Eichten:1989zv} we take the expansion point to be
the static limit, i.e.~$1/m^\mathrm{expand}_H = 0$.\footnote{For a linear fit,
  this amounts simply to a re-definition of the constant
  $f(0,m_\pi^\mathrm{phys},m_H^\mathrm{expand})$.} We again test the stability
of our fit result by setting the heavy quark mass using the PDG values $B$,
$B_s$ and $\eta_b$ as well as systematically applying cuts to the data that
enters the fit. For each choice to set the heavy mass, we carry out the
following variations:
\begin{enumerate}
\item baseline fit (inc)
\item excluding the heaviest mass point of each coarse ensemble (exc h/C)
\item excluding the heaviest mass point of all ensembles (exc h/all)
\item excluding the lightest mass point of each coarse ensemble (exc l/C)
\item excluding the lightest mass point of all ensembles (exc l/all)
\end{enumerate}
We again take the full spread of the central values as our systematic error for
the linear part of the heavy quark extrapolation, which is slightly larger than
one statistical standard deviation. The right hand panels of Figures
\ref{fig:GFstability_fBsfB}, \ref{fig:GFstability_bagrat} and \ref{fig:GFstability_xi} show the corresponding
results for $f_{B_s}/f_B$, $B_{B_s}/B_{B_d}$ and $\xi$, respectively.

We estimate neglected higher order terms to be of the form
\begin{equation}
  \mc{O}|_\mathrm{static} \left[1+\alpha \frac{\Lambda}{m_{B_s}} + \beta \left(\frac{\Lambda}{m_{B_s}}\right)^2 \right] = \mc{O}|_\mathrm{static} \left[1+\alpha \frac{\Lambda}{m_{B_s}} \left(1 + \frac{\beta}{\alpha} \frac{\Lambda}{m_{B_s}} \right) \right]
  \label{eq:mHpowercounting}
\end{equation}
for some scale $\Lambda$. Assuming that the coefficients of this expansion are
of similar order, we approximate the missing higher order contributions to be
the difference between our baseline fit result and the observable evaluated at
the physical heavy meson mass. At the physical pion mass, zero lattice spacing
and $\beta \equiv 0$, equation \eqref{eq:mHpowercounting} reproduces equation
\eqref{eq:GFansatz} if we identify $\alpha \equiv C_h/(\Lambda C)$. Taking
$\Lambda = 500\,\mathrm{MeV}$ and conservatively allowing for a large
coefficient (i.e.~$\beta/\alpha = 5$), we can substitute $C_H$, $C$ from the
fit. We obtain $\Delta f_{B_s}/f_B = \fBsfBsysmH$ and $\Delta \xi = \xisysmH$,
which we assign as a (sub-leading) systematic error for higher order
extrapolation terms (labelled ``H.O. heavy'' in Table
\ref{tab:errorbudget}).

Our strategy to asses the systematic errors due to strong isospin
breaking and to estimate higher order discretisation errors which are
not included in our fit form closely follows
Ref.~\cite{Boyle:2017jwu}. Since our simulations are done with
degenerate light quark masses ($m_u = m_d = m_l$), we need to account
for the missing strong isospin corrections in our error budget. We
estimate this, by considering the difference between using the charged
or neutral pion, $D$ and $B$ meson masses. The corrections due to the
pion mass are given by $\sim C_\chi \left(m_{\pi^\pm}^2 -
m_{\pi^0}^2\right)$. Using $C_\chi$ from Table \ref{tab:GFcoefs}, this
amounts to $0.0009$ for the ratio of decay constants and to $0.0010$
for $\xi$. Similarly, using the slope $C_h$ with the inverse $D$
($B$) meson mass and applying it to the difference between the charged
and the neutral one gives an error of $7\times10^{-5}$ for
$f_{D_s}/f_D$, $6\times 10^{-7}$ for $f_{B_s}/f_B$ and $5\times
10^{-7}$ for $\xi$. We add the relevant terms in quadrature and list
them in Table \ref{tab:errorbudget} as ``$m_u \neq m_d$''. Assuming
$O(a^4)$ discretisation effects to be present, would lead to terms of
the form
\begin{equation}
  C_{CL} a^2 + D_{CL} a^4 = C_{CL} a^2 \left(1+ \frac{D_{CL}}{C_{CL}} a^2\right)
\end{equation}
in the fit ansatz. Since the leading order discretisation effects are accounted
for in our fit, it only remains to quantify the correction to them. Assuming
that discretisation effects grow as $a/\Lambda$ with
$\Lambda=500\,\mathrm{MeV}$, i.e.~$D_{CL}/C_{CL} = (0.5\mathrm{GeV})^2$, we can
simply substitute the values for $a$ and $C_{CL}$ (compare Table
\ref{tab:GFcoefs}) to obtain the corrections such a term would cause. From
this, we find the $O(a^4)$ corrections on the finest (coarsest) ensemble to be
0.0001 (0.0009) for the ratios of decay constants and 0.0003 (0.0021) for
$\xi$. We conservatively take the error on the estimated corrections on
the coarse ensembles and list these errors as ``H.O. disc.'' in Table
\ref{tab:errorbudget}.  Finally, for the finite size effects, we evaluate the
one-loop finite-volume HM$\chi$PT expressions given in
Ref.~\cite{Albertus:2010nm} for our choice of pion masses and volumes. For a
reasonable choice of parameters we find the maximal deviation to be less than
0.18\%, which we assign as the finite size error as listed in Table
\ref{tab:errorbudget}.

\begin{table}
  \begin{center}
    \resizebox{\textwidth}{!}{
      \begin{tabular}{|l||cc|cc|cc|cc|}
\hline
& \multicolumn{2}{c|}{$f_{D_s}/f_D$} & \multicolumn{2}{c|}{$f_{B_s}/f_B$} & \multicolumn{2}{c|}{$\xi$} & \multicolumn{2}{c|}{$B_{B_s}/B_{B_d}$} \\
& absolute & relative& absolute & relative& absolute & relative& absolute & relative \\\hline\hline
central & \multicolumn{2}{c|}{1.1740}& \multicolumn{2}{c|}{1.1949}& \multicolumn{2}{c|}{1.1939}& \multicolumn{2}{c|}{0.9984}\\\hline
stat   & 0.0051  &0.43\%  & 0.0060  &0.50\%  & 0.0067  &0.56\%  & 0.0045  &0.45\%  \\\hline
fit chiral-CL& $^{+0.0036}_{-0.0037}$ & $^{+0.31}_{-0.32}$\%   & $^{+0.0040}_{-0.0065}$ & $^{+0.34}_{-0.54}$\%   & $^{+0.0045}_{-0.0054}$ & $^{+0.38}_{-0.45}$\%   & $^{+0.0042}_{-0.0001}$ & $^{+0.42}_{-0.01}$\%   \\[2mm]
fit heavy mass& $^{+0.0009}_{-0.0006}$ & $^{+0.07}_{-0.05}$\%   & $^{+0.0000}_{-0.0098}$ & $^{+0.00}_{-0.82}$\%   & $^{+0.0000}_{-0.0103}$ & $^{+0.00}_{-0.87}$\%   & $^{+0.0027}_{-0.0022}$ & $^{+0.27}_{-0.22}$\%   \\[2mm]
H.O. heavy& $0.0000$ & $0.00$\%   & $0.0056$ & $0.47$\%   & $0.0042$ & $0.35$\%   & $0.0021$ & $0.21$\%   \\
H.O. disc.& $0.0001$ & $0.01$\%   & $0.0001$ & $0.01$\%   & $0.0014$ & $0.12$\%   & $0.0017$ & $0.17$\%   \\
$m_u \neq m_d$& $0.0009$ & $0.08$\%   & $0.0009$ & $0.07$\%   & $0.0009$ & $0.08$\%   & $0.0001$ & $0.01$\%   \\
finite size& $0.0021$ & $0.18$\%   & $0.0022$ & $0.18$\%   & $0.0021$ & $0.18$\%   & $0.0018$ & $0.18$\%   \\\hline\hline
total systematic& $^{+0.0045}_{-0.0044}$ & $^{+0.38}_{-0.38}$\%   & $^{+0.0073}_{-0.0165}$ & $^{+0.61}_{-1.38}$\%   & $^{+0.0067}_{-0.0164}$ & $^{+0.56}_{-1.37}$\%   & $^{+0.0066}_{-0.0045}$ & $^{+0.66}_{-0.45}$\%   \\[2mm]
total sys+stat& $^{+0.0068}_{-0.0068}$ & $^{+0.58}_{-0.58}$\%   & $^{+0.0095}_{-0.0175}$ & $^{+0.79}_{-1.47}$\%   & $^{+0.0095}_{-0.0177}$ & $^{+0.80}_{-1.48}$\%   & $^{+0.0080}_{-0.0063}$ & $^{+0.80}_{-0.63}$\%   \\[1mm] \hline
\end{tabular}

    }
    \caption{Summary of central values, statistical errors and all sources of
      systematic errors. The total systematic is found by adding the respective
      errors in quadrature. For ease we separately list the absolute and the
      relative errors, where the latter are presented in \%.}
    \label{tab:errorbudget}  
  \end{center}
\end{table}

\section{Results and comparison with the literature} \label{sec:comparison}
The results of our analysis are summarised in Table \ref{tab:errorbudget}. We
will now compare our values with those published in the literature.
\subsection{Ratio of decay constants}
\begin{figure}
  \includegraphics[width=.49\textwidth]{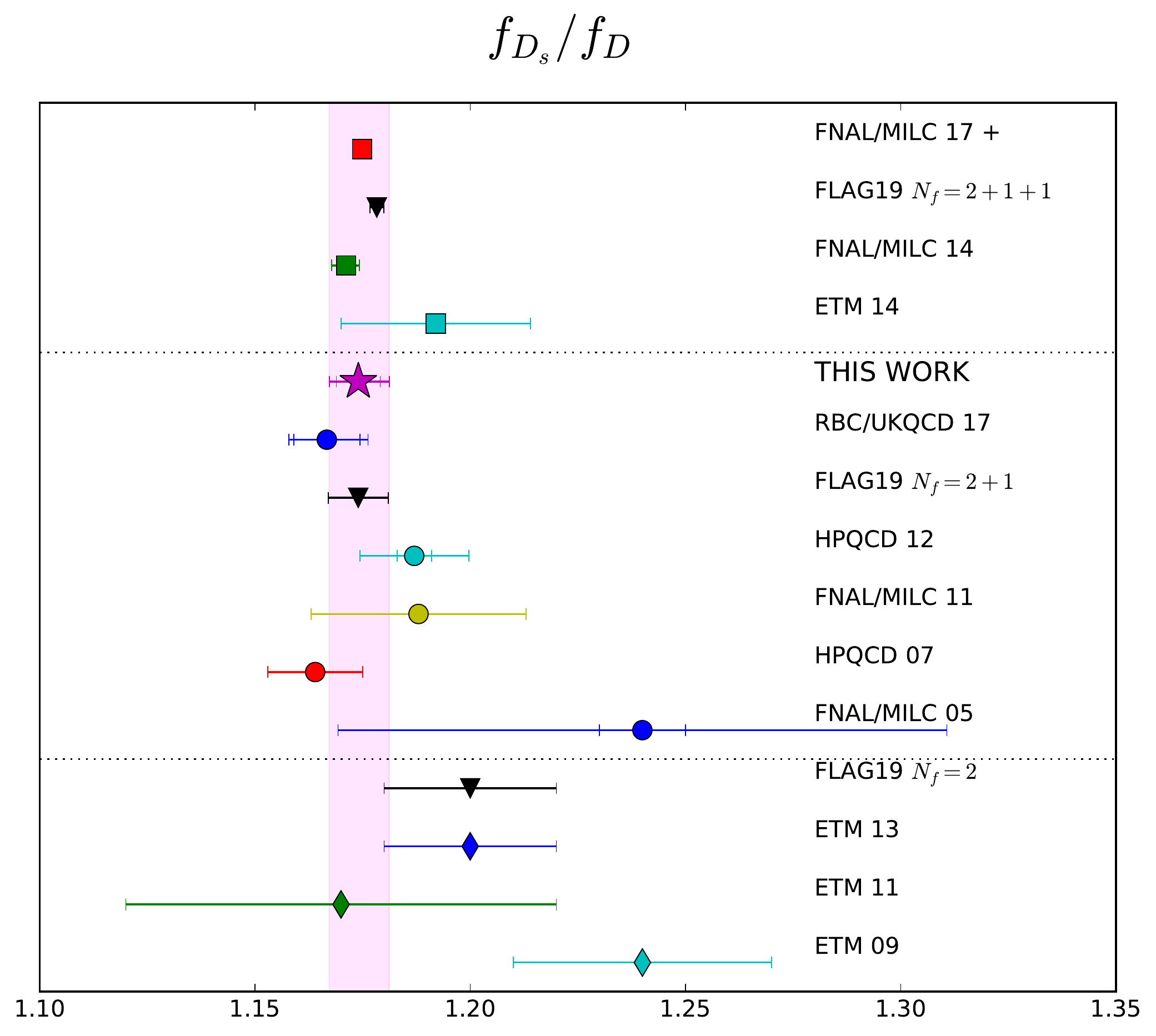}
  \includegraphics[width=.49\textwidth]{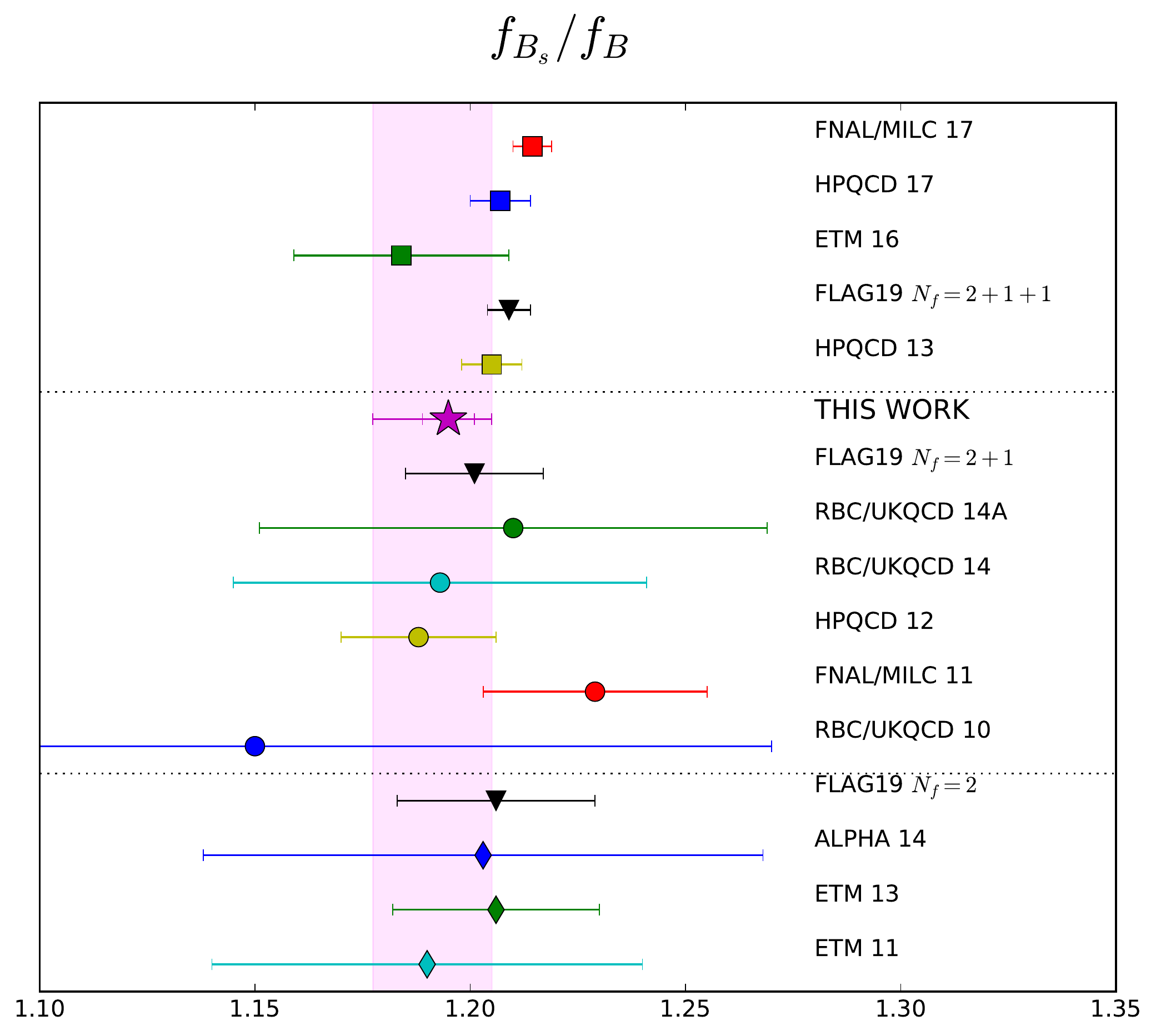}
  \caption{Comparison of our result (magenta star and band) with results from
    the literature for the (isospin symmetric) ratios of decay constants
    $f_{D_s}/f_D$ (left) and $f_{B_s}/f_B$ (right). The squares, circles and
    diamonds correspond to $N_f=2+1+1$, $N_f=2+1$ and $N_f=2$ flavour
    calculations. The black triangles show the averages published in the 2019
    FLAG report~\cite{Aoki:2019cca} for the given number of sea quark flavours with the
    results entering this average shown below these black triangles. References
    for all the displayed data points are given in the text.}
  \label{fig:decrat_compare}
\end{figure}

Figure \ref{fig:decrat_compare} shows a comparison of our results with the
literature for the ratios $f_{D_s}/f_D$ (left) and $f_{B_s}/f_B$ (right). The
result obtained in this work is shown as the magenta star and the vertical
magenta band.

For $f_{D_s}/f_D$ we find excellent agreement with our previous
result~\cite{Boyle:2017jwu} which was obtained on the same ensembles but with a
different choice of discretisation for the charm quarks. There is also no
significant tension with the published literature~\cite{Bazavov:2017lyh,
  Bazavov:2014wgs, Carrasco:2014poa, Boyle:2017jwu, Na:2012iu, Bazavov:2011aa,
  Follana:2007uv, Aubin:2005ar, Carrasco:2013zta, Dimopoulos:2011gx,
  Blossier:2009bx} or the averaged values presented by FLAG~\cite{Aoki:2019cca}.
We note that other than in this work, there are still only very few computations
including data directly calculated at the physical pion
mass~\cite{Bazavov:2017lyh, Bazavov:2014wgs, Boyle:2017jwu}.

For the ratio $f_{B_s}/f_B$ there are a variety of different results using
different methods in the literature~\cite{Bazavov:2017lyh, Hughes:2017spc,
  Bussone:2016iua, Dowdall:2013tga, Christ:2014uea, Aoki:2014nga, Na:2012kp,
  Bazavov:2011aa, Gamiz:2009ku, Bernardoni:2014fva, Carrasco:2013zta,
  Dimopoulos:2011gx}. We note that some of the results in Refs.
\cite{Christ:2014uea,Aoki:2014nga} have been carried out on a subset of the
ensembles (C1/2 and M1/2/3) used in this study, however using different choices
for the heavy quark discretisation. Besides the use of a fully relativistic
formulation, our results improve upon these by the inclusion of physical pion
mass ensembles and a third lattice spacing, leading to a more than three-fold
reduction in error.

We note that in FNAL/MILC 17~\cite{Bazavov:2017lyh} and RBC/UKQCD
14A~\cite{Christ:2014uea}, no isospin symmetric result for the ratio
$f_{B_s}/f_B$ is quoted. For the comparison in Figure \ref{fig:decrat_compare}
we instead take the correlated average of the results quoted for
$f_{B_s}/f_{B^\pm}$ and $f_{B_s}/f_{B^0}$ which are plotted as the red
square~\cite{Bazavov:2017lyh} and the blue circle~\cite{Christ:2014uea} in the
right panel of Figure \ref{fig:decrat_compare}. Prior to this work, only two
fully relativistic fermion actions have been employed as heavy quark
discretisation, namely the HISQ action in Ref.~\cite{Bazavov:2017lyh} and the
twisted mass action in Refs.~\cite{Bussone:2016iua, Dowdall:2013tga,
  Carrasco:2013zta, Dimopoulos:2011gx}. Other than the result presented here,
only one computation~\cite{Bazavov:2017lyh} with physical pion masses is
currently available for $f_{B_s}/f_B$.

\subsection{Neutral meson mixing}

\begin{figure}
  \center
  \includegraphics[width=.49\textwidth]{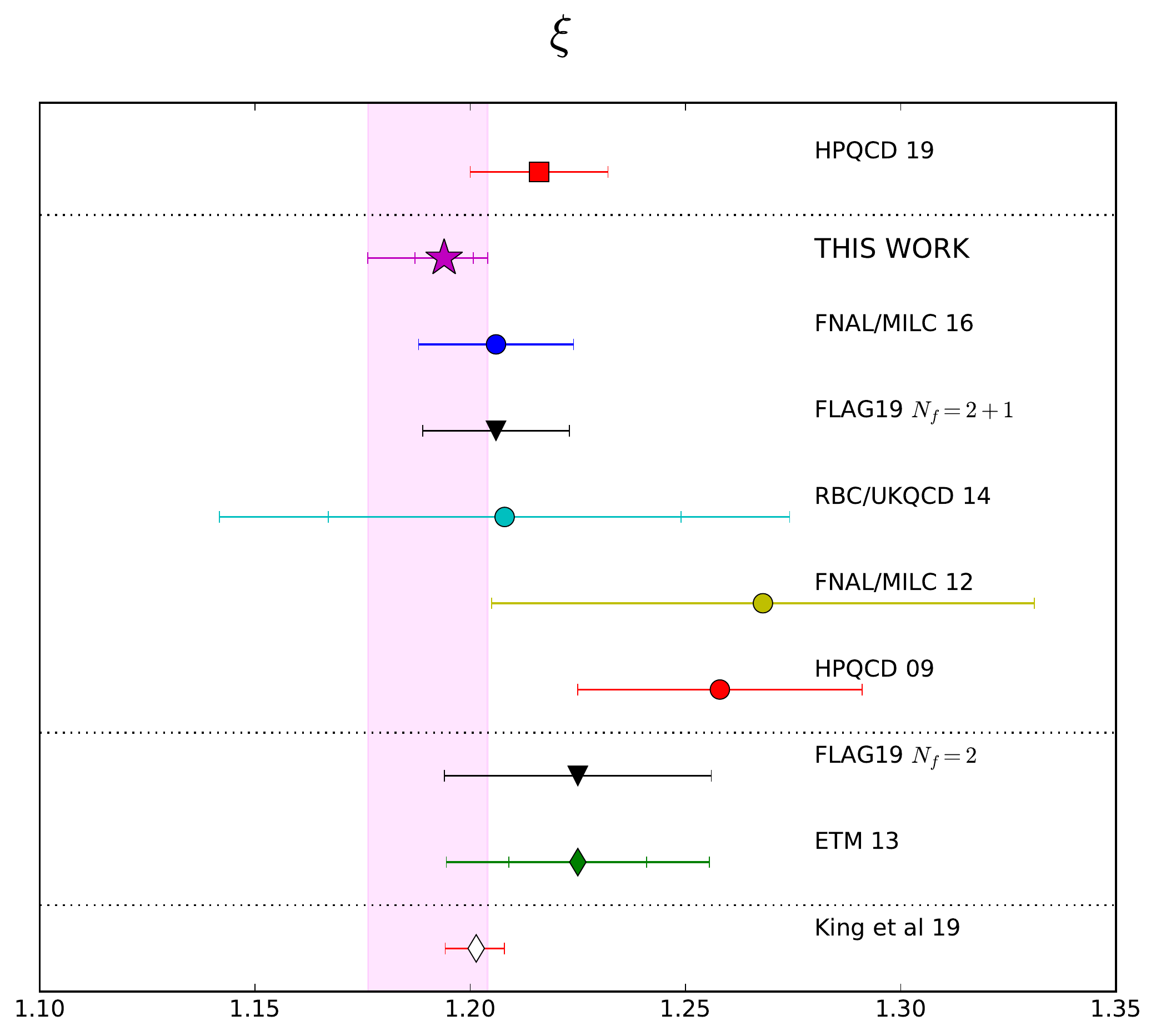}
  \includegraphics[width=.49\textwidth]{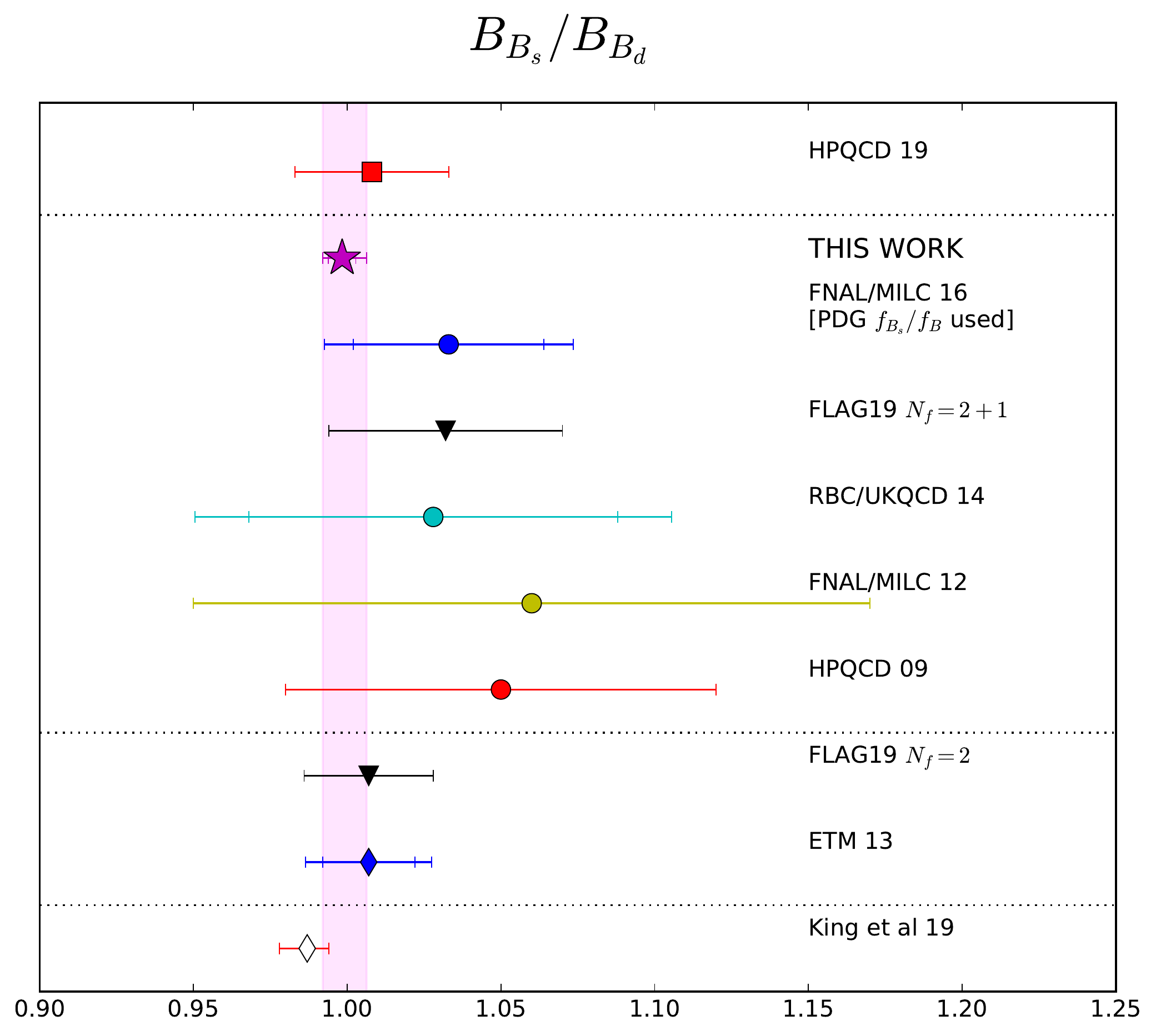}
  \caption{Comparison of our result (magenta star and band) with results from
    the literature for the $SU(3)$ breaking ratios $\xi$ (left) and the ratio of
    bag parameters $B_{B_s}/B_B$ (right) analogous to Figure
    \ref{fig:decrat_compare}. Closed symbols refer to lattice QCD computations,
    whilst the open symbol shows a recent QCD sum-rule result. References for the
    displayed data points are given in the text.}
  \label{fig:xi_compare}
\end{figure}

Figure \ref{fig:xi_compare} summarises the current status of the
literature for the mixing parameter $\xi$ and the ratios of bag
parameters $B_{B_s}/B_{B_d}$~\cite{Carrasco:2013zta, Aoki:2014nga,
  Bazavov:2016nty, Gamiz:2009ku,
  Bazavov:2012zs,Dowdall:2019bea,King:2019lal}.\footnote{Refs.~\cite{Dowdall:2019bea,King:2019lal}
  appeared after version 1 of this paper was posted.} We note that
compared to the ratio of decay constants, there are far fewer
computations for these observables. This is the first calculation for
$\xi$ and $B_{B_s}/B_{B_d}$ which includes ensembles with physical
pion masses.\footnote{Ref.~\cite{Dowdall:2019bea} appeared after the
  first version of this paper.} The only other
result~\cite{Carrasco:2013zta} that employs a fully relativistic
set-up is presented in the $N_f=2$ calculation using twisted mass
fermions.  We obtain a similar error with a somewhat smaller central
value for the quantity $\xi$ compared to
Ref.~\cite{Bazavov:2016nty}. Ref.~\cite{Bazavov:2016nty} used the
PDG~\cite{Rosner:2015wva} average of the decay constants $f_{B_s}/f_B$
to obtain the ratio of bag parameters, resulting in a larger error for
this quantity. For the ratio $B_{B_s}/B_{B_d}$, our result is two
times more precise than the previously most precise lattice QCD value
obtained by~\cite{Carrasco:2013zta}.

We stress that our systematic errors differ from most other lattice computation
since due to the use of a chiral action the decay constants and bag parameter
renormalise multiplicatively and therefore cancel in the considered
ratios. As a consequence, our computation is free from lattice renormalisation
uncertainties.

\section{Phenomenological implications and extraction of CKM matrix elements} \label{sec:CKM}
Having obtained SM predictions for the ratios of decay constants
$f_{D_s}/f_D$ and $\xi$, we are now in a position to combine these
with experimentally observed quantities to obtain ratios of CKM matrix
elements. Treating the experimental averages from the first two lines
of \eqref{eq:expdecayconstants} as uncorrelated we obtain
$\abs{V_{cd}/V_{cs}} = \CKMfDsfDexccentral(\CKMfDsfDexcexperrint)
\times f_{D_s}/f_D$. If we choose the average including the new BESIII
results~\cite{Ablikim:2018jun} (see equation
\eqref{eq:expdecayconstantsBES}), this changes to
$\abs{V_{cd}/V_{cs}} = \CKMfDsfDinccentral(\CKMfDsfDincexperrint)
\times f_{D_s}/f_D$.\footnote{At the time version 1 of this work was
  posted, ~\cite{Ablikim:2018jun} only existed as a
  pre-print.} Similarly, combining the experimental averages $\Delta
m_d$ and $\Delta m_s$ (see \eqref{eq:deltamexp}) with the PDG values
for $m_{B_s^0}$ and $m_{B^0}$ (see \eqref{eq:PDGheavymasses}) yields
$\abs{V_{td}/V_{ts}} = \CKMxicentral(\CKMxiexperrint) \times \xi$.

Were we to consider the decay rates of the individual charged decays or the
individual mass differences $\Delta m_q$ for $q=s,d$, we would have to correct
for electromagnetic effects before extracting $V_{cd}$, $V_{cs}$, $V_{ub}$,
$V_{td}$ or $V_{ts}$ from the pure QCD entities $f_{D_{(s)}}$ and $f_{B_{(s)}}
\sqrt{\hat{B}_{B_{(s)}}}$. However, given that $D_{s}^+$ and $D^+$ and
respectively $B_s^0$ and $B^0$ are identical when replacing the $s$ by the $d$
quark, and both of these have the same charge, we assume that these effects are
highly suppressed in the ratios we consider.

Inserting the lattice results, propagating the errors and assuming that there
are no new physics contributions in the experimental measurements leads to the
ratios
\begin{equation}
  \begin{aligned}
    \abs{V_{cd}/V_{cs}} &= \VcdVcsinccentral \left( \VcdVcsincexperrint
    \right)_\mathrm{exp}\left(^{\VcdVcsinclaterrpint}_{\VcdVcsinclaterrmint}\right)_\mathrm{lat},\\
    \abs{V_{td}/V_{ts}} &= \VtdVtscentral \left( \VtdVtsexperrint
    \right)_\mathrm{exp}\left(^{\VtdVtslaterrpint}_{\VtdVtslaterrmint}\right)_\mathrm{lat}.
  \end{aligned}
\end{equation}
where we included the new BESIII result~\cite{Ablikim:2018jun} in our
determination. Had we not included the new BESIII result we would obtain
$\abs{V_{cd}/V_{cs}} = \VcdVcsexccentral \left( \VcdVcsexcexperrint
\right)_\mathrm{exp}\left(^{\VcdVcsexclaterrpint}_{\VcdVcsexclaterrmint}\right)_\mathrm{lat}$.
These results are slightly lower than those currently reported (compare equation
\eqref{eq:CKMfitterratios}). We anticipate, that the global fit values will
change as a result of this work.

The error on the ratio $\abs{V_{cd}/V_{cs}}$ is currently dominated by the
experimental uncertainty. For $\abs{V_{td}/V_{ts}}$, the situation is reversed
and the theoretical uncertainty dominates the error. This work improves on this
by providing a first computation based on chiral fermions with physical pion
mass ensembles.
\section{Conclusion and outlook} \label{sec:summary}
We have, for the first time, predicted the $SU(3)$ breaking ratios
$B_{B_s}/B_{B_d}$ and $\xi$ in a calculation based on ensembles with physical
pion masses, therefore eliminating any large chiral extrapolations.
Furthermore, we present for the first time, results for $SU(3)$ breaking ratios
in the $B_{(s)}$ mesons systems obtained from an all-domain wall calculation.
We have illustrated that such ratios display a very benign behaviour from below
the charm mass to $\sim$ half the bottom quark mass and that lattice artefacts
in our choice of discretisation are small for these observables. We found that
nearly all of the $SU(3)$ breaking effects observed in the difference of $\xi$
from unity, arise from the ratio of decay constants $f_{B_s}/f_B$.  This yields
the to-date most precise computation of the ratio of CKM matrix elements
$\abs{V_{td}/V_{ts}}$.

Looking forwards, we anticipate the generation of a third ensemble
with physical pion masses, at the same lattice spacing as our
currently finest ensemble (F1M). This will address our leading
systematic error, namely the chiral-continuum limit and heavy quark
mass extrapolation. It will allow to lower the pion mass cut to $\sim
250\,\mathrm{MeV}$ whilst still constraining the continuum limit with
three lattice spacings.

Based on the presented dataset we are also working on the mixed action
renormalisation, to deduce the decay constants $f_{D_{(s)}}$, $f_{B_{(s)}}$ and
the standard model bag parameters $B_{B_{(s)}}$. We will also address the full
set of beyond the SM four-quark operators for $B_{(s)}$-mixing and the short
distance contribution to $D-\bar{D}$ mixing. This will be analogous to the
computation presented in \cite{Boyle:2018eor} for the Kaon sector.

\acknowledgments The authors would like to thank the members of the
RBC and UKQCD collaborations for many valuable discussions with
special thanks to Felix Erben.  The authors further thank the
CKMfitter and the UTfit groups for providing their current estimates
for $\abs{V_{cd}/V_{cs}}$ and $\abs{V_{td}/V_{ts}}$.

P.A.B., L.D.D. and J.T.T. are supported by UK STFC grants ST/L000458/1
and ST/P000630/1. L.D.D. is also supported by the Royal Society
Wolfson Research Merit Award WM140078.  A.J. received funding from
STFC consolidated grant ST/P000711/1 and from the European Research
Council under the European Union's Seventh Framework Program
(FP7/2007- 2013) / ERC Grant agreement 279757.  A.S. is supported in
part by the USDOE grant de-sc0012704. J.T.T. was additionally
supported by DFF Research project 1. Grant
n. 8021-00122B. O.W. acknowledges support by DOE grant
DE-SC0010005. This project has received funding from the European
Union's Horizon 2020 research and innovation programme under the Marie
Sk{\l}odowska-Curie grant agreement No 659322.

This work used the DiRAC Blue Gene Q Shared Petaflop system and the DiRAC
Extreme Scaling service at the University of Edinburgh, operated by the
Edinburgh Parallel Computing Centre on behalf of the STFC DiRAC HPC Facility
(www.dirac.ac.uk). The equipment for the former was funded by BIS National
E-infrastructure capital grant ST/K000411/1, STFC capital grant ST/H008845/1,
and STFC DiRAC Operations grants ST/K005804/1 and ST/K005790/1 and of the latter
by BEIS capital funding via STFC capital grants ST/R00238X/1 and ST/S002537/1
and STFC DiRAC Operations grant ST/R001006/1. DiRAC is part of the National
e-Infrastructure. Code development has been supported by ST/P002447/1. P.A.B is
a Wolfson Fellow WM/60035 and a Turing Fellow.
\appendix
\clearpage
\section{Ensemble properties of F1M}
\label{sec:F1M}
We noticed that the Shamir action approximation was used for the
ensemble generation of the original F1 ensemble (presented in
Ref.~\cite{Boyle:2017jwu}), whilst valence measurements were carried
out with the M\"obius action approximation. To avoid any ambiguity,
this ensemble will henceforth be referred to as F1S. At fixed ensemble
parameters this results in a larger residual chiral symmetry breaking
(residual mass) compared the the M\"obius action approximation which
in turn results in slightly heavier pion masses.  Previously, the
valence measurements were carried out with the M\"obius action
approximation on the F1S ensemble which is a partial quenching effect
in the quark masses. The pion mass and residual mass that were
determined in Ref.~\cite{Boyle:2017jwu} were $am_\pi = 0.08446(18)$
and $am^l_\mathrm{res} = 0.0002290(19)$, respectively. The scale
setting fit gave $a^{-1} = 2.774(10)\,\mathrm{GeV}$ and
$am_s^\mathrm{phys} = 0.02132(17)$.
      
To resolve this, we performed valence measurements with the Shamir
action approximation on the original ensemble and additionally
generated an ensemble with identical parameters but using the M\"obius
approximation to the sign function (referred to as F1M). The results
for the unitary pion and residual masses are
\begin{equation}
  \begin{aligned}
    \mathrm{F1S}:\qquad & am_\pi = 0.09608(27), & am^l_\mathrm{res} =0.0009679(21),\\
    \mathrm{F1M}:\qquad & am_\pi = 0.08575(16), & am^l_\mathrm{res} =0.0002356(16).
  \end{aligned}
\end{equation}

By repeating the scale setting analysis of
Refs.~\cite{Blum:2014tka,Boyle:2017jwu} under inclusion of both of
these ensembles we obtained determinations of the inverse lattice
spacing and the physical strange quark mass
\begin{equation}
  \begin{aligned}
    \mathrm{F1S}:\qquad & a^{-1}=2.785(11)\,\mathrm{GeV}, & am_s^\mathrm{phys}=0.02167(20),\\
    \mathrm{F1M}:\qquad & a^{-1}=2.708(10)\,\mathrm{GeV}, & am_s^\mathrm{phys}=0.02217(16).
  \end{aligned}
\end{equation}

We note that the shift in lattice spacing at fixed value of $\beta$
when changing from Shamir to M\"obius is in agreement with previous
work~\cite{Blum:2014tka}. Combining the above we obtain pion masses of
$232\,\mathrm{MeV}$ (F1M) and $267\,\mathrm{MeV}$ (F1S). In this paper
only the F1M ensemble is used.

\section{Results of correlation function fits}
\label{sec:corrfits}
In this section we list the relevant results of the correlation function fits
for the two point functions (Table \ref{tab:corrfits}) and the three point
functions (Table \ref{tab:bagfits}). The fit strategy is discussed in more
detail in the text.
\begin{table}[h]
  \begin{center}
    \resizebox{\textwidth}{!}{
      \begin{tabular}{|ll||llll||llll||l||l|}
\hline
Name & $am_h$ & range & $am_{hl}$ & $af_{hl}$ & $\chi^2/\mathrm{dof}$ & range & $am_{hs}$ & $af_{hs}$ & $\chi^2/\mathrm{dof}$ & $am_{hh}$ & $f_{sh}/f_{lh}$ \\\hline
\hline
C0 & 0.51 & [ 7,17) &  0.90759(67) &  0.14135(57) & 0.041  & [ 7,24) &  0.96797(13) &  0.16485(13) & 0.033 &  1.413572(53) &  1.1663(47) \\
C0 & 0.57 & [ 7,17) &  0.97440(79) &  0.14208(69) & 0.053  & [ 7,24) &  1.03298(15) &  0.16599(16) & 0.058 &  1.525400(50) &  1.1682(56) \\
C0 & 0.63 & [ 7,17) &  1.03899(94) &  0.14172(84) & 0.069  & [ 8,24) &  1.09592(18) &  0.16573(21) & 0.051 &  1.632449(49) &  1.1694(69) \\
C0 & 0.69 & [ 7,17) &  1.1006(12) &  0.1397(10) & 0.094  & [ 8,24) &  1.15600(21) &  0.16337(27) & 0.098 &  1.732940(49) &  1.1692(88) \\
\hline
C1 & 0.50 & [ 6,17) &  0.90826(78) &  0.14842(64) & 0.029  & [ 6,24) &  0.95316(40) &  0.16324(35) & 0.029 &  1.39276(20) &  1.0998(38) \\
C1 & 0.58 & [ 6,17) &  0.99656(95) &  0.14935(84) & 0.044  & [ 6,24) &  1.03965(44) &  0.16454(43) & 0.037 &  1.54135(19) &  1.1018(53) \\
C1 & 0.64 & [ 6,17) &  1.0601(11) &  0.1486(11) & 0.065  & [ 7,24) &  1.10187(52) &  0.16372(58) & 0.034 &  1.64697(18) &  1.1018(75) \\
C1 & 0.69 & [ 6,17) &  1.1106(14) &  0.1465(13) & 0.094  & [ 7,24) &  1.15137(59) &  0.16136(70) & 0.047 &  1.72972(18) &  1.1012(98) \\
\hline
C2 & 0.51 & [ 7,18) &  0.93015(72) &  0.15441(64) & 0.016  & [ 7,20) &  0.96602(43) &  0.16537(39) & 0.012 &  1.41243(20) &  1.0710(27) \\
C2 & 0.59 & [ 7,18) &  1.01801(88) &  0.15548(85) & 0.021  & [ 8,20) &  1.05216(56) &  0.16646(60) & 0.013 &  1.55999(18) &  1.0706(37) \\
C2 & 0.64 & [ 8,18) &  1.0708(13) &  0.1549(15) & 0.022  & [ 8,20) &  1.10400(64) &  0.16581(74) & 0.017 &  1.64761(18) &  1.0705(70) \\
C2 & 0.68 & [ 8,18) &  1.1115(15) &  0.1534(18) & 0.028  & [ 9,20) &  1.14377(85) &  0.1639(11) & 0.013 &  1.71429(18) &  1.0684(94) \\
\hline
\hline
M0 & 0.41 & [ 9,22) &  0.74184(60) &  0.10269(52) & 0.056  & [10,24) &  0.78458(11) &  0.120553(99) & 0.018 &  1.170921(44) &  1.1739(57) \\
M0 & 0.50 & [ 9,22) &  0.84158(80) &  0.10247(72) & 0.061  & [10,24) &  0.88275(12) &  0.12096(12) & 0.021 &  1.344482(41) &  1.1804(80) \\
M0 & 0.59 & [ 9,22) &  0.9350(11) &  0.10017(97) & 0.070  & [10,24) &  0.97514(14) &  0.11896(15) & 0.034 &  1.507094(39) &  1.188(11) \\
M0 & 0.68 & [ 9,22) &  1.0198(14) &  0.0948(13) & 0.090  & [10,24) &  1.05931(18) &  0.11338(19) & 0.085 &  1.654201(37) &  1.195(15) \\
\hline
M1 & 0.41 & [ 8,22) &  0.74931(91) &  0.10867(65) & 0.030  & [ 8,24) &  0.78445(36) &  0.12142(27) & 0.016 &  1.17063(15) &  1.1174(54) \\
M1 & 0.50 & [ 8,22) &  0.8489(11) &  0.10873(82) & 0.053  & [ 8,24) &  0.88259(39) &  0.12183(31) & 0.028 &  1.34397(14) &  1.1206(70) \\
M1 & 0.59 & [ 8,22) &  0.9423(13) &  0.1067(10) & 0.086  & [10,24) &  0.97496(47) &  0.11990(41) & 0.030 &  1.50613(13) &  1.1241(94) \\
M1 & 0.68 & [ 8,22) &  1.0272(17) &  0.1014(13) & 0.124  & [11,24) &  1.05894(59) &  0.11425(59) & 0.041 &  1.65226(13) &  1.126(13) \\
\hline
M2 & 0.41 & [ 9,22) &  0.7531(10) &  0.11027(96) & 0.031  & [10,24) &  0.78463(48) &  0.12139(43) & 0.023 &  1.17078(19) &  1.1009(77) \\
M2 & 0.50 & [ 9,22) &  0.8524(14) &  0.1102(14) & 0.041  & [10,24) &  0.88280(56) &  0.12179(54) & 0.031 &  1.34407(17) &  1.105(11) \\
M2 & 0.59 & [ 9,22) &  0.9456(18) &  0.1079(19) & 0.050  & [11,24) &  0.97500(85) &  0.1196(10) & 0.035 &  1.50625(16) &  1.109(17) \\
M2 & 0.68 & [ 9,22) &  1.0299(25) &  0.1021(26) & 0.061  & [11,24) &  1.0588(11) &  0.1137(13) & 0.046 &  1.65232(15) &  1.114(25) \\
\hline
M3 & 0.41 & [ 9,22) &  0.75830(78) &  0.11308(71) & 0.016  & [ 9,24) &  0.78609(41) &  0.12240(33) & 0.017 &  1.17153(17) &  1.0824(50) \\
M3 & 0.50 & [ 9,22) &  0.85767(100) &  0.1131(10) & 0.017  & [ 9,24) &  0.88421(48) &  0.12270(45) & 0.020 &  1.34477(16) &  1.0848(73) \\
M3 & 0.59 & [ 9,22) &  0.9509(13) &  0.1109(14) & 0.022  & [10,24) &  0.97626(74) &  0.12019(92) & 0.022 &  1.50691(15) &  1.084(10) \\
M3 & 0.68 & [ 9,22) &  1.0357(17) &  0.1053(19) & 0.032  & [10,24) &  1.0599(10) &  0.1139(13) & 0.029 &  1.65305(15) &  1.082(15) \\
\hline
\hline
F1M & 0.32 & [ 9,24) &  0.61852(51) &  0.09136(39) & 0.054  & [10,24) &  0.65291(17) &  0.10409(16) & 0.025 &  0.967402(95) &  1.1394(51) \\
F1M & 0.41 & [ 9,24) &  0.72222(64) &  0.09205(52) & 0.067  & [10,24) &  0.75503(19) &  0.10532(19) & 0.034 &  1.149482(87) &  1.1442(68) \\
F1M & 0.50 & [ 9,24) &  0.81987(79) &  0.09117(64) & 0.075  & [10,24) &  0.85160(22) &  0.10468(24) & 0.046 &  1.321898(82) &  1.1482(87) \\
F1M & 0.59 & [ 9,24) &  0.91076(96) &  0.08835(77) & 0.090  & [10,24) &  0.94167(27) &  0.10172(29) & 0.059 &  1.482450(77) &  1.151(11) \\
F1M & 0.68 & [ 9,24) &  0.9922(12) &  0.08272(87) & 0.141  & [10,24) &  1.02245(33) &  0.09534(35) & 0.090 &  1.625929(74) &  1.153(13) \\
\hline
\end{tabular}

    }
  \end{center}
  \caption{Results of the correlation function fits.}
  \label{tab:corrfits}
\end{table}

\begin{table}
  \begin{center}
    \resizebox{\textwidth}{!}{
      \begin{tabular}{|ll||llll||llll||l|}
\hline
Name & $am_h$ & $\Delta T$ & $t_\mathrm{min}$ & $B_{hl}$ & $\chi^2/\mathrm{dof}$ & $\Delta T$ & $t_\mathrm{min}$ & $B_{hs}$ & $\chi^2/\mathrm{dof}$ & $B_{sh}/B_{lh}$ \\\hline
\hline
C0 & 0.51 & 20 &  6 &  0.7875(32) & 0.059  & 32 &  9 &  0.81071(31) & 0.026 &  1.0295(42) \\
C0 & 0.57 & 20 &  6 &  0.7971(39) & 0.073  & 32 &  9 &  0.82014(39) & 0.033 &  1.0289(50) \\
C0 & 0.63 & 20 &  5 &  0.8066(44) & 0.085  & 32 &  7 &  0.82935(46) & 0.066 &  1.0282(56) \\
C0 & 0.69 & 20 &  4 &  0.8161(50) & 0.091  & 32 &  6 &  0.83880(58) & 0.079 &  1.0278(63) \\
\hline
C1 & 0.50 & 22 &  5 &  0.7916(25) & 0.204  & 26 &  8 &  0.80895(74) & 0.198 &  1.0219(32) \\
C1 & 0.58 & 20 &  6 &  0.8054(32) & 0.343  & 26 &  7 &  0.82186(90) & 0.220 &  1.0204(42) \\
C1 & 0.64 & 20 &  6 &  0.8144(39) & 0.285  & 26 &  6 &  0.8312(11) & 0.147 &  1.0206(50) \\
C1 & 0.69 & 20 &  5 &  0.8228(43) & 0.283  & 26 &  5 &  0.8392(13) & 0.193 &  1.0200(56) \\
\hline
C2 & 0.51 & 22 &  7 &  0.7978(19) & 0.103  & 24 &  8 &  0.81038(85) & 0.028 &  1.0158(23) \\
C2 & 0.59 & 22 &  6 &  0.8107(23) & 0.121  & 24 &  7 &  0.8229(10) & 0.046 &  1.0150(28) \\
C2 & 0.64 & 22 &  5 &  0.8185(26) & 0.115  & 24 &  7 &  0.8304(12) & 0.025 &  1.0145(32) \\
C2 & 0.68 & 22 &  5 &  0.8245(30) & 0.121  & 24 &  7 &  0.8364(14) & 0.025 &  1.0145(37) \\
\hline
\hline
M0 & 0.41 & 22 &  8 &  0.7905(21) & 0.146  & 32 & 12 &  0.80410(30) & 0.276 &  1.0172(27) \\
M0 & 0.50 & 22 &  8 &  0.8091(29) & 0.074  & 32 & 12 &  0.82039(39) & 0.173 &  1.0140(36) \\
M0 & 0.59 & 22 &  6 &  0.8275(36) & 0.098  & 32 & 11 &  0.83506(49) & 0.130 &  1.0091(43) \\
M0 & 0.68 & 22 &  5 &  0.8456(46) & 0.017  & 32 &  9 &  0.84962(61) & 0.044 &  1.0047(54) \\
\hline
M1 & 0.41 & 20 &  7 &  0.7932(22) & 0.147  & 28 &  9 &  0.80395(85) & 0.053 &  1.0135(30) \\
M1 & 0.50 & 18 &  7 &  0.8105(26) & 0.198  & 28 &  9 &  0.8203(11) & 0.050 &  1.0121(33) \\
M1 & 0.59 & 18 &  7 &  0.8246(35) & 0.154  & 28 &  8 &  0.8350(14) & 0.128 &  1.0126(44) \\
M1 & 0.68 & 18 &  7 &  0.8377(47) & 0.122  & 28 &  8 &  0.8496(20) & 0.179 &  1.0143(58) \\
\hline
M2 & 0.41 & 20 &  8 &  0.7914(21) & 0.272  & 28 &  8 &  0.80290(87) & 0.429 &  1.0145(28) \\
M2 & 0.50 & 18 &  7 &  0.8097(24) & 0.144  & 28 &  8 &  0.8190(11) & 0.297 &  1.0115(30) \\
M2 & 0.59 & 18 &  7 &  0.8241(30) & 0.087  & 28 &  8 &  0.8335(14) & 0.178 &  1.0114(39) \\
M2 & 0.68 & 18 &  6 &  0.8387(35) & 0.146  & 28 &  8 &  0.8480(19) & 0.082 &  1.0112(48) \\
\hline
M3 & 0.41 & 20 &  8 &  0.7940(18) & 0.093  & 22 &  9 &  0.80460(87) & 0.000 &  1.0134(23) \\
M3 & 0.50 & 20 &  8 &  0.8110(22) & 0.031  & 22 &  7 &  0.82136(96) & 0.073 &  1.0128(29) \\
M3 & 0.59 & 18 &  7 &  0.8253(24) & 0.033  & 22 &  7 &  0.8357(12) & 0.095 &  1.0126(33) \\
M3 & 0.68 & 18 &  7 &  0.8398(30) & 0.001  & 22 &  6 &  0.8498(15) & 0.310 &  1.0120(41) \\
\hline
\hline
F1M & 0.32 & 28 & 12 &  0.7777(20) & 0.023  & 34 & 15 &  0.78754(40) & 0.010 &  1.0127(25) \\
F1M & 0.41 & 28 & 11 &  0.7989(24) & 0.016  & 34 & 15 &  0.80743(48) & 0.010 &  1.0107(30) \\
F1M & 0.50 & 28 & 10 &  0.8164(30) & 0.009  & 34 & 15 &  0.82342(61) & 0.009 &  1.0085(36) \\
F1M & 0.59 & 28 & 10 &  0.8327(38) & 0.063  & 34 & 15 &  0.83771(82) & 0.006 &  1.0060(46) \\
F1M & 0.68 & 28 &  8 &  0.8493(40) & 0.190  & 34 & 14 &  0.8519(10) & 0.006 &  1.0031(48) \\
\hline
\end{tabular}

    }
  \end{center}
  \caption{Results of the bag parameter fits.}
  \label{tab:bagfits}
\end{table}
\clearpage
\section{Results of the global fit}
\label{sec:globalfits}
Here we list the results of the global fits for the $SU(3)$ breaking ratios
$f_{D_s}/f_D$ (Table \ref{tab:GFfDsfD}), $f_{B_s}/f_B$ (Table
\ref{tab:GFfBsfB}), $B_{B_s}/B_{B_d}$ (Table \ref{tab:GFBBsBB}) and $\xi$ (Table
\ref{tab:GFxi}).
\begin{table}[h]
  \begin{center}
    \resizebox{\textwidth}{!}{
      \begin{tabular}{|lll||l||llll|rcc|}
\hline
cut name & $m_\pi^\mathrm{max}/\mathrm{MeV}$ & $m_H$ & $f_{D_s}/f_D$ & $C_{CL}/\mathrm{GeV^2}$ & $C_\chi/\mathrm{GeV}^{-2}$ & $C_{H}/\mathrm{GeV}$  & $C_s$& d.o.f & $\chi^2/\mathrm{dof}$ & $p$ \\\hline\hline
inc & 450 & $D_s$&  1.1685(40) &  0.001(13) &  -0.556(15) &  -0.0611(57) &  0.12(10)& 28 & 2.41 & 0.000\\
exc h/C & 450 & $D_s$&  1.1683(42) &  0.002(14) &  -0.563(15) &  -0.0581(68) &  0.14(10)& 25 & 2.11 & 0.001\\
exc h/all & 450 & $D_s$&  1.1664(43) &  0.004(14) &  -0.556(16) &  -0.0544(76) &  0.10(11)& 20 & 2.17 & 0.002\\
phys mh cut & 450 & $D_s$&  1.1706(54) &  0.002(17) &  -0.585(19) &  -0.069(13) &  0.18(13)& 14 & 1.55 & 0.084\\
\hline
inc & 430 & $D_s$&  1.1809(46) &  -0.032(15) &  -0.627(20) &  -0.0661(61) &  0.23(11)& 24 & 1.36 & 0.112\\
exc h/C & 430 & $D_s$&  1.1792(48) &  -0.029(16) &  -0.624(21) &  -0.0614(70) &  0.24(11)& 22 & 1.33 & 0.138\\
exc h/all & 430 & $D_s$&  1.1773(50) &  -0.027(16) &  -0.617(21) &  -0.0573(79) &  0.21(11)& 17 & 1.30 & 0.180\\
phys mh cut & 430 & $D_s$&  1.1746(62) &  -0.012(19) &  -0.605(24) &  -0.068(13) &  0.23(14)& 11 & 1.73 & 0.060\\
\hline
inc & 400 & $D_s$&  1.1761(48) &  -0.006(17) &  -0.683(26) &  -0.0654(62) &  0.14(11)& 20 & 1.02 & 0.430\\
exc h/C & 400 & $D_s$&  1.1747(51) &  -0.004(17) &  -0.684(27) &  -0.0610(71) &  0.15(11)& 18 & 0.92 & 0.551\\
exc h/all & 400 & $D_s$&  1.1722(52) &  0.000(18) &  -0.679(28) &  -0.0572(80) &  0.10(12)& 14 & 0.68 & 0.795\\
phys mh cut & 400 & $D_s$&  1.1725(63) &  0.011(21) &  -0.683(32) &  -0.077(13) &  0.12(14)& 9 & 0.68 & 0.729\\
\hline
inc & 350 & $D_s$&  1.1740(51) &  0.003(19) &  -0.697(29) &  -0.0650(62) &  0.10(11)& 16 & 1.17 & 0.282\\
exc h/C & 350 & $D_s$&  1.1724(53) &  0.006(19) &  -0.699(30) &  -0.0606(72) &  0.11(11)& 14 & 1.05 & 0.400\\
exc h/all & 350 & $D_s$&  1.1703(54) &  0.008(19) &  -0.690(30) &  -0.0563(80) &  0.07(12)& 11 & 0.76 & 0.681\\
phys mh cut & 350 & $D_s$&  1.1704(65) &  0.020(22) &  -0.699(35) &  -0.077(14) &  0.08(14)& 7 & 0.66 & 0.705\\
\hline
inc & 350 & $\eta_c$&  1.1734(51) &  0.003(19) &  -0.697(29) &  -0.0781(75) &  0.09(11)& 16 & 1.22 & 0.243\\
exc h/C & 350 & $\eta_c$&  1.1717(53) &  0.007(19) &  -0.699(30) &  -0.0724(86) &  0.09(11)& 14 & 1.08 & 0.370\\
exc h/all & 350 & $\eta_c$&  1.1696(54) &  0.009(19) &  -0.689(30) &  -0.0678(97) &  0.06(12)& 11 & 0.82 & 0.615\\
phys mh cut & 350 & $\eta_c$&  1.1710(66) &  0.020(22) &  -0.701(35) &  -0.103(20) &  0.18(24)& 6 & 0.65 & 0.687\\
\hline
inc & 350 & $D$&  1.1737(51) &  0.004(19) &  -0.706(29) &  -0.0563(54) &  0.10(11)& 16 & 1.16 & 0.289\\
exc h/C & 350 & $D$&  1.1720(53) &  0.007(19) &  -0.707(30) &  -0.0522(62) &  0.10(11)& 14 & 1.05 & 0.402\\
exc h/all & 350 & $D$&  1.1699(55) &  0.009(19) &  -0.697(30) &  -0.0488(69) &  0.06(12)& 11 & 0.77 & 0.674\\
phys mh cut & 350 & $D$&  1.1701(65) &  0.021(22) &  -0.707(35) &  -0.068(12) &  0.07(14)& 7 & 0.61 & 0.747\\
\hline
inc & 330 & $D_s$&  1.1776(67) &  -0.010(24) &  -0.739(58) &  -0.0662(72) &  0.10(11)& 12 & 1.45 & 0.135\\
exc h/C & 330 & $D_s$&  1.1752(69) &  -0.004(24) &  -0.732(58) &  -0.0609(78) &  0.11(11)& 11 & 1.24 & 0.256\\
exc h/all & 330 & $D_s$&  1.1712(72) &  0.004(25) &  -0.703(61) &  -0.0552(88) &  0.07(12)& 8 & 0.94 & 0.479\\
phys mh cut & 330 & $D_s$&  1.1705(84) &  0.021(29) &  -0.697(69) &  -0.080(14) &  0.07(15)& 4 & 0.91 & 0.459\\
\hline
inc & 250 & $D_s$&  1.1784(68) &  -0.010(24) &  -0.83(12) &  -0.0681(77) & - & 9 & 1.70 & 0.084\\
exc h/C & 250 & $D_s$&  1.1758(70) &  -0.005(24) &  -0.84(12) &  -0.0623(83) & - & 8 & 1.47 & 0.161\\
exc h/all & 250 & $D_s$&  1.1719(73) &  0.003(25) &  -0.77(13) &  -0.0576(95) & - & 6 & 1.11 & 0.351\\
phys mh cut & 250 & $D_s$&  1.1712(85) &  0.022(30) &  -0.75(17) &  -0.085(16) & - & 3 & 0.97 & 0.407\\
\hline
\end{tabular}

    }
  \end{center}
  \caption{Results of the global fit for $f_{D_s}/f_D$.}
  \label{tab:GFfDsfD}
\end{table}

\begin{table}
  \begin{center}
    \resizebox{\textwidth}{!}{
      \begin{tabular}{|lll||l||llll|rcc|}
\hline
cut name & $m_\pi^\mathrm{max}/\mathrm{MeV}$ & $m_H$ & $f_{B_s}/f_B$ & $C_{CL}/\mathrm{GeV^2}$ & $C_\chi/\mathrm{GeV}^{-2}$ & $C_{H}/\mathrm{GeV}$  & $C_s$& d.o.f & $\chi^2/\mathrm{dof}$ & $p$ \\\hline\hline
inc & 450 & $B_s$&  1.1881(50) &  0.001(13) &  -0.556(15) &  -0.0611(57) &  0.12(10)& 28 & 2.41 & 0.000\\
exc h/C & 450 & $B_s$&  1.1870(56) &  0.002(14) &  -0.563(15) &  -0.0581(68) &  0.14(10)& 25 & 2.11 & 0.001\\
exc h/all & 450 & $B_s$&  1.1839(57) &  0.004(14) &  -0.556(16) &  -0.0544(76) &  0.10(11)& 20 & 2.17 & 0.002\\
exc l/C & 450 & $B_s$&  1.1903(68) &  -0.002(16) &  -0.591(18) &  -0.0586(78) &  0.18(10)& 25 & 1.41 & 0.084\\
exc l/all & 450 & $B_s$&  1.1872(77) &  0.022(21) &  -0.598(22) &  -0.068(13) &  -0.02(15)& 20 & 1.40 & 0.111\\
\hline
inc & 430 & $B_s$&  1.2022(55) &  -0.032(15) &  -0.627(20) &  -0.0661(61) &  0.23(11)& 24 & 1.36 & 0.112\\
exc h/C & 430 & $B_s$&  1.1990(61) &  -0.029(16) &  -0.624(21) &  -0.0614(70) &  0.24(11)& 22 & 1.33 & 0.138\\
exc h/all & 430 & $B_s$&  1.1957(63) &  -0.027(16) &  -0.617(21) &  -0.0573(79) &  0.21(11)& 17 & 1.30 & 0.180\\
exc l/C & 430 & $B_s$&  1.1956(73) &  -0.019(18) &  -0.616(22) &  -0.0596(78) &  0.23(11)& 22 & 1.40 & 0.098\\
exc l/all & 430 & $B_s$&  1.1942(82) &  0.007(22) &  -0.651(32) &  -0.071(13) &  0.03(15)& 17 & 1.27 & 0.200\\
\hline
inc & 400 & $B_s$&  1.1971(58) &  -0.006(17) &  -0.683(26) &  -0.0654(62) &  0.14(11)& 20 & 1.02 & 0.430\\
exc h/C & 400 & $B_s$&  1.1943(64) &  -0.004(17) &  -0.684(27) &  -0.0610(71) &  0.15(11)& 18 & 0.92 & 0.551\\
exc h/all & 400 & $B_s$&  1.1906(65) &  0.000(18) &  -0.679(28) &  -0.0572(80) &  0.10(12)& 14 & 0.68 & 0.795\\
exc l/C & 400 & $B_s$&  1.1937(75) &  0.000(19) &  -0.683(30) &  -0.0612(80) &  0.14(11)& 18 & 1.07 & 0.374\\
exc l/all & 400 & $B_s$&  1.1905(84) &  0.026(23) &  -0.693(36) &  -0.072(13) &  -0.06(15)& 14 & 1.05 & 0.394\\
\hline
inc & 350 & $B_s$&  1.1949(60) &  0.003(19) &  -0.697(29) &  -0.0650(62) &  0.10(11)& 16 & 1.17 & 0.282\\
exc h/C & 350 & $B_s$&  1.1919(66) &  0.006(19) &  -0.699(30) &  -0.0606(72) &  0.11(11)& 14 & 1.05 & 0.400\\
exc h/all & 350 & $B_s$&  1.1884(67) &  0.008(19) &  -0.690(30) &  -0.0563(80) &  0.07(12)& 11 & 0.76 & 0.681\\
exc l/C & 350 & $B_s$&  1.1918(76) &  0.009(20) &  -0.704(33) &  -0.0612(81) &  0.10(11)& 14 & 1.24 & 0.241\\
exc l/all & 350 & $B_s$&  1.1885(87) &  0.033(24) &  -0.704(37) &  -0.071(13) &  -0.09(16)& 11 & 1.23 & 0.259\\
\hline
inc & 350 & $\eta_b$&  1.1913(58) &  0.003(19) &  -0.697(29) &  -0.0781(75) &  0.09(11)& 16 & 1.22 & 0.243\\
exc h/C & 350 & $\eta_b$&  1.1883(63) &  0.007(19) &  -0.699(30) &  -0.0724(86) &  0.09(11)& 14 & 1.08 & 0.370\\
exc h/all & 350 & $\eta_b$&  1.1851(65) &  0.009(19) &  -0.689(30) &  -0.0678(97) &  0.06(12)& 11 & 0.82 & 0.615\\
exc l/C & 350 & $\eta_b$&  1.1878(73) &  0.011(20) &  -0.701(33) &  -0.0728(97) &  0.09(11)& 14 & 1.29 & 0.207\\
exc l/all & 350 & $\eta_b$&  1.1853(83) &  0.035(24) &  -0.702(37) &  -0.089(16) &  -0.10(16)& 11 & 1.28 & 0.229\\
\hline
inc & 350 & $B$&  1.1932(60) &  0.004(19) &  -0.706(29) &  -0.0563(54) &  0.10(11)& 16 & 1.16 & 0.289\\
exc h/C & 350 & $B$&  1.1901(65) &  0.007(19) &  -0.707(30) &  -0.0522(62) &  0.10(11)& 14 & 1.05 & 0.402\\
exc h/all & 350 & $B$&  1.1868(66) &  0.009(19) &  -0.697(30) &  -0.0488(69) &  0.06(12)& 11 & 0.77 & 0.674\\
exc l/C & 350 & $B$&  1.1899(75) &  0.010(20) &  -0.709(33) &  -0.0527(70) &  0.10(11)& 14 & 1.24 & 0.236\\
exc l/all & 350 & $B$&  1.1869(85) &  0.034(24) &  -0.710(38) &  -0.063(11) &  -0.10(16)& 11 & 1.23 & 0.259\\
\hline
inc & 330 & $B_s$&  1.1989(78) &  -0.010(24) &  -0.739(58) &  -0.0662(72) &  0.10(11)& 12 & 1.45 & 0.135\\
exc h/C & 330 & $B_s$&  1.1948(82) &  -0.004(24) &  -0.732(58) &  -0.0609(78) &  0.11(11)& 11 & 1.24 & 0.256\\
exc h/all & 330 & $B_s$&  1.1890(85) &  0.004(25) &  -0.703(61) &  -0.0552(88) &  0.07(12)& 8 & 0.94 & 0.479\\
exc l/C & 330 & $B_s$&  1.1948(86) &  -0.001(25) &  -0.733(58) &  -0.0620(82) &  0.10(11)& 11 & 1.48 & 0.130\\
exc l/all & 330 & $B_s$&  1.193(10) &  0.022(30) &  -0.751(84) &  -0.074(13) &  -0.10(16)& 8 & 1.54 & 0.137\\
\hline
inc & 250 & $B_s$&  1.2002(80) &  -0.010(24) &  -0.83(12) &  -0.0681(77) & - & 9 & 1.70 & 0.084\\
exc h/C & 250 & $B_s$&  1.1958(84) &  -0.005(24) &  -0.84(12) &  -0.0623(83) & - & 8 & 1.47 & 0.161\\
exc h/all & 250 & $B_s$&  1.1904(88) &  0.003(25) &  -0.77(13) &  -0.0576(95) & - & 6 & 1.11 & 0.351\\
exc l/C & 250 & $B_s$&  1.1962(90) &  -0.002(26) &  -0.83(12) &  -0.0638(89) & - & 8 & 1.81 & 0.071\\
exc l/all & 250 & $B_s$&  1.192(10) &  0.022(30) &  -0.66(17) &  -0.072(14) & - & 6 & 2.02 & 0.059\\
\hline
\end{tabular}

    }
  \end{center}
  \caption{Results of the global fit for $f_{B_s}/f_B$.}
  \label{tab:GFfBsfB}
\end{table}

\begin{table}
  \begin{center}
    \resizebox{\textwidth}{!}{
      \begin{tabular}{|lll||l||llll|rcc|}
\hline
cut name & $m_\pi^\mathrm{max}/\mathrm{MeV}$ & $m_H$ & $B_{B_s}/B_{B_d}$ & $C_{CL}/\mathrm{GeV^2}$ & $C_\chi/\mathrm{GeV}^{-2}$ & $C_{H}/\mathrm{GeV}$  & $C_s$& d.o.f & $\chi^2/\mathrm{dof}$ & $p$ \\\hline\hline
inc & 450 & $B_s$&  1.0040(31) &  0.041(11) &  -0.053(12) &  0.0205(50) &  0.098(72)& 28 & 0.89 & 0.625\\
exc h/C & 450 & $B_s$&  1.0040(31) &  0.043(11) &  -0.052(12) &  0.0196(51) &  0.094(72)& 25 & 0.94 & 0.553\\
exc h/all & 450 & $B_s$&  1.0049(32) &  0.046(11) &  -0.053(13) &  0.0161(58) &  0.066(76)& 20 & 1.01 & 0.449\\
exc l/C & 450 & $B_s$&  1.0045(37) &  0.035(13) &  -0.046(14) &  0.0205(55) &  0.095(73)& 25 & 0.95 & 0.534\\
exc l/all & 450 & $B_s$&  1.0033(42) &  0.039(14) &  -0.038(18) &  0.0175(80) &  0.059(84)& 20 & 1.09 & 0.352\\
\hline
inc & 430 & $B_s$&  1.0011(41) &  0.051(14) &  -0.042(15) &  0.0209(55) &  0.069(76)& 24 & 0.97 & 0.504\\
exc h/C & 430 & $B_s$&  1.0009(41) &  0.052(14) &  -0.042(15) &  0.0209(55) &  0.066(76)& 22 & 1.00 & 0.458\\
exc h/all & 430 & $B_s$&  1.0023(42) &  0.054(14) &  -0.045(15) &  0.0170(64) &  0.043(80)& 17 & 1.13 & 0.319\\
exc l/C & 430 & $B_s$&  1.0024(46) &  0.044(18) &  -0.041(15) &  0.0205(56) &  0.075(77)& 22 & 1.03 & 0.415\\
exc l/all & 430 & $B_s$&  0.9992(56) &  0.054(19) &  -0.023(22) &  0.0172(84) &  0.009(93)& 17 & 1.18 & 0.273\\
\hline
inc & 400 & $B_s$&  1.0007(43) &  0.054(14) &  -0.067(19) &  0.0225(63) &  0.064(76)& 20 & 0.90 & 0.593\\
exc h/C & 400 & $B_s$&  1.0005(43) &  0.055(14) &  -0.065(19) &  0.0224(63) &  0.062(76)& 18 & 0.94 & 0.531\\
exc h/all & 400 & $B_s$&  1.0026(45) &  0.058(14) &  -0.069(20) &  0.0166(77) &  0.037(81)& 14 & 1.08 & 0.371\\
exc l/C & 400 & $B_s$&  1.0018(48) &  0.049(18) &  -0.068(21) &  0.0224(63) &  0.073(77)& 18 & 0.97 & 0.486\\
exc l/all & 400 & $B_s$&  0.9990(56) &  0.055(19) &  -0.068(30) &  0.0239(91) &  0.031(94)& 14 & 1.04 & 0.405\\
\hline
inc & 350 & $B_s$&  0.9984(45) &  0.062(15) &  -0.087(26) &  0.0246(66) &  0.033(80)& 16 & 0.87 & 0.607\\
exc h/C & 350 & $B_s$&  0.9982(45) &  0.063(15) &  -0.085(26) &  0.0246(66) &  0.033(80)& 14 & 0.93 & 0.525\\
exc h/all & 350 & $B_s$&  1.0003(47) &  0.066(16) &  -0.092(27) &  0.0185(83) &  0.004(85)& 11 & 1.05 & 0.402\\
exc l/C & 350 & $B_s$&  0.9987(50) &  0.061(20) &  -0.099(30) &  0.0252(67) &  0.035(81)& 14 & 0.92 & 0.535\\
exc l/all & 350 & $B_s$&  0.9962(59) &  0.065(21) &  -0.101(39) &  0.029(10) &  0.002(96)& 11 & 1.10 & 0.359\\
\hline
inc & 350 & $\eta_b$&  0.9994(43) &  0.062(15) &  -0.086(26) &  0.0306(82) &  0.036(80)& 16 & 0.89 & 0.585\\
exc h/C & 350 & $\eta_b$&  0.9993(43) &  0.062(15) &  -0.084(26) &  0.0305(82) &  0.035(80)& 14 & 0.95 & 0.504\\
exc h/all & 350 & $\eta_b$&  1.0011(45) &  0.066(16) &  -0.092(27) &  0.023(10) &  0.006(86)& 11 & 1.06 & 0.386\\
exc l/C & 350 & $\eta_b$&  0.9998(49) &  0.061(20) &  -0.099(30) &  0.0313(84) &  0.038(81)& 14 & 0.94 & 0.509\\
exc l/all & 350 & $\eta_b$&  0.9972(58) &  0.064(21) &  -0.101(39) &  0.037(13) &  0.003(96)& 11 & 1.12 & 0.339\\
\hline
inc & 350 & $B$&  0.9988(44) &  0.062(15) &  -0.084(26) &  0.0218(58) &  0.034(80)& 16 & 0.88 & 0.598\\
exc h/C & 350 & $B$&  0.9987(44) &  0.062(15) &  -0.082(26) &  0.0217(58) &  0.034(80)& 14 & 0.94 & 0.516\\
exc h/all & 350 & $B$&  1.0006(46) &  0.066(16) &  -0.090(27) &  0.0163(74) &  0.005(85)& 11 & 1.05 & 0.397\\
exc l/C & 350 & $B$&  0.9991(50) &  0.061(20) &  -0.097(30) &  0.0223(59) &  0.036(81)& 14 & 0.93 & 0.526\\
exc l/all & 350 & $B$&  0.9966(58) &  0.064(21) &  -0.099(38) &  0.0261(92) &  0.003(96)& 11 & 1.10 & 0.353\\
\hline
inc & 330 & $B_s$&  0.9992(60) &  0.056(22) &  -0.095(39) &  0.0259(67) &  0.042(84)& 12 & 0.97 & 0.476\\
exc h/C & 330 & $B_s$&  0.9993(60) &  0.055(23) &  -0.095(39) &  0.0260(67) &  0.043(84)& 11 & 1.05 & 0.401\\
exc h/all & 330 & $B_s$&  1.0017(63) &  0.057(23) &  -0.105(40) &  0.0203(85) &  0.021(89)& 8 & 1.28 & 0.249\\
exc l/C & 330 & $B_s$&  0.9972(76) &  0.066(32) &  -0.094(39) &  0.0264(68) &  0.029(88)& 11 & 1.04 & 0.408\\
exc l/all & 330 & $B_s$&  0.9943(96) &  0.068(38) &  -0.097(59) &  0.032(11) &  -0.01(12)& 8 & 1.30 & 0.239\\
\hline
inc & 250 & $B_s$&  0.9990(62) &  0.056(22) &  -0.135(96) &  0.0266(75) & - & 9 & 0.96 & 0.473\\
exc h/C & 250 & $B_s$&  0.9991(62) &  0.055(23) &  -0.137(96) &  0.0267(75) & - & 8 & 1.06 & 0.384\\
exc h/all & 250 & $B_s$&  1.0005(65) &  0.055(23) &  -0.13(10) &  0.0239(90) & - & 6 & 1.34 & 0.237\\
exc l/C & 250 & $B_s$&  0.9968(78) &  0.066(32) &  -0.12(10) &  0.0273(76) & - & 8 & 1.05 & 0.394\\
exc l/all & 250 & $B_s$&  0.9929(99) &  0.064(39) &  -0.10(15) &  0.039(16) & - & 6 & 1.25 & 0.278\\
\hline
\end{tabular}

    }
  \end{center}
  \caption{Results of the global fit for $B_{B_s}/B_{B_d}$.}
  \label{tab:GFBBsBB}
\end{table}

\begin{table}
  \begin{center}
    \resizebox{\textwidth}{!}{
      \begin{tabular}{|lll||l||llll|rcc|}
\hline
cut name & $m_\pi^\mathrm{max}/\mathrm{MeV}$ & $m_H$ & $\xi$ & $C_{CL}/\mathrm{GeV^2}$ & $C_\chi/\mathrm{GeV}^{-2}$ & $C_{H}/\mathrm{GeV}$  & $C_s$& d.o.f & $\chi^2/\mathrm{dof}$ & $p$ \\\hline\hline
inc & 450 & $B_s$&  1.1788(65) &  0.035(17) &  -0.591(17) &  -0.0282(92) &  0.25(13)& 28 & 1.16 & 0.256\\
exc h/C & 450 & $B_s$&  1.1797(70) &  0.033(17) &  -0.592(18) &  -0.029(10) &  0.25(13)& 25 & 1.17 & 0.251\\
exc h/all & 450 & $B_s$&  1.1794(72) &  0.033(17) &  -0.589(18) &  -0.029(11) &  0.26(13)& 20 & 1.21 & 0.233\\
exc l/C & 450 & $B_s$&  1.1882(82) &  0.024(19) &  -0.616(21) &  -0.039(11) &  0.26(13)& 25 & 1.03 & 0.422\\
exc l/all & 450 & $B_s$&  1.1936(92) &  0.045(24) &  -0.643(26) &  -0.063(17) &  0.24(17)& 20 & 1.06 & 0.384\\
\hline
inc & 430 & $B_s$&  1.1929(82) &  0.008(20) &  -0.637(25) &  -0.039(11) &  0.32(14)& 24 & 0.99 & 0.471\\
exc h/C & 430 & $B_s$&  1.1914(83) &  0.006(21) &  -0.635(25) &  -0.036(11) &  0.33(14)& 22 & 1.00 & 0.462\\
exc h/all & 430 & $B_s$&  1.1913(85) &  0.007(21) &  -0.632(25) &  -0.036(12) &  0.33(14)& 17 & 1.02 & 0.429\\
exc l/C & 430 & $B_s$&  1.1922(88) &  0.010(22) &  -0.635(25) &  -0.039(11) &  0.31(14)& 22 & 1.07 & 0.371\\
exc l/all & 430 & $B_s$&  1.2013(99) &  0.026(27) &  -0.697(37) &  -0.064(18) &  0.31(18)& 17 & 0.95 & 0.516\\
\hline
inc & 400 & $B_s$&  1.1909(84) &  0.038(22) &  -0.723(36) &  -0.046(11) &  0.16(14)& 20 & 0.60 & 0.917\\
exc h/C & 400 & $B_s$&  1.1895(85) &  0.036(22) &  -0.720(36) &  -0.043(12) &  0.18(14)& 18 & 0.58 & 0.916\\
exc h/all & 400 & $B_s$&  1.1907(87) &  0.039(23) &  -0.720(37) &  -0.047(13) &  0.15(15)& 14 & 0.44 & 0.961\\
exc l/C & 400 & $B_s$&  1.1889(91) &  0.043(24) &  -0.729(38) &  -0.044(12) &  0.17(14)& 18 & 0.64 & 0.869\\
exc l/all & 400 & $B_s$&  1.195(10) &  0.053(28) &  -0.760(42) &  -0.062(18) &  0.21(18)& 14 & 0.50 & 0.935\\
\hline
inc & 350 & $B_s$&  1.1887(85) &  0.051(24) &  -0.749(39) &  -0.048(12) &  0.09(15)& 16 & 0.56 & 0.912\\
exc h/C & 350 & $B_s$&  1.1874(86) &  0.049(24) &  -0.746(39) &  -0.045(12) &  0.11(15)& 14 & 0.54 & 0.910\\
exc h/all & 350 & $B_s$&  1.1883(88) &  0.052(24) &  -0.745(40) &  -0.049(13) &  0.08(15)& 11 & 0.34 & 0.976\\
exc l/C & 350 & $B_s$&  1.1859(93) &  0.059(26) &  -0.761(41) &  -0.045(12) &  0.09(15)& 14 & 0.59 & 0.878\\
exc l/all & 350 & $B_s$&  1.192(11) &  0.064(29) &  -0.776(44) &  -0.060(18) &  0.16(19)& 11 & 0.50 & 0.905\\
\hline
inc & 350 & $\eta_b$&  1.1860(82) &  0.052(24) &  -0.747(39) &  -0.058(14) &  0.08(15)& 16 & 0.59 & 0.892\\
exc h/C & 350 & $\eta_b$&  1.1848(82) &  0.050(24) &  -0.745(39) &  -0.054(15) &  0.10(15)& 14 & 0.57 & 0.888\\
exc h/all & 350 & $\eta_b$&  1.1858(84) &  0.053(24) &  -0.743(40) &  -0.059(16) &  0.08(16)& 11 & 0.37 & 0.969\\
exc l/C & 350 & $\eta_b$&  1.1834(89) &  0.060(26) &  -0.759(41) &  -0.054(15) &  0.08(15)& 14 & 0.62 & 0.852\\
exc l/all & 350 & $\eta_b$&  1.190(10) &  0.066(29) &  -0.775(44) &  -0.077(23) &  0.16(19)& 11 & 0.49 & 0.910\\
\hline
inc & 350 & $B$&  1.1874(84) &  0.052(24) &  -0.754(40) &  -0.042(10) &  0.09(15)& 16 & 0.57 & 0.908\\
exc h/C & 350 & $B$&  1.1861(85) &  0.050(24) &  -0.751(40) &  -0.039(10) &  0.10(15)& 14 & 0.55 & 0.904\\
exc h/all & 350 & $B$&  1.1870(86) &  0.053(24) &  -0.749(40) &  -0.043(11) &  0.08(16)& 11 & 0.35 & 0.975\\
exc l/C & 350 & $B$&  1.1847(91) &  0.060(26) &  -0.765(42) &  -0.039(11) &  0.09(15)& 14 & 0.60 & 0.868\\
exc l/all & 350 & $B$&  1.191(10) &  0.065(29) &  -0.781(44) &  -0.054(16) &  0.16(19)& 11 & 0.50 & 0.906\\
\hline
inc & 330 & $B_s$&  1.193(10) &  0.035(30) &  -0.809(78) &  -0.049(12) &  0.09(15)& 12 & 0.63 & 0.822\\
exc h/C & 330 & $B_s$&  1.192(10) &  0.033(30) &  -0.806(78) &  -0.046(12) &  0.11(15)& 11 & 0.60 & 0.832\\
exc h/all & 330 & $B_s$&  1.192(10) &  0.041(31) &  -0.789(81) &  -0.050(13) &  0.08(16)& 8 & 0.39 & 0.929\\
exc l/C & 330 & $B_s$&  1.190(11) &  0.045(33) &  -0.805(78) &  -0.045(13) &  0.09(15)& 11 & 0.64 & 0.794\\
exc l/all & 330 & $B_s$&  1.200(13) &  0.035(40) &  -0.867(94) &  -0.060(18) &  0.19(19)& 8 & 0.44 & 0.895\\
\hline
inc & 250 & $B_s$&  1.193(10) &  0.035(30) &  -0.90(16) &  -0.047(12) & - & 9 & 0.67 & 0.734\\
exc h/C & 250 & $B_s$&  1.191(10) &  0.033(30) &  -0.91(16) &  -0.043(13) & - & 8 & 0.63 & 0.754\\
exc h/all & 250 & $B_s$&  1.193(11) &  0.041(31) &  -0.86(17) &  -0.051(14) & - & 6 & 0.36 & 0.902\\
exc l/C & 250 & $B_s$&  1.189(11) &  0.046(33) &  -0.90(16) &  -0.043(14) & - & 8 & 0.69 & 0.703\\
exc l/all & 250 & $B_s$&  1.199(13) &  0.033(40) &  -1.06(22) &  -0.055(21) & - & 6 & 0.54 & 0.777\\
\hline
\end{tabular}

    }
  \end{center}
  \caption{Results of the global fit for $\xi$.}
  \label{tab:GFxi}
\end{table}

\begin{figure}
  \begin{center}
    \includegraphics[width=.49\textwidth]{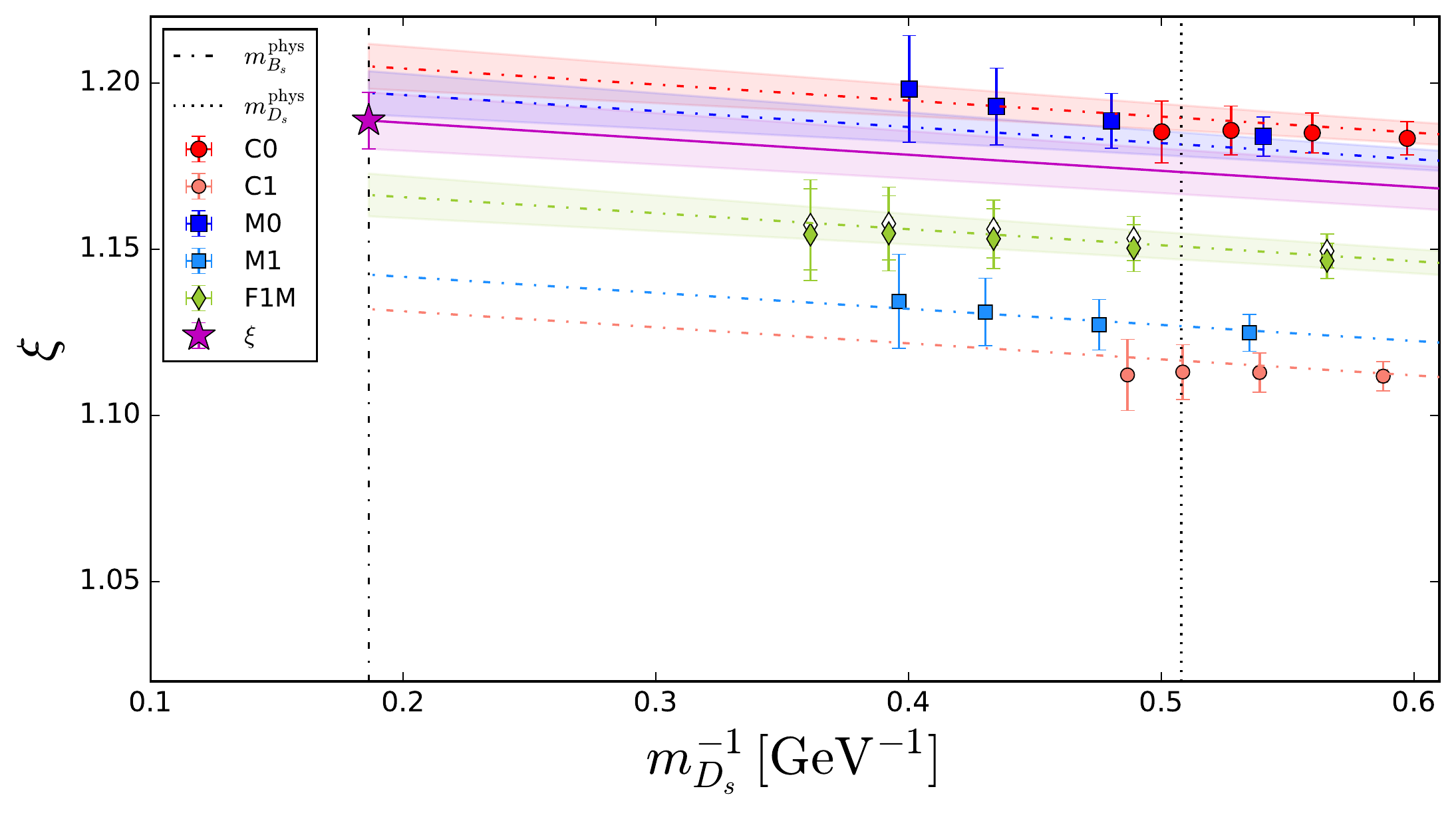}
    \includegraphics[width=.49\textwidth]{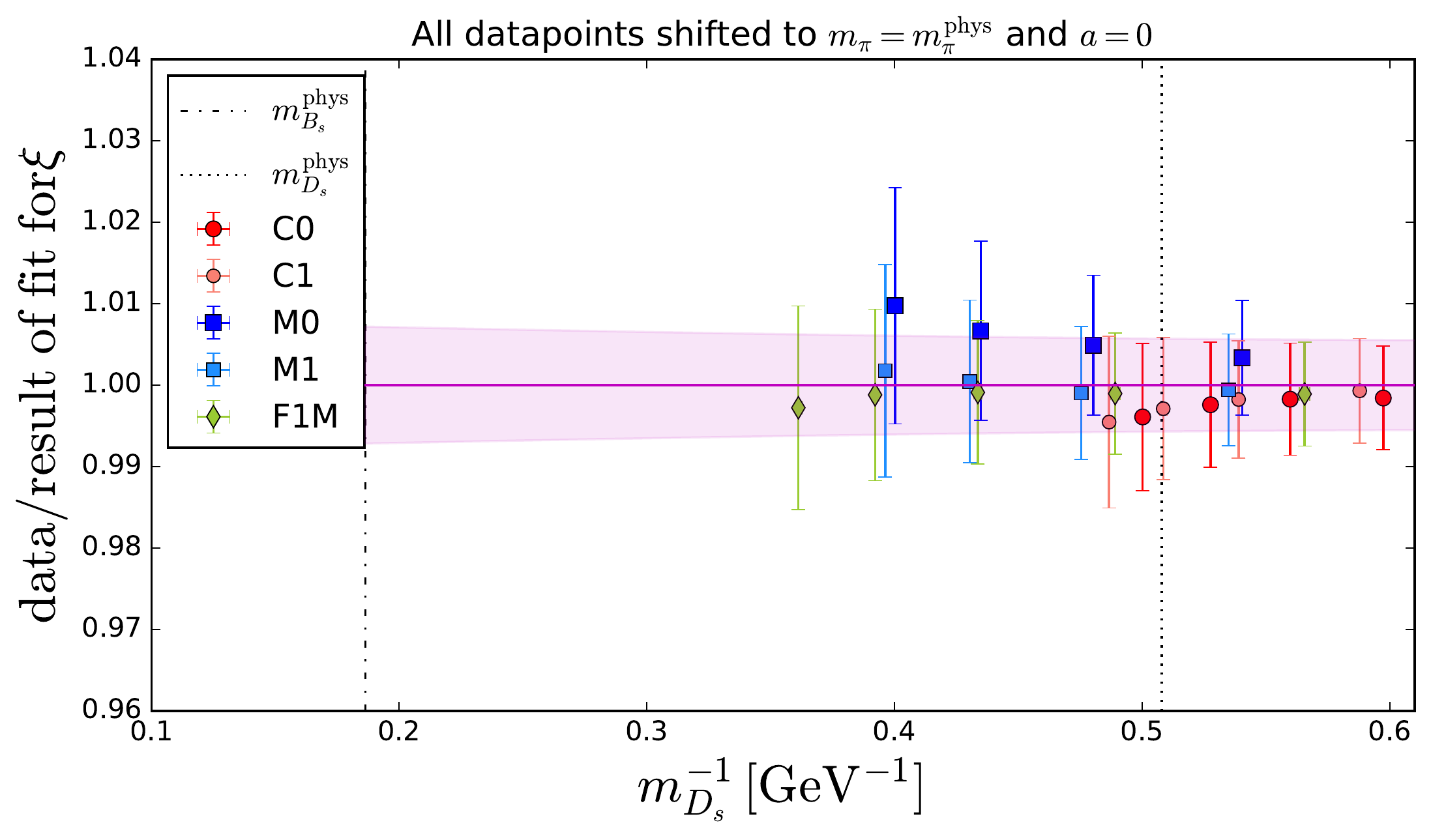}
  \end{center}
  \caption{Analogous plots to Figure \ref{fig:GFdecrat_notproj} for $\xi$.}
  \label{fig:GFbagrat_notproj}
  \label{fig:GFxi_notproj}
\end{figure}

\begin{figure}
  \begin{center}
    \includegraphics[width=.49\textwidth]{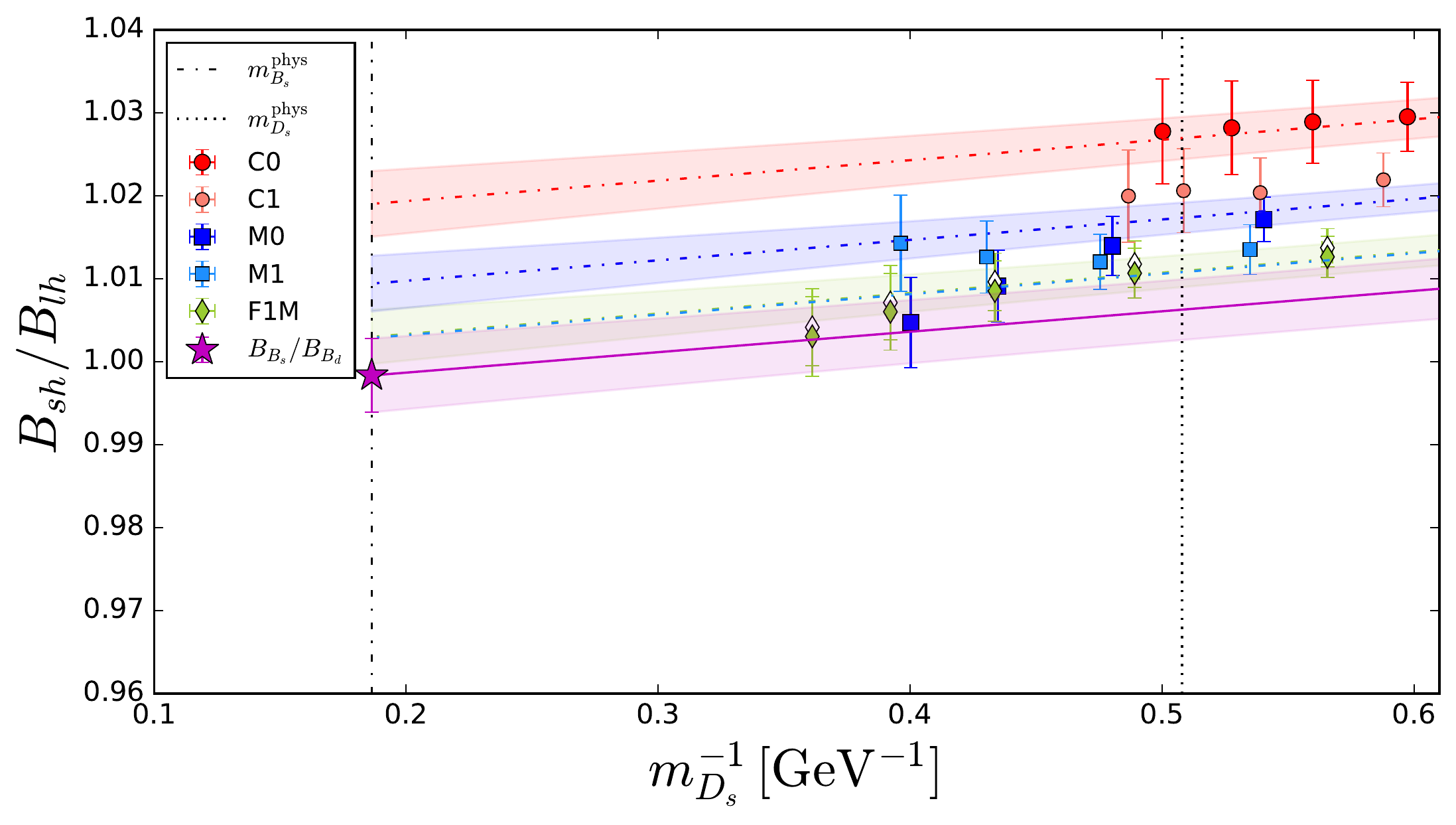}
    \includegraphics[width=.49\textwidth]{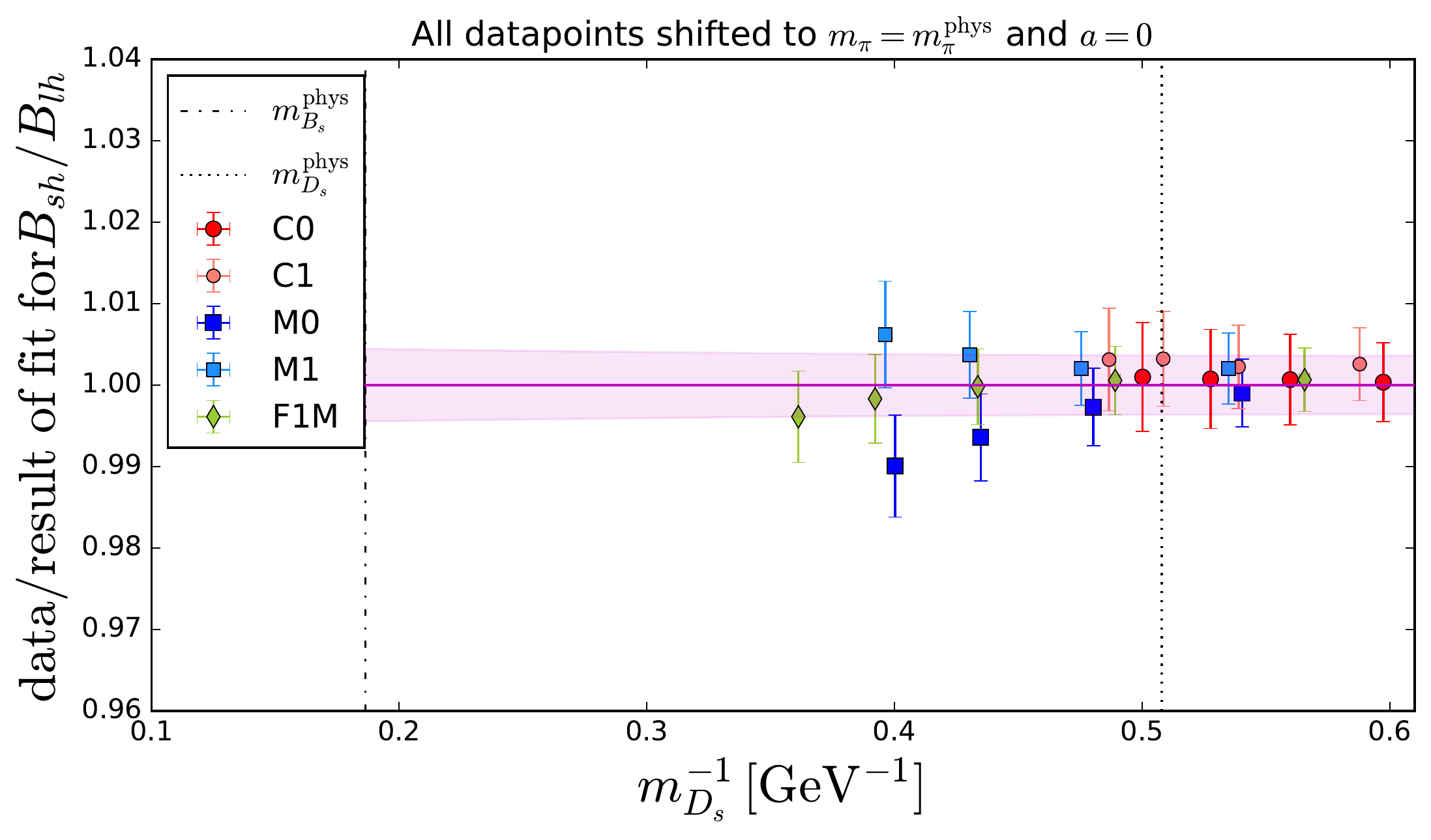}
  \end{center}
  \caption{Analogous plots to Figure \ref{fig:GFdecrat_notproj} for $B_{B_s}/B_{B_d}$.}
  \label{fig:GFbagrat_notproj}
  \label{fig:GFxi_notproj}
\end{figure}

\begin{figure}
  \begin{center}
    \includegraphics[width=.45\textwidth]{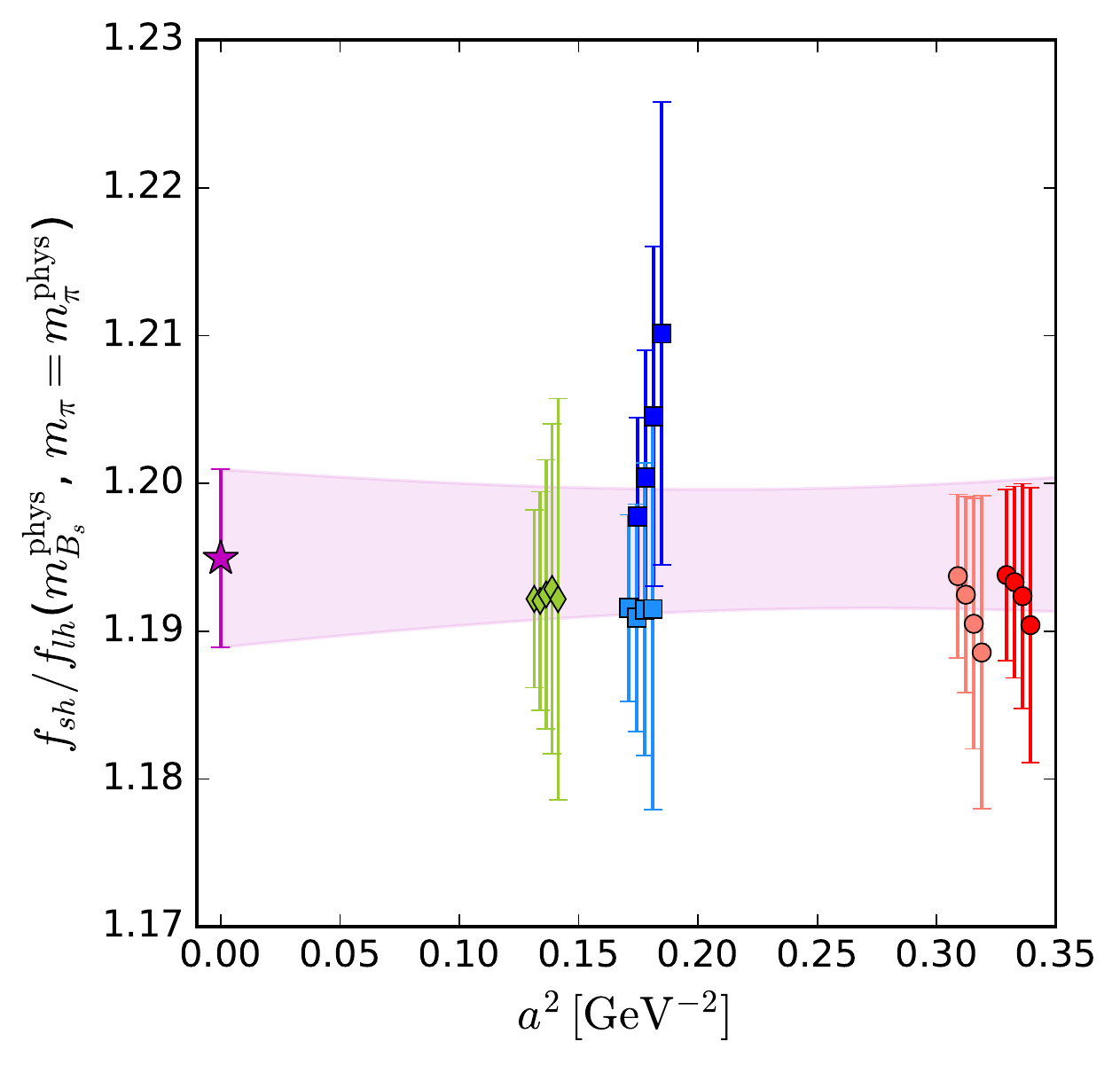}
    \includegraphics[width=.45\textwidth]{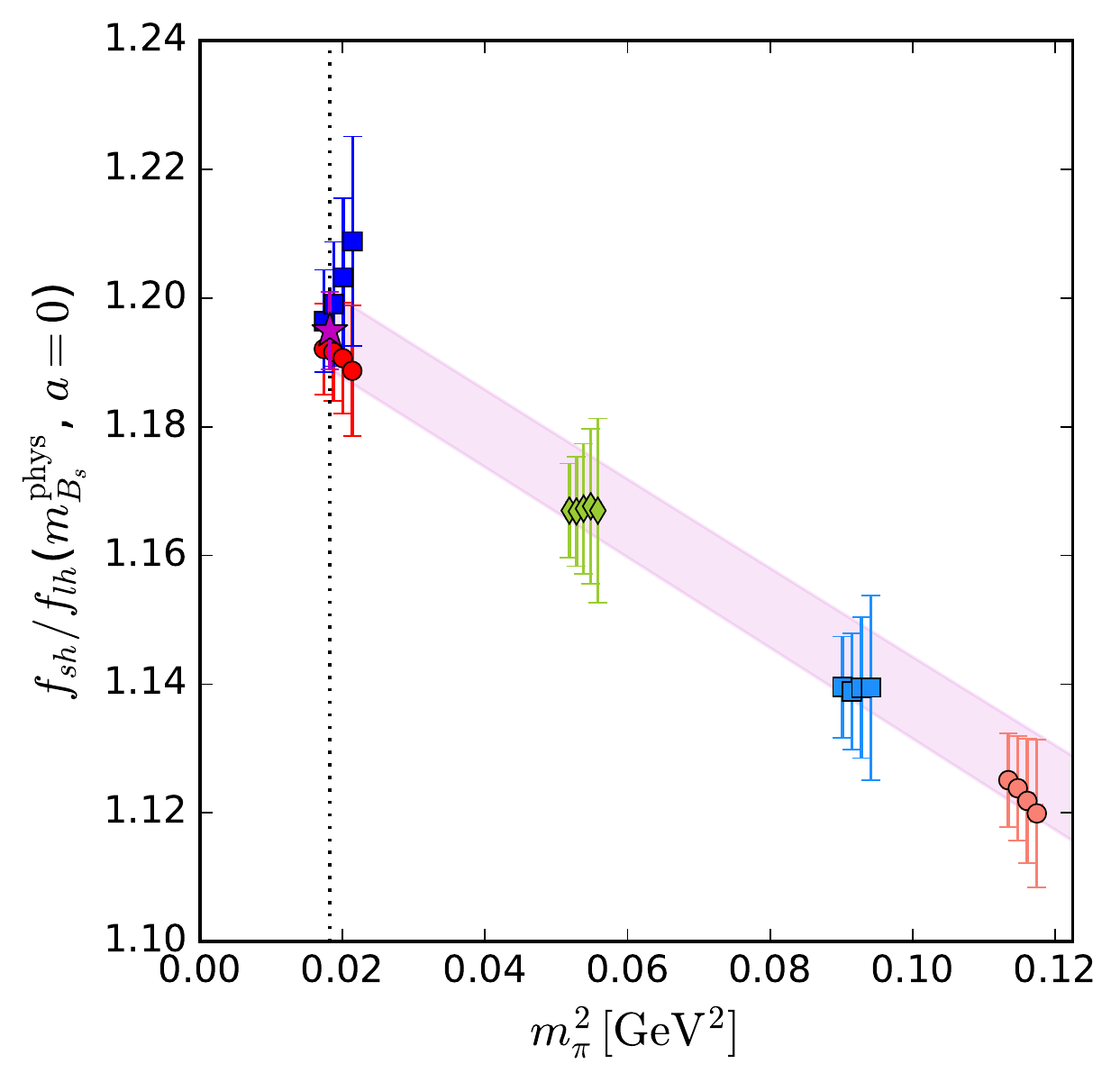}\\
  \end{center}
  \caption{The analogous plots to Figure \ref{fig:GFdecratprojection},
    displaying the chiral and continuum limit behaviour for the ratio $f_{B_s}/f_B$.}
  \label{fig:GFdecratprojection_atb}
\end{figure}

\clearpage

{\small
  \bibliographystyle{JHEP}
  \bibliography{xi}}

\end{document}